\documentclass[12pt, letterpaper]{JHEP3}

\usepackage{array}

\usepackage{amssymb, amsmath, amsopn, amsthm}

\usepackage{epsfig}

\newcommand{\eref}[1]{(\ref{#1})}

\newcommand{\fref}[1]{Figure~\ref{#1}}
\newcommand{\cref}[1]{Chapter~\ref{#1}}
\newcommand{\beq}{\begin{equation}}
\newcommand{\eeq}{\end{equation}}
\newcommand{\ba}{\begin{array}}
\newcommand{\ea}{\end{array}}
\newcommand{\bcenter}{\begin{center}}
\newcommand{\ecenter}{\end{center}}

\def\IB{\relax\hbox{$\inbar\kern-.3em{\rm B}$}}
\def\IC{\relax\hbox{$\inbar\kern-.3em{\rm C}$}}
\def\ID{\relax\hbox{$\inbar\kern-.3em{\rm D}$}}
\def\IE{\relax\hbox{$\inbar\kern-.3em{\rm E}$}}
\def\IF{\relax\hbox{$\inbar\kern-.3em{\rm F}$}}
\def\IG{\relax\hbox{$\inbar\kern-.3em{\rm G}$}}
\def\IGa{\relax\hbox{${\rm I}\kern-.18em\Gamma$}}
\def\IH{\relax{\rm I\kern-.18em H}}
\def\IK{\relax{\rm I\kern-.18em K}}
\def\IL{\relax{\rm I\kern-.18em L}}
\def\IP{\relax{\rm I\kern-.18em P}}
\def\IR{\relax{\rm I\kern-.18em R}}
\def\IZ{\relax\ifmmode\mathchoice
{\hbox{\cmss Z\kern-.4em Z}}{\hbox{\cmss Z\kern-.4em Z}}
{\lower.9pt\hbox{\cmsss Z\kern-.4em Z}}
{\lower1.2pt\hbox{\cmsss Z\kern-.4em Z}}\else{\cmss Z\kern-.4em Z}\fi}
\def\II{\relax{\rm I\kern-.18em I}}

\def\sCC{{\kern 0.27em\vrule height1.45ex width0.03em depth0em
          \kern-0.30em\rm C}}
\def\C{{\mathchoice
  {\sCC}
  {\sCC}
  {\kern 0.225em \vrule height1.05ex width0.025em depth0em \kern-0.25em \rm C}
  {\kern 0.180em \vrule height0.78ex width0.02em depth0em \kern-0.2em \rm C}
        }}
\def\sHH{{\rm I\kern-.16em{}H}}
\def\H{{\mathchoice
  {\sHH}
  {\sHH}
  {\rm I\kern-.13em{}H}
  {\rm I\kern-.13em{}H} }}
\def\sNN{{\rm I\kern-.16em{}N}}
\def\N{{\mathchoice
  {\sNN}
  {\sNN}
  {\rm I\kern-.12em{}N}
  {\rm I\kern-.10em{}N} }}
\def\sPP{{\rm I\kern-.16em{}P}}
\def\P{{\mathchoice
  {\sPP}
  {\sPP}
  {\rm I\kern-.12em{}P}
  {\rm I\kern-.10em{}P} }}
\def\sQQ{{\kern 0.27em \vrule height1.45ex width0.03em depth0em
          \kern-0.30em \rm Q}}
\def\Q{{\mathchoice
        {\sQQ}
        {\sQQ}
  {\kern 0.225em \vrule height1.05ex width0.025em depth0em \kern-0.25em \rm Q}
  {\kern 0.180em \vrule height0.78ex width0.020em depth0em \kern-0.20em \rm Q}
        }}
\def\sRR{{\rm I\kern-0.16em{}R}}
\def\R{{\mathchoice
  {\sRR}
  {\sRR}
  {\rm I\kern-0.12em{}R}
  {\rm I\kern-0.10em{}R} }}
\def\sZZ{{\rm Z\kern-0.32em{}Z}}
\def\Z{{\mathchoice
  {\sZZ}
  {\sZZ} 
  {\rm Z\kern-0.3em{}Z}     
  {\rm Z\kern-0.25em{}Z} }}  
\def\ZZZ{{\rm Z\kern-0.24em{}Z}}
\def\sII{{\rm I\kern-0.16em{}I}}
\def\I{{\mathchoice
  {\sII}
  {\sII}
  {\rm I\kern-0.12em{}I}
  {\rm I\kern-0.10em{}I} }}

\def\inbar{\,\vrule height1.5ex width.4pt depth0pt}
\font\cmss=cmss10 \font\cmsss=cmss10 at 7pt

\def\smiley{\hbox{\large$\bigcirc$\hspace{-0.80em}\raise.2ex
\hbox{$\cdot\cdot$}\kern-.61em\lower.2ex\hbox{\scriptsize$\smile$}}\ }
\def\frowny{\hbox{\large$\bigcirc$\hspace{-0.80em}\raise.2ex
\hbox{$\cdot\cdot$}\kern-.635em\lower.2ex\hbox{\scriptsize$\frown$}}\ }

\def\I{{\rlap{1} \hskip 1.6pt \hbox{1}}}

\makeatletter
\let\hangafter\@hangfrom
\makeatother

\def\makeatletter{\catcode`\@=11}
\makeatletter
\def\mathbox#1{\hbox{$\m@th#1$}}%
\def\math@ccstyles#1#2#3#4#5#6#7{{\leavevmode
     \setbox0\mathbox{#6#7}%
     \setbox2\mathbox{#4#5}%
     \dimen@ #3%
     \baselineskip\z@\lineskiplimit#1\lineskip\z@
     \vbox{\ialign{##\crcr
            \hfil \kern #2\box2 \hfil\crcr
            \noalign{\kern\dimen@}%
            \hfil\box0\hfil\crcr}}}}
\def\mathaccstyles{\math@ccstyles\maxdimen}
\def\maththroughstyles{\math@ccstyles{-\maxdimen}}
\def\unity%
{\maththroughstyles{.45\ht0}\z@\displaystyle {\mathchar"006C}\displaystyle 1}

\title{\hspace{2.65cm} Bipartite Field Theories: from \\
\hspace{.45cm} D-Brane Probes to Scattering Amplitudes}

\author{Sebasti\'an Franco$^{1,2}$

\\

\vspace{0.2cm}

~\\
$^1$ Theory Group, SLAC National Accelerator Laboratory \\
Menlo Park, CA 94309, USA \\
\vspace{0.25cm}

$^2$ Institute for Particle Physics Phenomenology, Department of Physics \\
Durham University, Durham DH1 3LE, United Kingdom \\
\vspace{0.3cm}

\email{sfranco@slac.stanford.edu}\\

}

\abstract{We introduce and initiate the investigation of a general class of 4d, $\mathcal{N} = 1$ quiver gauge theories whose Lagrangian is defined by a bipartite graph on a Riemann surface, with or without boundaries. We refer to such class of theories as Bipartite Field Theories (BFTs). BFTs underlie a wide spectrum of interesting physical systems, including: D3-branes probing toric Calabi-Yau 3-folds, their mirror configurations of D6-branes, cluster integrable systems in (0+1) dimensions and leading singularities in scattering amplitudes for $\mathcal{N}=4$ SYM. While our discussion is fully general, we focus on models that are relevant for scattering amplitudes. We investigate the BFT perspective on graph modifications, the emergence of Calabi-Yau manifolds (which arise as the master and moduli spaces of BFTs), the translation between square moves in the graph and Seiberg duality and the identification of dual theories by means of the underlying Calabi-Yaus, the phenomenon of loop reduction and the interpretation of the boundary operator for cells in the positive Grassmannian as higgsing in the BFT. We develop a technique based on generalized Kasteleyn matrices that permits an efficient determination of the Calabi-Yau geometries associated to arbitrary graphs. Our techniques allow us to go beyond the planar limit by both increasing the number of boundaries of the graphs and the genus of the underlying Riemann surface. Our investigation suggests a central role for Calabi-Yau manifolds in the context of leading singularities, whose full scope is yet to be uncovered.
}

\preprint{SLAC-PUB-15127 \\ IPPP/12/47 \\ DCPT/12/94}

\def\be{\begin{equation}}

\def\ee{\end{equation}}

\def\bea{\begin{eqnarray}}

\def\eea{\end{eqnarray}}

\begin{document}

\tableofcontents

\section{Introduction}

\label{section_introduction}

In this paper we introduce and initiate the investigation of a general class of $4d$, $\mathcal{N}=1$ quiver gauge theories whose Lagrangian is defined by a bipartite graph on a Riemann surface, with or without boundaries. We refer to such class of theories as Bipartite Field Theories (BFTs).

The motivation for studying these theories follows from the fact that they underlie a wide range of interesting physical systems.\footnote{Here we do not mention the obvious, and extremely interesting, applications of bipartite graphs to condensed matter physics.} One of their first appearances has been in the context of D3-branes probing toric Calabi-Yau 3-folds in Type IIB string theory. The superconformal field theory on the worldvolume of the D3-branes is indeed encoded by a bipartite graph on a 2-torus \cite{Franco:2005rj}. When investigating the Type IIA configurations of D6-branes related to the previous setup by mirror symmetry, bipartite graphs on higher genus Riemann surfaces surprisingly emerge as fundamental objects \cite{Feng:2005gw}. More recently, bipartite graphs on a 2-torus have been shown to give rise to an infinite class of quantum mechanical integrable systems \cite{GK}.

In the latest addition to this string of applications, bipartite graphs have been used in the context of scattering amplitudes in quantum field theory (QFT) \cite{Nima}. In recent years, we have witnessed tremendous progress in our understanding of scattering amplitudes in gauge theory, most notably for $\mathcal{N}=4$ super Yang-Mills in the planar limit. These developments were originally triggered by Witten's twistor string \cite{Witten:2003nn} and have resulted in efficient tools for the computation of scattering amplitudes at tree level, such as CSW diagrams \cite{Cachazo:2004kj} and BCFW recursion relations \cite{Britto:2004ap,Britto:2005fq,Brandhuber:2008pf,ArkaniHamed:2008gz}, and loop level \cite{Bern:2007dw,Berger:2008sj,Bern:2004cz,Bern:2007ct}. At the same time, the hidden dual superconformal symmetry of planar $\mathcal{N}=4$ SYM was unveiled \cite{Drummond:2006rz, Alday:2007hr}, and it was realized that superconformal and dual superconformal symmetries combine to give rise to the infinite dimensional Yangian symmetry \cite{Drummond:2009fd}. In order to better understand the role of this symmetry, it was suggested one should focus on the leading singularities of scattering amplitudes, which in turn arise as residues of a contour integral over the Grassmannian \cite{ArkaniHamed:2009dn}. At a fundamental level, the new insights have led to the idea that a new formulation of QFT might exist, displaying its otherwise hidden simplicity by abandoning manifest locality and unitarity in favor of making the infinite Yangian symmetry explicit. The fundamental structure behind this new formulation might take several tightly related disguises: the Grassmannian, algebraic geometry, certain graphs or, as we advocate in this paper, certain quiver gauge theories.

For all the systems mentioned above, BFTs are not merely a different interpretation of the same underlying graph. In fact, every statement and computation in any of these systems has a counterpart in the corresponding BFT. As usual, having an alternative perspective on the same physics, in this case in terms of a gauge theory, is extremely valuable. It not only allows us to understand known facts in a new light, but it provides intuition, suggests new approaches and might eventually become useful for answering new questions. The dynamics, duality and natural connection to geometry in the form of moduli spaces of the gauge theory have fruitful applications to the other systems.

The appearance of quivers in connection to all these systems is not a coincidence. To some extent, it can be understood as the result of a powerful mathematical structure that underlies all of them: {\it cluster algebras} \cite{MR1887642}. Quivers indeed provide the natural physical arena for cluster algebras.

In this paper we will study BFTs in full generality, but our examples will mainly focus on graphs with boundaries, which is the sub-class that has been less explored so far and the one that is relevant for leading singularities in scattering amplitudes. Our results will also shed light in the relatively unexplored area of the combinatorics of bipartite graphs on general Riemann surfaces with boundaries. 

This paper is organized as follows. BFTs are introduced in Section \ref{section_BFTs}. Section \ref{section_graph_modifications} explains the BFT interpretation of various modifications of the underlying graph. Section \ref{section_BFTs_everywhere} provides an overview of several areas in which BFTs arise. In Section \ref{section_Kasteleyn}, we generalize the method based on the Kasteleyn matrix for the determination of perfect matchings to deal with bipartite graphs with boundaries. Section \ref{section_BFTs_and_CYs} discusses two toric Calabi-Yau (CY) manifolds that are associated to any BFT, its master and moduli spaces. Section \ref{section_examples} collects explicit examples of BFTs. The examples presented are related to leading singularities and go beyond the planar limit in two directions, increasing both the number of boundaries of the graph and the genus of the underlying Riemann surface. Section \ref{section_Seiberg_duality} studies the action of Seiberg duality on BFTs, which translates to square moves in the underlying graph. We analyze both planar and non-planar graphs, explicitly illustrating that the moduli is invariant under Seiberg duality and can be efficiently exploited for identifying theories related by square moves. In Section \ref{section_loop_reduction}, we explain how the number of loops in certain diagrams can be reduced and explain this process in terms of the dynamics of the associated BFT. The equivalence between different multi-loop diagrams has a striking manifestation in terms of a single underlying Calabi-Yau manifold, arising as the moduli space of the BFTs. Section \ref{section_Higgsing} explains how the boundary operator on a cell in the positive Grassmannian maps to the Higgs mechanism in the corresponding BFT. We show how a new bipartite graph, related to the original one by the untwisting map on zig-zag paths, is an efficient tool for identifying consistent higgsings. We conclude and summarize some open questions for further research in Section \ref{section_conclusions}.

\bigskip
\noindent {\bf Note added:} while this paper was being finalized, we became aware of \cite{Xie:2012mr}, which has some overlap with this work.
\bigskip

\section{Bipartite Field Theories}

\label{section_BFTs}

In this section we introduce the concept of a Bipartite Field Theory. A BFT is a 4d, $\mathcal{N}=1$ quiver gauge theory whose Lagrangian is defined by a {\it bipartite} graph $G$ on a Riemann surface $\Sigma$, which can contain boundaries.
BFTs are natural generalizations of toric quivers, which are defined by bipartite graphs without boundaries on $T^2$ \cite{Franco:2005rj}.

\begin{figure}[h]
\begin{center}
\includegraphics[width=6.5cm]{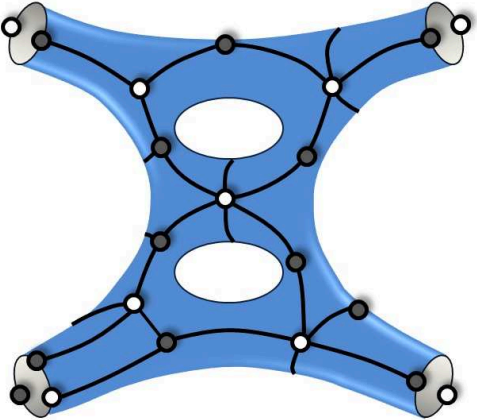}
\caption{An artistic representation of a bipartite graph on a Riemann surface defining a BFT.}
\label{artistic_BFT}
\end{center}
\end{figure}

We can separate the nodes of $G$ into internal and external (those on boundaries) ones. $G$ is bipartite if nodes can be colored in black and white such that:
\begin{itemize}
\item[{\bf 1)}] Every edge in $G$ connects nodes of different colors.
\item[{\bf 2)}] Every boundary node is connected to a single edge.
\end{itemize} 

\bigskip

\subsection{The Dictionary}

\label{section_dictionary}

Let us now describe the translation between the elements of the bipartite graph on $\Sigma$ and a 4d, $\mathcal{N}=1$ quiver gauge theory. 

\bigskip

\begin{figure}[h]
\begin{center}
\includegraphics[width=9cm]{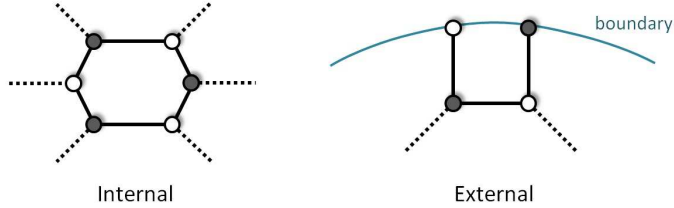}
\caption{Examples of internal and external faces in a BFT.}
\label{Internal_external_faces}
\end{center}
\end{figure}

\begin{itemize}
\item {\bf \underline{Faces}:} to every face in the graph, we associate an $SU(N)$ group.\footnote{In fact, each face corresponds to a $U(N)$ group. When gauging some of them, the $U(1)$ pieces are IR free and become global symmetries at low energies.} We can identify faces as internal or external. Internal faces are those whose entire perimeter is given by edges in $G$. On the other hand, external faces are those in which part of its perimeter overlaps with the boundary of the graph. We show examples of both types of faces in \fref{Internal_external_faces}.

\item {\bf \underline{Edges}:} every edge is identified with a chiral multiplet transforming in the bifundamental representation of the two faces it separates. The bipartiteness of $G$ introduces a natural orientation of the bifundamentals dual to edges which, without loss of generality, can be taken to be oriented clockwise around white nodes and counterclockwise around black nodes. 

\item {\bf \underline{Nodes}:} every internal node corresponds to a monomial in the superpotential, given by the product of all chiral fields associated with the edges terminating on it. The valence of the node, i.e. the number of edges terminating on the node, corresponds to the order of the superpotential term. We assign signs to superpotential terms such that white and black nodes correspond to plus and minus signs, respectively. External nodes do not have any superpotential interpretation, and simply follow from the existence of edges that have one endpoint on an internal node and terminate on the boundary.
\end{itemize}

\bigskip

Due to the bipartiteness of $G$, the number of edges around an internal face is even. Furthermore, the numbers of white and black nodes around an internal face are equal. In contrast, the number of edges on the perimeter of an external face can be either even or odd. Given the interpretation of edges as bifundamental chiral fields with an orientation dictated by the color of nodes we conclude that, from a quiver perspective, internal faces have an equal number of incoming and outgoing bifundamentals arrows, which in turn implies that they are free of anomalies. As a result, the $SU(N)$ groups associated to internal faces can be consistently gauged. From now on, we include in the definition of a BFT the fact that internal faces are gauged, while external faces correspond to global symmetry groups.\footnote{Following the previous discussion, there exists an alternative natural gauging, in which a basis for all closed loops in the graph whose perimeter is made out of internal edges is gauged. This alternative becomes relevant for non-planar graphs and is useful in some contexts such as leading singularities. While we will not consider this gauging any further, the resulting theories can be studied with exactly the same tools we develop in this paper.}$^,$\footnote{Internal faces might contain punctures inside them. A possible way of distinguishing such faces at the level of the BFT is by not gauging the corresponding anomaly-free $SU(N)$ group. While in this paper we will gauge the symmetries associated to all internal faces, it is important to keep this possibility in mind.}

This interpretation of faces is rather natural in light of similar systems that admit an interpretation as brane configurations in string theory. In such systems, faces correspond to stacks of D-branes suspended from a web of branes associated with the graph $G$, which extends over dimensions transverse to the ones in which the gauge theory lives. Internal faces have a finite extension along the directions transverse to the field theory ones and hence give rise to gauge symmetries. External faces can instead have an infinite extension in these directions and thus lead to global symmetries. BFTs associated to graphs with no boundaries on $T^2$ indeed arise as configurations of D5 and NS5-branes \cite{Franco:2005rj}. The question of whether general BFTs can arise as systems of branes is a very interesting one, but we postpone it for future studies.

According to the map between faces and gauge or global symmetry groups, edges can correspond to: bifundamentals of the gauge group (when they sit between two internal faces), fundamental or antifundamental flavors (internal/external) or gauge singlets transforming in a bifundamental representation of the global symmetry group (external/external). 

\begin{figure}[h]
\begin{center}
\includegraphics[width=12cm]{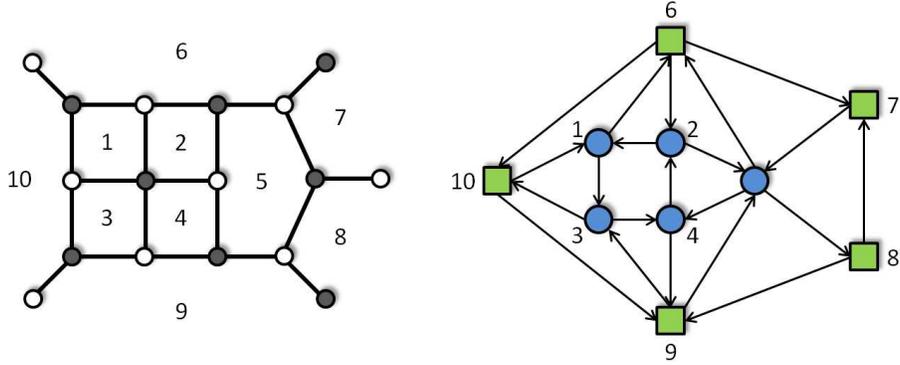}
\caption{A bipartite graph and its quiver dual. Blue nodes in the quiver represent gauge groups and green squares represent global symmetry groups. Plaquettes in the quiver correspond to superpotential terms.}
\label{Example_graph_BFT}
\end{center}
\end{figure}

The correspondence we have introduced implies that every BFT has a quiver diagram living on $\Sigma$, which is indeed dual to the bipartite graph $G$ as illustrated with an example in \fref{Example_graph_BFT}. Faces, edges and nodes in $G$ are mapped to symmetry group nodes (gauged or global), arrows and plaquettes (superpotential terms with sign determined by their clockwise or counterclockwise orientation) in the quiver, respectively. Below we summarize the dictionary between graphs and BFTs.

\bigskip
\bigskip

\hspace{-.7cm}\begin{tabular}{|l|l|}
\hline
{\bf Graph} & {\bf BFT} \\ \hline \hline
Internal face ($2n$-sided) & Gauge group with $n$ flavors \\ \hline
External face & Global symmetry group \\ \hline \hline
Edge between two faces $i$ and $j$ & Chiral multiplet in the bifundamental \\ & representation of the groups $i$ and $j$. The \\ & orientation of the corresponding arrow is such \\ & that it goes clockwise around white nodes and \\ & counterclockwise around black nodes \\ \hline \hline
$k$-valent node & Monomial in the superpotential involving $k$ chiral \\ & multiplets. The signs of the terms are (+/-) for \\ & (white/black) nodes \\ \hline
\end{tabular}

\bigskip
\bigskip

Let us briefly comment on the scale dependence properties of BFTs.  A natural representation of bipartite graphs is given by the so called {\it isoradial embedding}. We refer the reader to \cite{Kennaway:2007tq} for details about this construction. Interestingly, in this embedding it is possible to map the scaling dimensions of chiral fields to the angles subtended by the corresponding edges. Vanishing of the beta functions for gauge and superpotential couplings then translates into local flatness of the graph \cite{Franco:2005rj,FT}. As a result, BFTs on curved Riemann surfaces are not conformal.

In the remainder of this section, we discuss two important concepts in the study of bipartite graphs: perfect matchings and zig-zag paths.

\bigskip

\subsection{Perfect Matchings}

An {\it almost perfect matching} $p$ is a subset of the edges in $G$ such that:

\begin{itemize}
\item Every internal node is the endpoint of exactly on edge in $p$.
\item Every external node belongs to either one or zero edges in $p$.
\end{itemize}
Almost perfect matchings can be regarded as regular perfect matchings of a larger bipartite graph that contains $G$ and extends it beyond its boundaries. For brevity, we simply refer to them as {\it perfect matchings} in what follows. Perfect matchings connect bipartite graphs to gauge theory, toric geometry and integrable systems, as we explain later.

The map between chiral fields in the quiver $X_i$, equivalently edges in $G$, and perfect matchings $p_\mu$ is given by
\beq
X_i = \prod_{\mu=1}^c p_\mu^{P_{i\mu}},
\label{X_pm_map}
\eeq
where $c$ is the total number of perfect matchings, and $P_{i\mu}$ is equal to $1$ if the edge in the bipartite graph associated to the chiral field $X_i$ is contained in $p_\mu$ and zero otherwise \cite{Franco:2005rj}, i.e.

\beq
P_{i\mu}=\left\{ \begin{array}{ccccc} 1 & \rm{ if } & X_i  & \in & p_\mu \\
0 & \rm{ if } & X_i  & \notin & p_\mu
\end{array}\right.
\label{Xi_to_pmu}
\eeq
In Section \ref{section_Kasteleyn}, we will introduce efficient methods for determining the matrix $P$. It will play an important role in Section \ref{section_BFTs_and_CYs}, when computing the moduli spaces of the BFTs.\footnote{Throughout the paper, we will also use a notation for chiral fields involving two subindices, explicitly indicating the gauge groups under which they are charged, as opposed to the single subindex notation in \eref{Xi_to_pmu}. We are confident that these two alternative notations will not generate any confusion.} 

\bigskip

\subsection{Zig-Zag Paths}

\label{section_zig_zags}

\begin{figure}[h]
\begin{center}
\includegraphics[width=7cm]{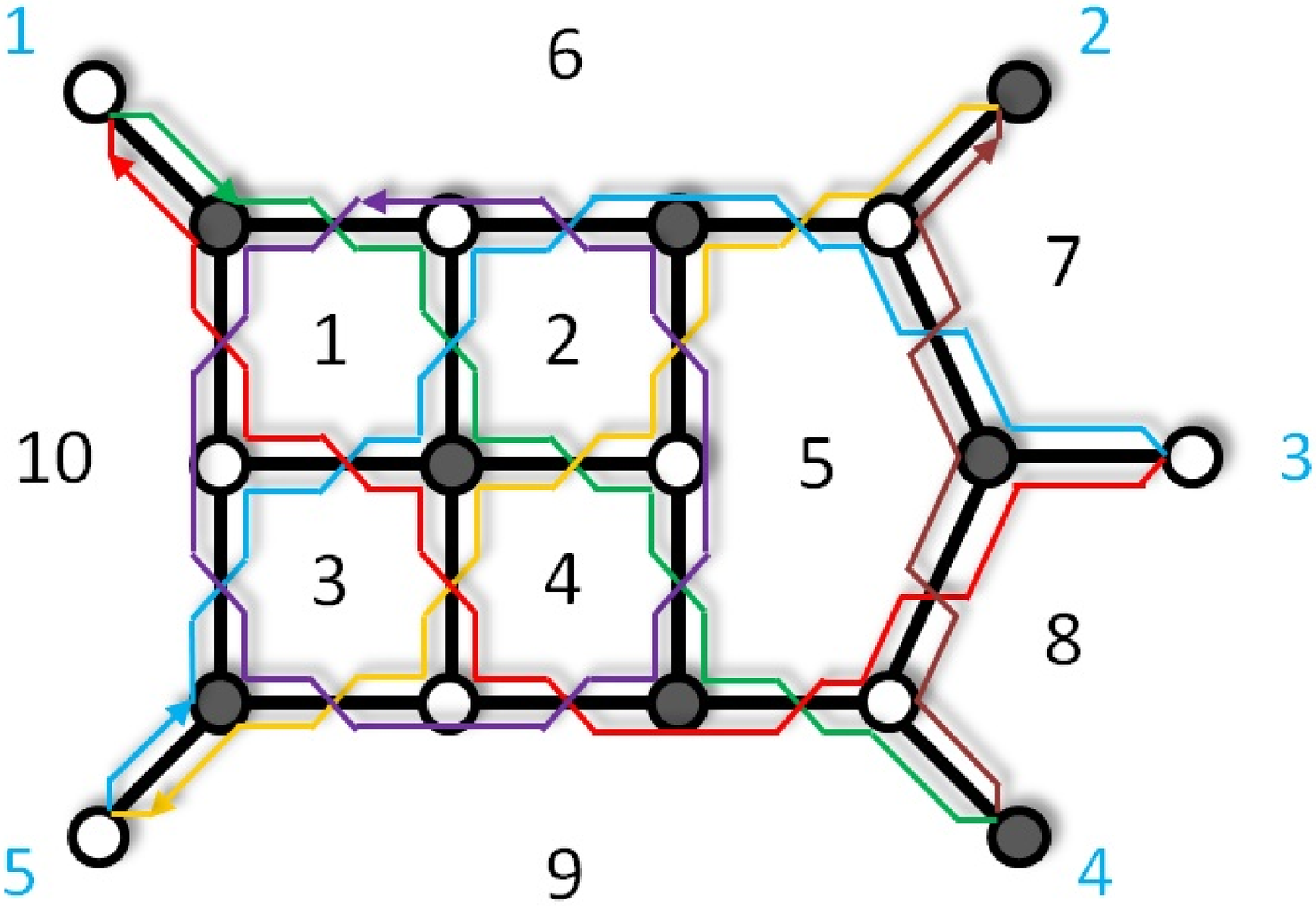}
\caption{Zig-zag paths in double line notation. This example has five zig-zag paths going through external legs and an internal one, shown in purple.}
\label{zig_zags_example}
\end{center}
\end{figure}

{\it Zig-zag paths}, also denoted {\it alternating strands}, are oriented paths in a bipartite graph that alternate between turning maximally right and maximally left at every node. They can be efficiently implemented in terms of a double line notation for edges \cite{Feng:2005gw}, in which two zig-zag paths go over every edge in opposite directions, crossing at the middle point. \fref{zig_zags_example} shows the zig-zag paths for the example in \fref{Example_graph_BFT}. 

Zig-zag paths play a prominent role in the study of bipartite graphs for various reasons. First, it is possible to reconstruct the graph $G$ from knowledge of its zig-zags. There is an edge for every intersection between a pair of them, and black (white) nodes correspond to disks with clockwise (counterclockwise) oriented boundaries. Relative motion of the zig-zags results in different graphs, which translate to dual BFTs as we will briefly discuss in Section \ref{section_Seiberg_duality} (see also \cite{Hanany:2005ss}). In the case of graphs without boundaries on $T^2$, zig-zag paths are directly connected to the Calabi-Yau geometry of the moduli space of the associated BFT \cite{Feng:2005gw,Hanany:2005ss}. They are the legs in the $(p,q)$-web \cite{Aharony:1997ju,Aharony:1997bh,Leung:1997tw} dual to the corresponding toric diagram. Whether there also exist a direct link between zig-zags paths and moduli spaces in the generalized context of arbitrary BFTs is an extremely interesting question that we plan to revisit in the future.

\bigskip

\noindent {\bf \underline{Consistency}:} we will restrict to graphs that do not contain self-intersecting zig-zag paths. The existence of such paths has, in some sub-classes of BFTs, been linked to inconsistencies of the field theory \cite{Hanany:2005ss}. For brevity, we will refer to the resulting graphs and BFTs as {\it consistent}. A full study of the consequences of self-intersecting zig-zags on general BFTs is beyond the scope of this paper. We will often explicitly determine the zig-zags of graphs, showing this condition is met.

\bigskip

\begin{figure}[h]
\begin{center}
\includegraphics[width=9cm]{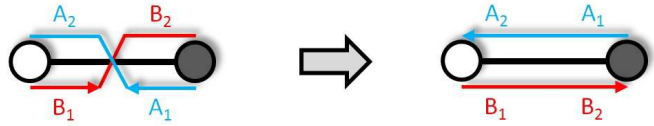}
\caption{The untwisting map.}
\label{untwisting_map}
\end{center}
\end{figure}

\noindent {\bf \underline{Untwisting Map}:} the action of the untwisting map is schematically shown in \fref{untwisting_map}. It interchanges:

\beq
\begin{array}{ccccc}
\mbox{$G$ on $\Sigma$} & & & & \mbox{$\tilde{G}$ on $\tilde{\Sigma}$} \\
\mbox{zig-zag path} & \ \ \ \ & \leftrightarrow & \ \ \ \ & \mbox{face}\\
\mbox{face} & \ \ \ \ & \leftrightarrow & \ \ \ \ & \mbox{zig-zag path}
\end{array}
\nonumber
\eeq 

\medskip

The graph $\tilde{G}$ on $\tilde{\Sigma}$ resulting from untwisting can be interpreted as a new bipartite field theory, which we denote $\widetilde{\mbox{BFT}}$. 
The concept of $\widetilde{\mbox{BFT}}$ will be used in Section \ref{section_boundary_operator} for keeping track of zig-zag paths under certain modifications of the graph.

\bigskip

\noindent {\bf \underline{Deligne Permutations}:} for bipartite graphs with boundaries, it is possible to introduce the concept of {\it Deligne permutations} \cite{Nima}, which correspond to a bijection $f_d(b_i)=b_j$ that maps every boundary point $b_i$ to another boundary point $b_j$.\footnote{Other works in which bipartite graphs have been discussed in conjunction with certain permutations and Calabi-Yau manifolds include \cite{Koch:2010zza,Jejjala:2010vb}.} More concretely, Deligne permutations are in one-to-one correspondence with zig-zag paths going through external legs as follows

\beq
f_d(b_i)=b_j \ \ \ \ \iff \ \ \ \ \mbox{The zig-zag path starting at $b_i$ ends at $b_j$}
\label{Deligne_permutation_zig_zag}
\eeq
For example, for the model in \fref{zig_zags_example} we have
\beq
\begin{array}{ccccccccccc}
f_d(1) & = & 4, & \ \ \ \ & f_d(2) & = & 5, & \ \ \ \ & f_d(3) & = & 1, \\
f_d(4) & = & 2, & \ \ \ \ & f_d(5) & = & 3. & & & &  
\end{array}
\eeq

\bigskip

\section{A BFT Perspective on Graph Modifications}

\label{section_graph_modifications}

In this section we discuss the BFT interpretation of various possible modifications of the bipartite graph.

\subsection{Reduction to 2 and 3-valent Graphs}

\label{section_2_and_3_valent}

\begin{figure}[h]
\begin{center}
\includegraphics[width=8cm]{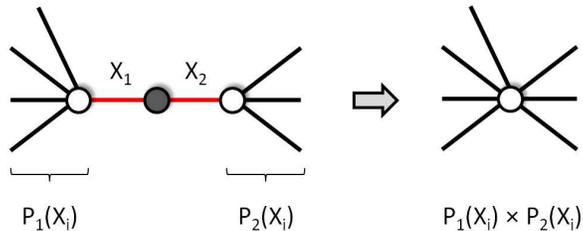}
\caption{Integrating out massive fields associated to 2-valent nodes corresponds to condensing nodes on the graph.}
\label{tiling_massive_fields}
\end{center}
\end{figure}

Nodes with valence $k$ in the graph correspond to order $k$ superpotential terms in the BFT. In particular, 2-valent nodes correspond to mass terms. Due to the specific structure of the BFT superpotentials, integrating out massive fields has a simple graphical implementation: it corresponds to condensing the two nodes at both endpoints of the 2-valent one, as illustrated in \fref{tiling_massive_fields}.

The origin of the condensation can be understood as follows. The superpotential takes the form

\beq
W=X_1 P_1(X_i) + X_2 P_2(X_i) -X_1 X_2 +\ldots
\label{W_massive_fields}
\eeq
where we have identified all terms in the superpotential containing $X_1$ and $X_2$. $P_1(X_i)$ and  $P_2(X_i)$ are products of bifundamentals fields. The equations of motion for the massive fields are

\beq
\partial_{X_1} W =0 \Leftrightarrow X_2 = P_1(X_i) \ \ \ \ \mbox{    and    } \ \ \ \ \partial_{X_2} W =0 \Leftrightarrow X_1 = P_2(X_i).
\eeq
Plugging them back into \eref{W_massive_fields}, the terms involving $X_1$ and $X_2$ are replaced by 

\beq
W=P_1(X_i)P_2(X_i)+\ldots , 
\eeq
which is precisely the interaction associated to merging the nodes at both sides of the mass term. I.e. two terms of order $k_1$ and $k_2$ are combined into a single one of order $k_1+k_2-2$.

\begin{figure}[h]
\begin{center}
\includegraphics[width=10cm]{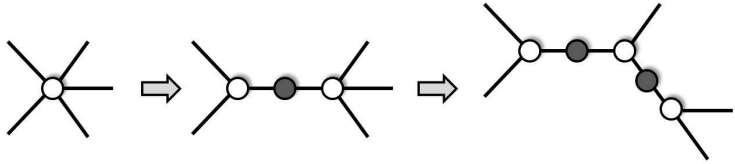}
\caption{A possible decomposition of a 5-valent into three 3-valent ones by the insertion of two 2-valent nodes. The field theory counterpart of this process corresponds to integrating-in massive fields.}
\label{decomposing_5-valent_node}
\end{center}
\end{figure}

Inverting this process, we can reduce the order of superpotential terms by inserting 2-valent nodes, i.e. by integrating-in massive fields. Iterating this process, it is possible to reduce any $k$-valent node to $(k-2)$ 3-valent and $(k-3)$ 2-valent ones. This process is clearly not unique, although the low energy physics is independent of how we perform it. \fref{decomposing_5-valent_node} shows a possible decomposition of a 5-valent node.

\subsection*{From Bicolored to Bipartite Graphs}

Bicolored graphs that are not bipartite, i.e. graphs that contain edges connecting nodes of the same color, can be studied with the same tools discussed in this paper. Plabic (i.e. planar bicolored) graphs are examples of this class of models \cite{Postnikov_plabic}. 

Whenever we encounter a graph with an edge connecting two nodes of the same color, we will interpret it as a bipartite one by introducing a 2-valent node of the opposite color in the middle of this edge. The corresponding massive fields can then be integrated out, resulting in the merging of the two original nodes. This field theoretic interpretation leads precisely to the two equivalent procedures that appear in the math literature for turning bicolored graphs into bipartite ones: introducing an intermediate 2-valent node or condensing the two nodes of the same color.  

\begin{figure}[h]
\begin{center}
\includegraphics[width=6cm]{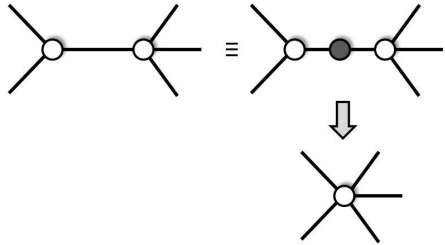}
\caption{A bicolored graph containing an edge between nodes of the same color can be turned into a bipartite one by introducing an intermediate 2-valent node or by node condensation.}
\label{bicolored_to_bipartite}
\end{center}
\end{figure}

\bigskip

\subsection{Seiberg Duality}

\label{section_Seiberg_duality}

Seiberg duality \cite{Seiberg:1994pq} is a remarkable property of $\mathcal{N}=1$ gauge theories that implies a full equivalence between two different theories (typically referred to as electric and magnetic) in the low energy limit. It plays an essential role in our understanding of the low energy dynamics of SUSY gauge theories \cite{Intriligator:1995au}. Being such a central concept in SUSY field theories, Seiberg duality has, in the BFT context, a natural implementation in terms of the underlying bipartite graph. 

We will focus on Seiberg dualities in which the electric and magnetic theories are BFTs on the same Riemann surface. This constraint implies that we can only dualize $N_f=2N_c$ gauge groups, where $N_f=N$, the common rank of all gauge and global symmetry groups. This class of gauge groups corresponds to internal square faces in the graph.  Let us briefly discuss the field theory side of the duality, focusing on the dualized gauge group, while considering all others as expectators. The electric theory has $SU(N_c)$ gauge groups and flavors $Q_1$, $Q_2$ in its fundamental representation and $\tilde{Q}_1$ and $\tilde{Q}_2$ in the antifundamental representation. More precisely, these flavors are in fact bifundamentals fields in the BFT but, for brevity, we omit discussion of the transformation properties under the additional symmetry groups. The dual gauge group is $SU(\tilde{N}_c)$, where the dual rank is $\tilde{N}_c=N_f-N_c=N_c$, i.e. it is equal to the original one. In addition, the duality implies the following transformations:

\bigskip

\begin{itemize}
\item Replace the electric quarks $Q_1$, $Q_2$, $\tilde{Q}_1$ and $\tilde{Q}_2$ by magnetic quarks $\tilde{q}_1$, $\tilde{q}_2$, $q_1$ and $q_2$ transforming in the conjugated (bifundamental) representations.
\item Introduce meson fields $M_{ij}$, which are composite from the perspective of the electric theory, i.e. $M_{ij}=\tilde{Q}_i Q_j$. Meson fields are singlets of the magnetic gauge group and transform in bifundamentals representations of the other groups, which are either gauged or global.
\item Introduce cubic superpotential couplings between the dual quarks and the mesons

\beq
\Delta W = \sum_{ij} q_i M_{ij} \tilde{q}_j , \nonumber
\eeq
and re-express any product of the electric quarks in the original superpotential in terms of the mesons.
\end{itemize}

\bigskip

\begin{figure}[h]
\begin{center}
\includegraphics[width=8.5cm]{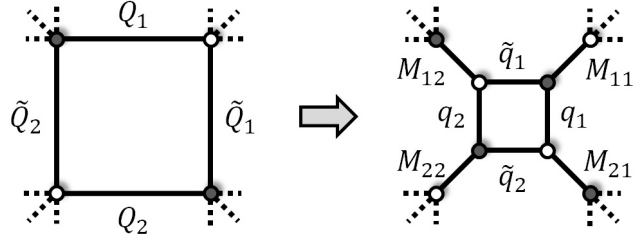}
\caption{Seiberg duality maps to square moves in the graph.}
\label{tiling_Seiberg_duality}
\end{center}
\end{figure}

In \fref{tiling_Seiberg_duality} we present the simple graph transformation that implements Seiberg duality. It was originally discovered in \cite{Franco:2005rj} in the context of BFTs without boundaries on $T^2$ but extends to generic BFTs without changes. It is easy to see that this modification of the graph automatically implements the three points discussed above. It is often referred to as {\it urban renewal}, {\it spider move} or {\it square move}. If 2-valent nodes are generated during this process, the corresponding massive fields can be integrated out as discussed in Section \ref{section_2_and_3_valent}. Repeating this operation twice on the same face obviously returns to the original graph.

\begin{figure}[h]
\begin{center}
\includegraphics[width=8.5cm]{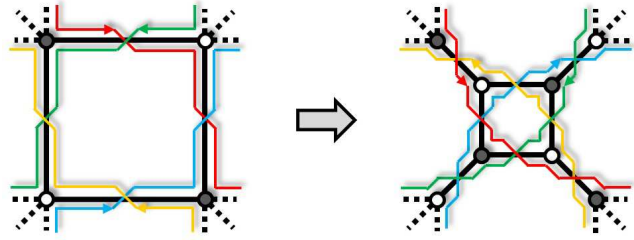}
\caption{The action of Seiberg duality on zig-zag paths.}
\label{zig_zags_Seiberg_duality}
\end{center}
\end{figure}

The effect of Seiberg duality on zig-zag paths is shown in \fref{zig_zags_Seiberg_duality}. It corresponds to a reorganization of the four zig-zag paths passing through the dualized square, in which zig-zags associated to opposite corners are pairwise interchanged. 

Seiberg duality is a full equivalence between the electric and magnetic theories in the IR limit. This implies agreement in their behavior under relevant deformations, matching of their moduli spaces, etc. The field theoretic perspective on bipartite graphs thus suggests natural invariants under square moves. This intuition can be exploited in the context of any other interpretation of the same graphs, such as leading singularities in scattering amplitudes. We will use this approach in Sections \ref{section_Seiberg_duality} and \ref{section_loop_reduction}.

\bigskip

\subsection{Higgsing}

\label{section_Higgsing}

\begin{figure}[h]
\begin{center}
\includegraphics[width=8.5cm]{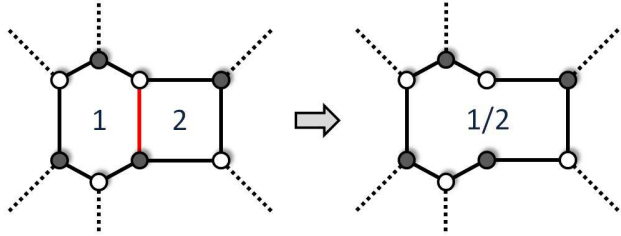}
\caption{Removing an edge in the graph corresponds to turning on a vev for a bifundamentals scalar, resulting in the merging of two faces.}
\label{tiling_higgsing}
\end{center}
\end{figure}

Edge removal is another natural operation on graphs. From a BFT viewpoint, it translates to the scalar in the corresponding bifundamentals chiral multiplet acquiring a non-zero vacuum expectation value (vev) \cite{Franco:2005rj}. The two faces on each side of the removed edge are combined into a single one as shown in \fref{tiling_higgsing}. Depending on the type of faces that are merged, we can have three different situations, which have the following BFT interpretation:

\begin{itemize}
\item {\bf Internal-internal:} higgsing of the corresponding $SU(N)\times SU(N)$ piece of the gauge group down to the diagonal $SU(N)$ subgroup.
\item  {\bf Internal-external:} color-flavor locking of the $SU(N)_{gauge}\times SU(N)_{global}$ symmetry associated to the faces.
\item {\bf External-external:} spontaneous breaking of an $SU(N)\times SU(N)$ subgroup of the global symmetry to the diagonal subgroup. This process results in massless Goldstone bosons.
\end{itemize}
If 2-valent nodes are generated in this process, the corresponding massive fields can be integrated out. Throughout the paper we will be interested in preserving external legs, which map to scattered particles, so we will no longer consider the third option.

In Section \ref{section_boundary_operator} we will investigate the connection between higgsing and the boundary operator in leading singularities and introduce an implementation of it that efficiently keeps track of zig-zag paths, ensuring that we only perform higgsings that result in consistent graphs.

\section{BFTs Everywhere}

\label{section_BFTs_everywhere}

BFTs and their dynamics play an important role in a wide spectrum of interesting physical systems. In this section we present a brief overview of some of them.

\subsection{D3-Branes over Toric Calabi-Yau 3-Folds}

\begin{figure}[h]
\begin{center}
\includegraphics[width=4cm]{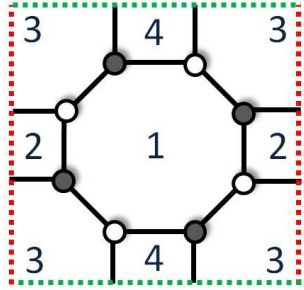}
\caption{Bipartite graph associated to the 4d, $\mathcal{N}=1$ SCFT on D3-branes over the complex cone over $F_0$. Opposite sides of the square are identified to form a $T^2$.}
\label{tiling_F0_2}
\end{center}
\end{figure}

The 4d, $\mathcal{N}=1$ superconformal field theory (SCFT) arising on a stack of D3-branes probing a CY 3-fold is a BFT on a 2-torus without boundaries \cite{Franco:2005rj}. The two fundamental directions of the $T^2$ correspond to a $U(1)^2$ flavor symmetry that follows from isometries of the toric CY. The remaining $U(1)$ isometry translates to the R-charge of the gauge theory. \fref{tiling_F0_2} shows an example, corresponding to the BFT on D3-branes at the complex cone over $F_0$ \cite{Franco:2005rj}.\footnote{In fact there is another BFT associated to D3-branes on the same geometry, which is connected to this one by Seiberg duality as discussed in Section \ref{section_Seiberg_duality}.}

The correspondence between this class of SCFTs and bipartite graphs has been extremely fruitful, fully answering the question of which gauge theory is associated to a given toric Calabi-Yau and vice versa.

\subsection{Mirror Symmetry}

Consider the configuration discussed in the previous section, with D3-branes on a toric singularity with characteristic polynomial $P(z_1, z_2) = \sum a_{n_1,n_2} z_1^{n_1} z_2^{n_2}$, where $(n_1, n_2)$ runs over points in the toric diagram.  On the D3-branes, we obtain a BFT theory described by a bipartite graph $G$ on a 2-torus. 

\begin{figure}[h]
\begin{center}
\includegraphics[width=10cm]{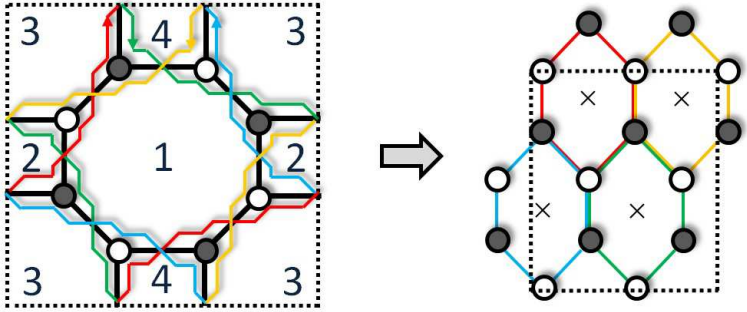}
\caption{The tiling of the mirror obtained by acting with the untwisting map on the $F_0$ model in \fref{tiling_F0_2}. The resulting Riemann surface on which the graph lives is a 2-torus with four punctures.}
\label{untwisting_mirror}
\end{center}
\end{figure}

The mirror manifold is given by $P(z_1, z_2) = W$,  $W = uv$. Let us consider the Riemann surface $\tilde{\Sigma}$ sitting at $W = 0$. The genus and number of punctures of $\Sigma$ are given by the number of internal points and the perimeter of the toric diagram, respectively. The original configuration of D3-branes is mapped to a set of intersecting D6-branes in the mirror, with one type of D6-branes for each gauge group in the quiver and bifundamentals chiral multiplets arising at their intersections \cite{Feng:2005gw}. The non-trivial information about the mirror configuration is encoded in a new bipartite graph $\tilde{G}$ on $\tilde{\Sigma}$, in which the D6-branes arise as zig-zag paths. The graph $\tilde{G}$ is obtained from $G$ by applying the untwisting map. We can associate a new $\widetilde{\mbox{BFT}}$ gauge theory to $\tilde{G}$, as mentioned in Section \ref{section_zig_zags}. The $\widetilde{\mbox{BFT}}$ has been referred to as the specular dual in \cite{Hanany:2012vc}. \fref{untwisting_mirror} shows $\tilde{G}$ obtained by untwisting the $F_0$ model in \fref{tiling_F0_2}. In this case, $\tilde{\Sigma}$ is a 2-torus with four punctures.

The discussion in this section makes it clear that mirror symmetry naturally gives rise to bipartite graphs, i.e. BFTs, on Riemann surface of arbitrarily high genus, providing an important motivation for their study.

\subsection{Cluster Integrable Systems}

\begin{figure}[h]
\begin{center}
\includegraphics[width=9cm]{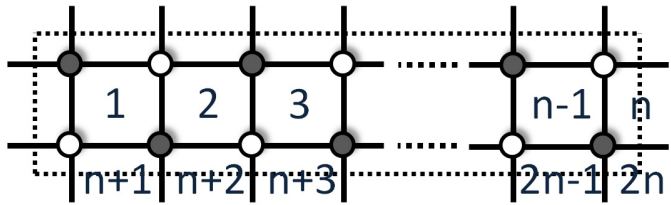}
\caption{Bipartite graph on a $T^2$ associated to the $n$-particle, relativistic, periodic Toda chain.}
\label{tiling_Toda}
\end{center}
\end{figure}

Bipartite graphs without boundaries on a 2-torus are also in one-to-one correspondence with an infinite class integrable systems in (0+1) dimensions, denoted {\it cluster integrable systems} \cite{GK}. The Poisson manifold of the integrable system is parametrized by loops on the graph, a useful basis for which is provided by the loops around faces and along the two fundamental directions of the $T^2$. The Poisson brackets between these variables are dictated by the number of edges over which the corresponding loops overlap, counted with orientation. Different patches of the Poisson manifold, typically described by different graphs, are connected by cluster transformations. In \cite{GK}, it was shown that the Hamiltonian and Casimir operators of the integrable system correspond to internal points and ratios of external points in the associated toric diagrams, respectively.  We refer the reader to \cite{GK} for a detailed explanation of how the integrals of motion are constructed in terms perfect matchings. \fref{tiling_Toda} shows a graph which, by means of this correspondence, is mapped to the $n$-particle, relativistic, periodic Toda chain. Physical implications of this correspondence and connections to other setups realizing the same integrable systems have been studied in \cite{Franco:2011sz,Eager:2011dp,Amariti:2012dd,Franco:2012hv}.

Motivated by generic BFTs, it is natural to ask whether bipartite graphs on Riemann surfaces other than $T^2$ also give rise to integrable systems. We expect it might be possible to extend the proof in \cite{GK} to at least graphs without external nodes that can be embedded in a $T^2$, i.e. disks and cylinders.

\subsection{Leading Singularities}

\label{section_leading_singularities}

The connection between bipartite graphs, and consequently BFTs, and the positive Grassmannian and leading singularities in planar $\mathcal{N}=4$ SYM is relatively more recent and less known to the general physics audience. For this reason, our discussion of these topics will be slightly lengthier. The material presented in this subsection is a brief summary, tailored for our specific needs, of the ideas in \cite{Nima,Postnikov_plabic} and references therein, to which we refer the interested reader for further details. 

Leading singularities can be found by an iterative procedure. Starting from a loop amplitude, we compute the discontinuity at branch cuts, next we determine the discontinuity at branch cuts of this result, and repeat the process until we are left with a rational function of the kinematical invariants. For brevity, this discontinuity across the leading singularity is often referred to as the leading singularity. We can regard leading singularities as calculable well-defined data associated to gauge theories. In addition it has been conjectured that, for maximally supersymmetric theories, leading singularities are sufficient for determining the perturbative S Matrix \cite{ArkaniHamed:2009dn}. Evidence supporting this proposal for $\mathcal{N}=4$ SYM has been given in \cite{Buchbinder:2005wp,Cachazo:2008vp,Cachazo:2008hp,Spradlin:2008uu}.

The Grassmannian $G(k,n)$ is the space of $k$-dimensional planes in $n$ dimensions. Points in $G(k,n)$ can thus be parametrized by a general $k\times n$ matrix $C$, whose rows correspond to $n$-dimensional vectors spanning a plane. We can take any linear combination of the rows without affecting the plane, so we conclude that this parametrization has a $GL(k)$ redundancy or `gauge symmetry'. The positive part of the Grassmannian $G_{\geq 0}(k,n)$ corresponds to the subspace in the Grassmannian in which the determinant of all $k\times k$ minors of $C$ are greater or equal to zero. In what follows, $n$ corresponds to the total number of scattered particles, with $k$ being the number of negative helicity ones. Amplitudes with $k=2$ are known as {\it maximally helicity violating} (MHV), while for $k>2$ they are denoted N$^{k-2}$MHV.

It has been proposed in \cite{ArkaniHamed:2009dn} that all leading singularities in planar $\mathcal{N}=4$ SYM arise as residues of the following contour integral over the Grassmannian

\beq
\mathcal{L}_{n,k}(\mathcal{W}_j)=\int {d^{k\times n} C_{ij}\over (12\cdots k)(23 \cdots (k+1)) \cdots (n 1 \cdots (-1))} \prod_{i=1}^k \delta^{4|4}(C_{ij}\mathcal{W}_j).
\eeq
Here $(i_1 \ldots i_k)$ indicates the determinant of the $k\times k$ matrix made out of the $i_1,\ldots,i_k$ columns of $C$, i.e. the denominator consists of the determinants of the $n$ sequential minors in $C$. $\mathcal{W}_j=(\tilde{\lambda}_j,\tilde{\mu}_j,\tilde{\eta}_j)$, $j=1,\ldots,n$, are the kinematic variables of the scattered particles in twistor space. We refer the reader to \cite{ArkaniHamed:2009dn} for a detailed explanation of this representation and of how to identify the contour integration, which determines the resulting leading singularity. Leading singularities correspond to certain subspaces, also denoted {\it cells}, of the Grassmmanian parametrized by a constrained matrix $C$.

On a parallel line of development, subspaces of the Grassmannian have been shown to be in one-to-one correspondence with bipartite graphs \cite{Postnikov_plabic}. More concretely, cells in $G(k,n)$ are associated to bipartite graphs on a disk with $n$ boundary points.\footnote{More generally, cells in the Grassmannian can be parametrized by plabic graphs, but it is straightforward to turn them into bipartite ones, as explained in Section \ref{section_2_and_3_valent}.} The determination of $k$ is explained below. We thus have a connection between the following objects:
\beq
\mbox{Leading singularities } \ \ \ \Leftrightarrow \ \ \ \mbox{Cells in the Grassmannian} \ \ \ \Leftrightarrow \ \ \ \mbox{Bipartite graphs}
\nonumber
\eeq

\begin{figure}[h]
\begin{center}
\includegraphics[width=7cm]{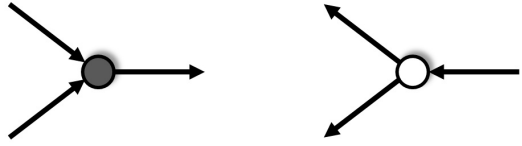}
\caption{Flow pattern on 3-valent nodes.}
\label{MHV_MHVbar_nodes}
\end{center}
\end{figure}

Let us explain how to go from a bipartite graph to a cell in the Grassmannian. The first step is to define certain `momentum flows' along edges of the graph, also called {\it perfect orientations}, which are in one-to-one correspondence with perfect matchings \cite{Postnikov_plabic}. These flows are such that there are two outgoing and one incoming arrows at each white node and two incoming and one outgoing arrows at each black node, as shown in \fref{MHV_MHVbar_nodes}. Flows go through 2-valent nodes without changing direction. Since every bipartite graph can be reduced to 2 and 3-valents nodes as explained in Section \ref{section_2_and_3_valent}, these rules are sufficient for determining perfect orientations.

\begin{figure}[h]
\begin{center}
\includegraphics[width=8.5cm]{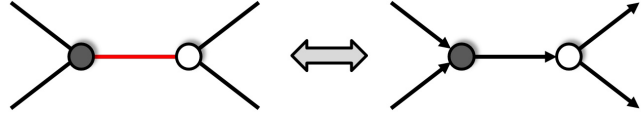}
\caption{There is a one-to-one correspondence between perfect matchings and perfect orientations. Here we indicate in red an edge in a perfect matching.}
\label{pm_to_flow}
\end{center}
\end{figure}

\begin{figure}[h]
 \centering
 \begin{tabular}[c]{ccc}
 \epsfig{file=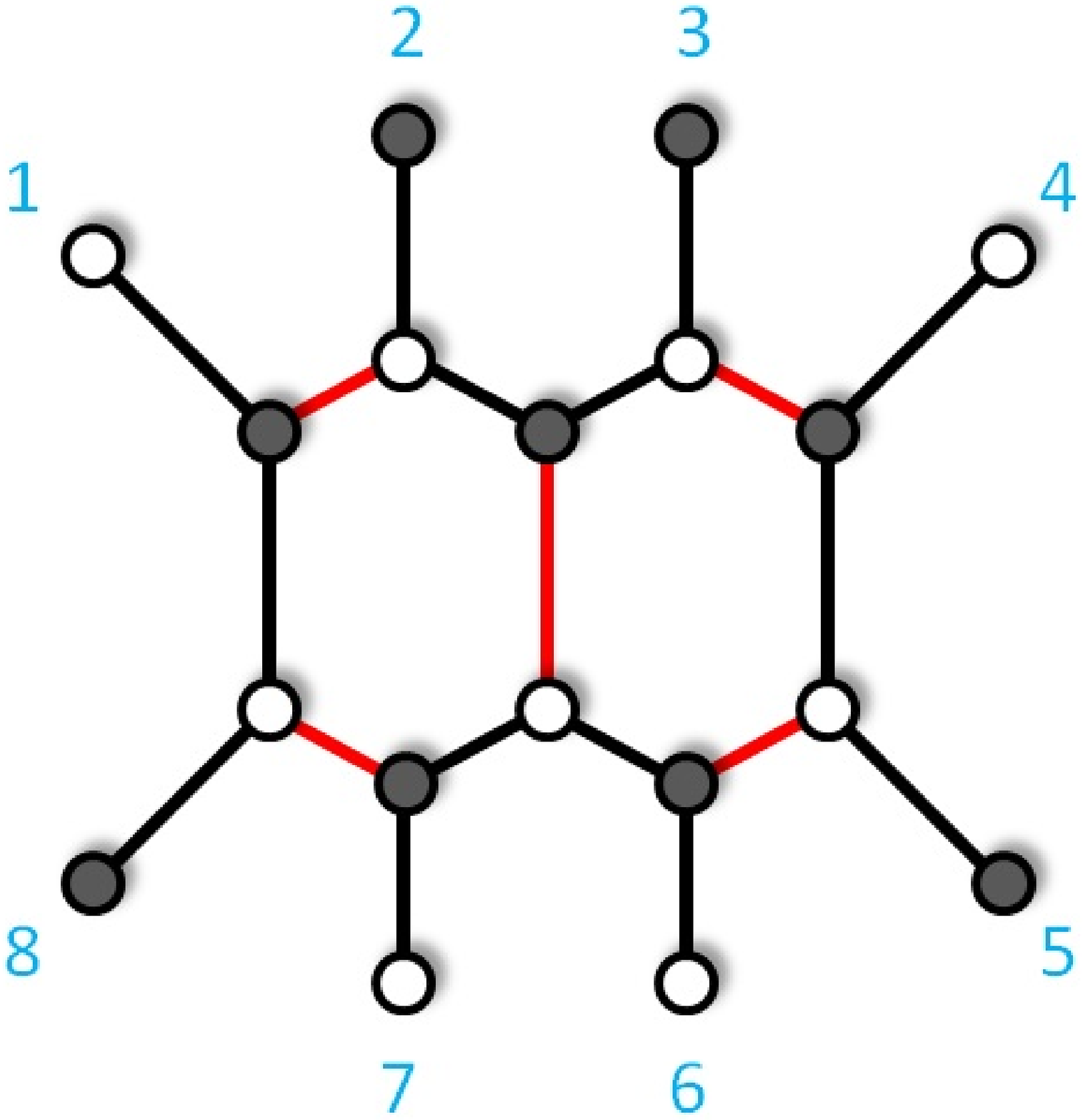,width=0.3\linewidth,clip=} & \ \ \ \ \ &
\epsfig{file=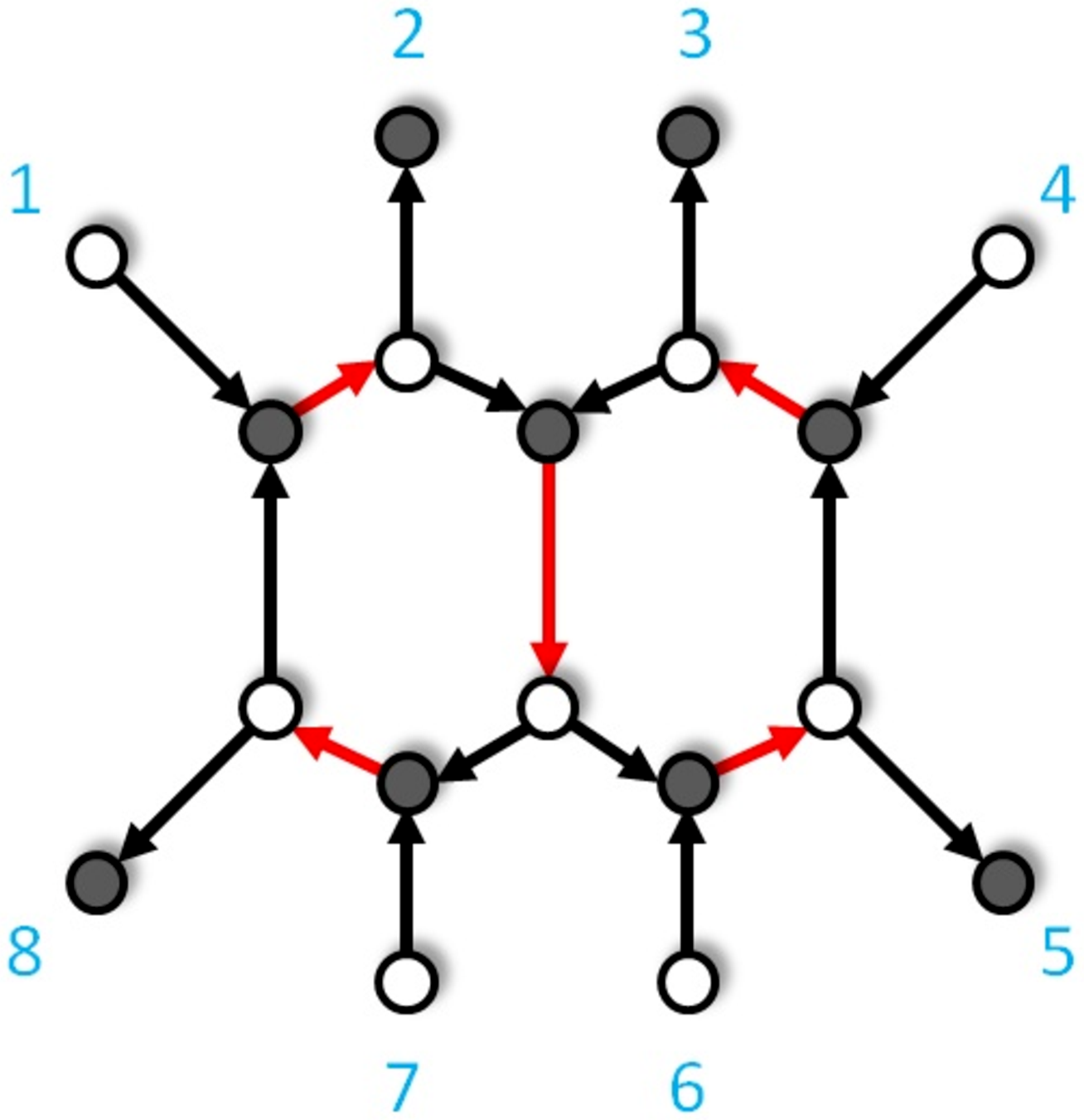,width=0.3\linewidth,clip=} \\ 
\mbox{(a)} & & \mbox{(b)}
 \end{tabular}
\caption{a) An example of a perfect matching and b) its corresponding perfect orientation.}
\label{pm_to_perfect_orientation} 
\end{figure}

The bijection between perfect matchings and perfect orientations works as follows. Given an edge contained in a perfect matching, we identify it with the incoming and outgoing arrows of the white and black nodes at its endpoints, respectively, as shown in \fref{pm_to_flow}. In \fref{pm_to_perfect_orientation} we present an example of a perfect matching and its corresponding perfect orientation.

We are ready for determining the restricted structure of $C$ associated to a graph, given a perfect orientation. $C$ is a $k \times n$ matrix in which rows correspond to the external nodes that are sources of the perfect orientation and columns correspond to all external nodes, i.e. negative and positive helicity particles are mapped to sources and sinks of a perfect orientation, respectively.\footnote{The explicit form of $C$ depends on how we arrange nodes within the rows and columns. This freedom can be taken care of by the existing $GL(k)$ `gauge' symmetry.} Entries for which the row and column correspond to the same node are set to 1. Entries associated to a pair of nodes that are not connected by an oriented path in the perfect orientation are set to 0. Finally, all other entries are determined in terms of edge weights via the so called {\it boundary measurement} \cite{Postnikov_plabic}, 

\beq
c_{ij}=\sum_{P:b_i \to b_j} \prod_{e \in P} x_e,
\label{boundary_measure}
\eeq
i.e. we sum over all directed paths $P$ starting from $b_i$ and terminating at $b_j$ in the graph with a perfect orientation, and the product is over all edges $e$ in $P$. When $P$ has self-intersections, we have to weigh the corresponding contribution by $(-1)^{wind(P)}$, where the winding $wind(P)$ is the signed number of full 360$^{\circ}$ turns $P$ makes. The edge weights $x_e$ are in one-to-one correspondence with the expectation values of the corresponding scalars in the BFT. The precise map between them will be clarified in Section \ref{section_boundary_operator}, when we discuss higgsing.

The procedure we have just outlined clearly depends on a choice of perfect orientation, equivalently on a choice of perfect matching. Different choices are physically equivalent. The result of this prescription coincides with the kinematical analysis of leading singularities, as the one presented in \cite{Kaplan:2009mh}. For example, for the configuration in \fref{pm_to_perfect_orientation}, the $C$ matrix becomes

{\small
\beq
C=\left(\begin{array}{c|cccccccc}
& \ \ 1 \ \ & \ \ 2 \ \ & \ \ 3 \ \ & \ \ 4 \ \ & \ \ 5 \ \ & \ \ 6 \ \ & \ \ 7 \ \ & \ \ 8 \ \ \\ \hline
1 \ & 1 & c_{12} & c_{13} & 0 & c_{15} & 0 & 0 & c_{18} \\ 
4 \ & 0 & c_{42} & c_{43} & 1 & c_{45} & 0 & 0 & c_{48} \\ 
6 \ & 0 & c_{62} & c_{63} & 0 & c_{65} & 1 & 0 & c_{68} \\ 
7 \ & 0 & c_{72} & c_{73} & 0 & c_{75} & 0 & 1 & c_{78}
\end{array}\right).
\label{C_matrix_example}
\eeq
}
The subspace parametrized by the constrained matrix $C$ associated to a bipartite graph when edge weights are restricted to be $\mathbb{R}\geq 0$, is a cell in the positive Grassmannian.

The previous discussion makes the connection between bipartite graphs and cells in the Grassmannian relatively natural. White and black nodes can be interpreted as MHV and $\overline{\rm{MHV}}$ 3-point amplitudes and, intuitively, the graph provides a picture of a scattering process in which external nodes represent scattered particles and internal faces correspond to loops.

\bigskip

\section{Kasteleyn Technology for General BFTs}

\label{section_Kasteleyn}

In this section we introduce an efficient method for finding the perfect matchings of a general bipartite graph, generalizing the approach based on the Kasteleyn matrix to graphs that might contain boundary nodes. These techniques will play an essential role in the efficient computation of moduli spaces.

We begin by defining the {\it master Kasteleyn matrix} $K_0$, as the adjacency matrix of the graph in which rows are indexed by white nodes and columns are indexed by black nodes, i.e. for every edge in the bipartite graph between nodes ${\bf w}_\mu$ and ${\bf b}_\nu$, we introduce a contribution to the $K_{0,\mu \nu}$ entry. We separate white nodes into two sets $W_i$ and $W_e$, corresponding to internal and external (i.e. boundary) nodes, respectively. Similarly, we split black nodes into $B_i$ and $B_e$. This separation is independent of the number of boundary components and of how external nodes are distributed among them. The individual numbers of internal and external nodes need not be the same for different colors. Furthermore, the total numbers of white and black nodes need not be equal, either. $K_0$ takes the general form

\beq
K_0 = \left(\begin{array}{c|c|c} & \ \ B_i \ \ & \ \ B_e \ \ \\ \hline
W_i & * & * \\ \hline
W_e & * & 0 \\
\end{array}
\right).
\eeq

\begin{figure}[h]
 \centering
 \begin{tabular}[c]{ccc}
 \epsfig{file=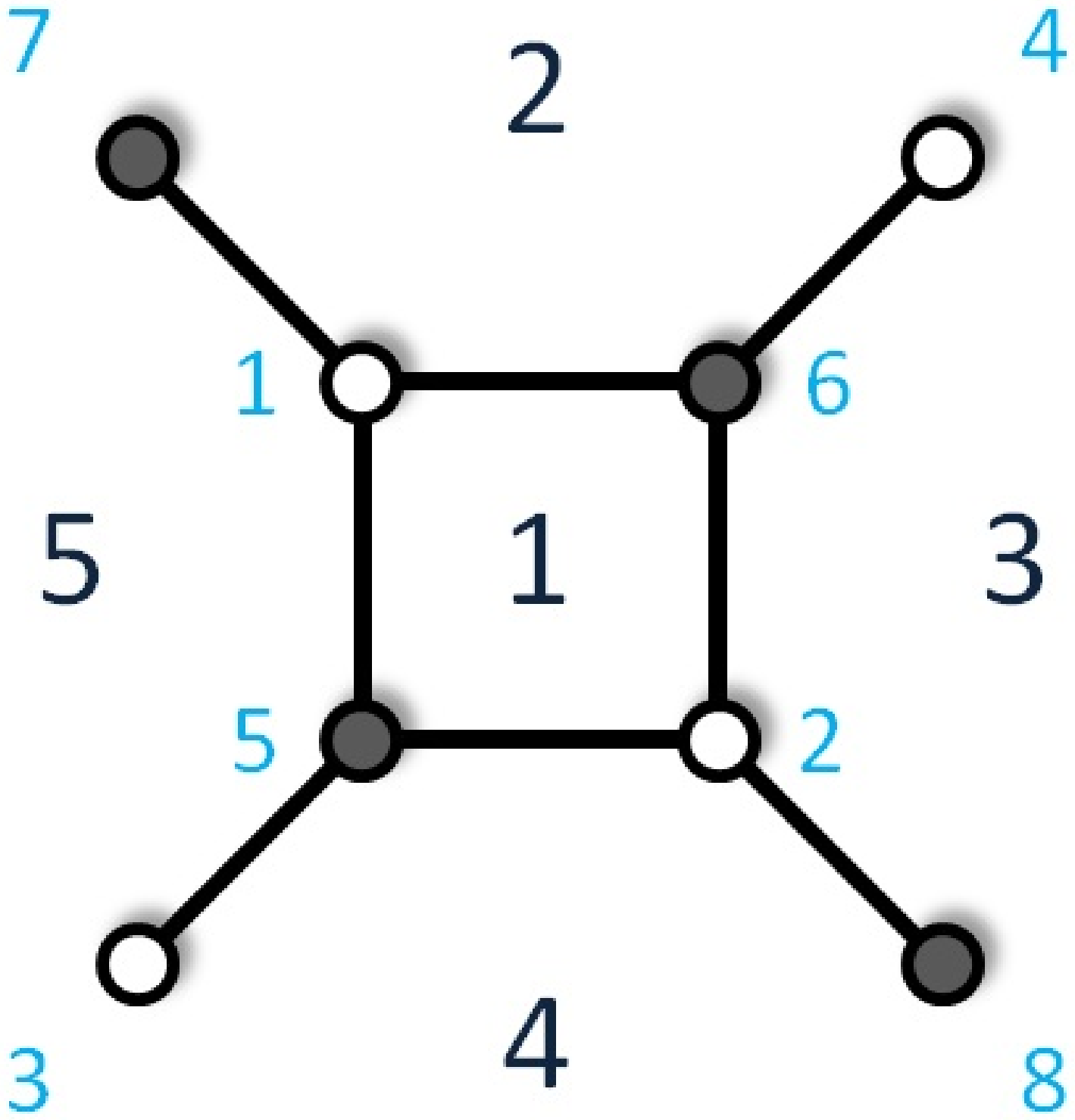,width=0.28\linewidth,clip=} & \ \ \ \ \ \ \ \ &
\epsfig{file=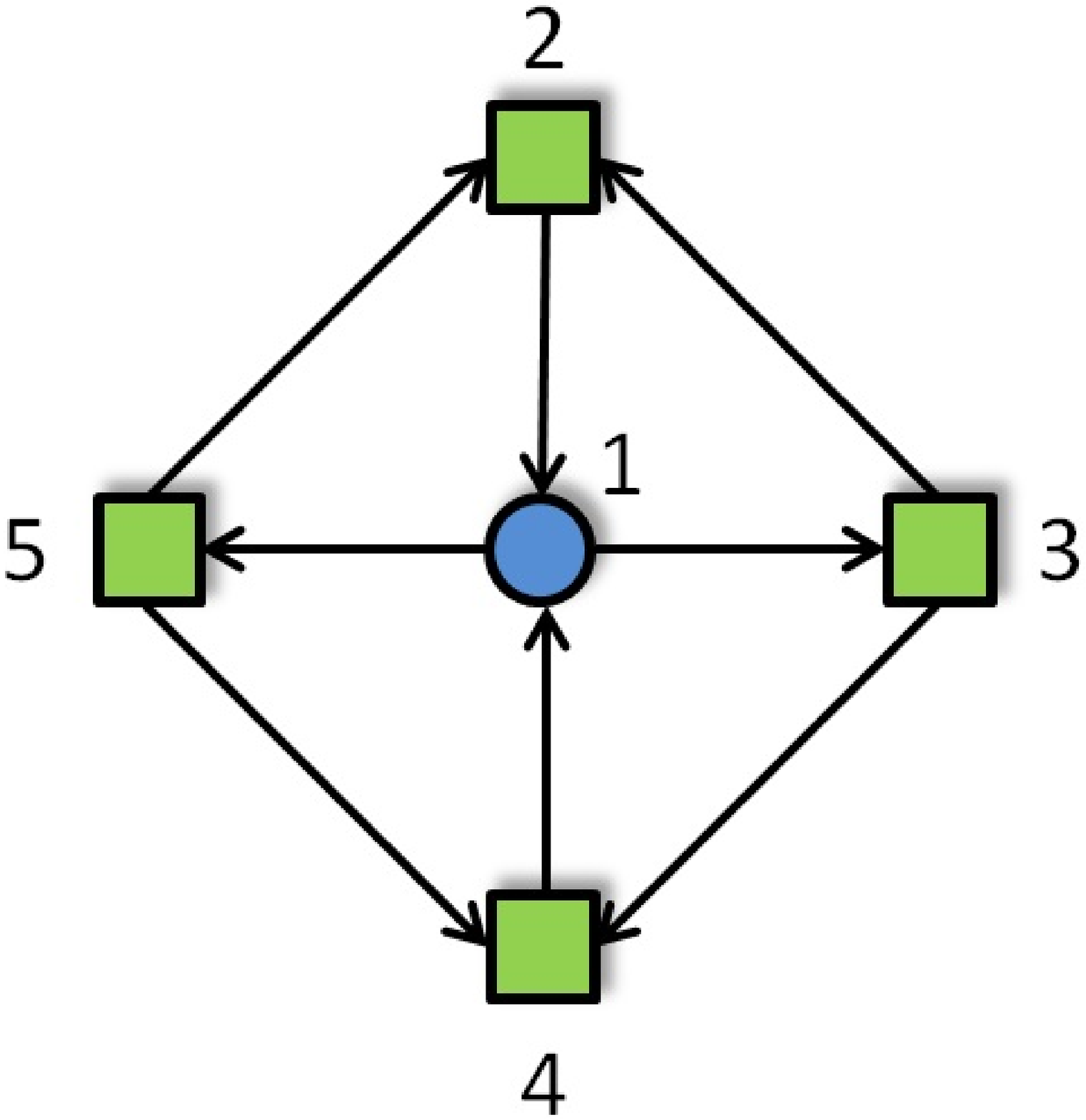,width=0.31\linewidth,clip=} \\ 
\mbox{(a)} & & \mbox{(b)}
 \end{tabular}
\caption{A BFT on a disk. a) The corresponding graph has one internal and four external faces. It also has four white nodes (two internal and two external) and four black nodes (two internal and two external). b) The associated quiver diagram.}
\label{toric_Gr24} 
\end{figure} 

Let us illustrate these ideas with the simple BFT shown in \fref{toric_Gr24}.a, which is related to a leading singularity in the scattering of 2 negative helicity and 2 positive helicity gluons at 1-loop. \fref{toric_Gr24}.b shows the corresponding quiver diagram, for which the superpotential is

\beq
W=X_{15} X_{52} X_{21}-X_{13} X_{32} X_{21}+X_{13} X_{34} X_{41}-X_{15} X_{54} X_{41},
\eeq
where color indices and their contractions are implicit. The master Kasteleyn matrix is
\beq
K_0 = \left(\begin{array}{c|cc|cc} 
 & \ \ 5 \ \ & \ \ 6 \ \ & \ \ 7 \ \ & \ \ 8 \ \ \\ \hline
\ 1 \ & \ X_{15} \ & \ X_{21} \ & \ X_{52} \ & 0 \\
\ 2 \ & X_{41} & X_{13} & 0 & \ X_{34} \ \\ \hline
\ 3 \ & X_{54} & 0 & 0 & 0 \\
\ 4 \ & 0 & X_{32} & 0 & 0 
\end{array}\right).
\label{K0_Gr24}
\eeq

For any subsets $W_{e,del} \subseteq W_e$ and $B_{e,del} \subseteq B_e$ of the boundary nodes, we define the reduced Kasteleyn matrix: 

\begin{eqnarray}
K_{\left( W_{e,del},B_{e,del} \right)} & \equiv & \mbox{matrix resulting from starting from $K_0$ and deleting the rows} \nonumber \\ 
& & \mbox{in $W_{e,del}$ and the columns in $B_{e,del}$}
\end{eqnarray}

All perfect matchings in the graph are then encoded in the polynomial
\beq
\mathcal{P}=\sum_{W_{e,del},B_{e,del}} \det K_{\left( W_{e,del},B_{e,del} \right)},
\label{generalized_Kasteleyn}
\eeq
where the sum runs over all possible subsets $W_{e,del}$ and $B_{e,del}$ of the external nodes such that the resulting reduced Kasteleyn matrices are square. Every term in this polynomial, which we denote $\mathcal{P}_\mu$, is interpreted as the product of edges in a perfect matching.

\begin{figure}[h]
\begin{center}
\includegraphics[width=3.2cm]{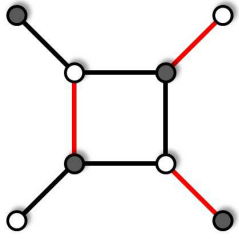}
\caption{Perfect matching generated by $\det K_{\left( W_{e,del}=\{3\},B_{e,del}=\{7\} \right)}$. Edges in the perfect matching are indicated in red.}
\label{pm_Gr24_1}
\end{center}
\end{figure}

Let us explain in more detail the reason for the sum over reduced Kasteleyn matrices in \eref{generalized_Kasteleyn}. The determinant of each $K_{\left( W_{e,del},B_{e,del}\right)}$ in \eref{generalized_Kasteleyn} generates all the perfect matchings containing the edges connected to the external nodes in $(W_e-W_{e,del})$ and $(B_e-B_{e,del})$. Once again, let us show how this works in the example in \fref{toric_Gr24}, for which $K_0$ is given in \eref{K0_Gr24}. Let us first consider $W_{e,del}=\{3\}$ and $B_{e,del}=\{7\}$, which results in

\beq
K_{\left( W_{e,del}=\{3\},B_{e,del}=\{7\} \right)} = \left(\begin{array}{c|cc|cc} 
 & \ \ 5 \ \ & \ \ 6 \ \ & \ \ 8 \ \ \\ \hline
\ 1 \ & \ X_{15} \ & \ X_{21} \ & 0 \\
\ 2 \ & X_{41} & X_{13} & \ X_{34} \ \\ \hline
\ 4 \ & 0 & X_{32} & 0 
\end{array}\right).
\eeq
Taking its determinant, we obtain 
\beq
\det K_{\left( W_{e,del}=\{3\},B_{e,del}=\{7\} \right)} = - X_{15} X_{32} X_{34}.
\eeq
This is the only perfect matching that contains the edges connected to the surviving external nodes 3 and 8, and we show it in \fref{pm_Gr24_1}. Generically, each reduced Kasteleyn matrix can give rise to multiple perfect matchings. 

Computing the full polynomial $\mathcal{P}$ in \eref{generalized_Kasteleyn}, we obtain
{\small
\beq
\mathcal{P} = X_{13} X_{15}  - X_{21} X_{41} + X_{32} X_{41} X_{52} - X_{15} X_{32} X_{34}  + X_{21} X_{34} X_{54} - X_{13} X_{52} X_{54} + X_{32} X_{34} X_{52} X_{54} ,
\eeq} \noindent which corresponds to the seven perfect matchings in \fref{pm_Gr24_1}. Given the definition in \eref{Xi_to_pmu}, it is very easy to find the matrix $P$ in terms of the polynomial $\mathcal{P}$. It is given by 

\beq
P_{i\mu}=\left. \left| {\partial \mathcal{P}_\mu \over \partial X_i} \right| \right|_{\mbox{all $X_j=1$}},
\eeq
where, as previously defined, $\mathcal{P}_\mu$ indicates the term in $\mathcal{P}$ associated to the perfect matching $p_\mu$. For our example, we obtain

{\small
\beq
P=
\left(
\begin{array}{c|ccccccc}
& \ p_1 \ & \ p_2 \ & \ p_3 \ & \ p_4 \ & \ p_5 \ & \ p_6 \ & \ p_7 \ \\ \hline
\ X_{13} \ & 1 & 0 & 0 & 0 & 0 & 1 & 0 \\
X_{15} & 1 & 0 & 0 & 1 & 0 & 0 & 0 \\
X_{21} & 0 & 1 & 0 & 0 & 1 & 0 & 0 \\
X_{32} & 0 & 0 & 1 & 1 & 0 & 0 & 1 \\
X_{34} & 0 & 0 & 0 & 1 & 1 & 0 & 1 \\
X_{41} & 0 & 1 & 1 & 0 & 0 & 0 & 0 \\
X_{52} & 0 & 0 & 1 & 0 & 0 & 1 & 1 \\
X_{54} & 0 & 0 & 0 & 0 & 1 & 1 & 1
\end{array}
\right).
\eeq}

\begin{figure}[h]
\begin{center}
\includegraphics[width=10.5cm]{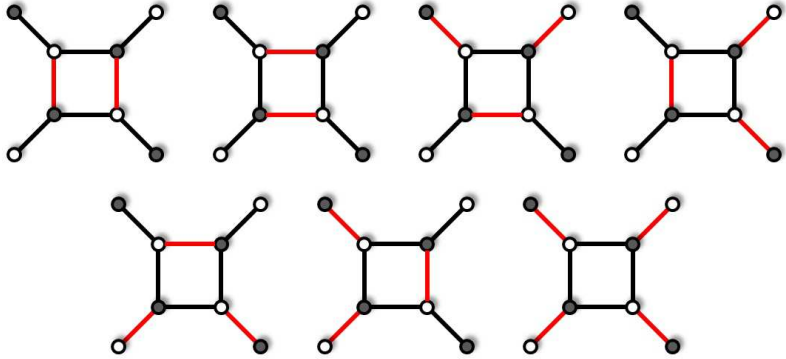}
\caption{The seven perfect matchings for the BFT in \fref{toric_Gr24}. Edges in the perfect matchings are indicated in red.}
\label{pms_Gr24}
\end{center}
\end{figure}

\section{BFTs and Calabi-Yau's: Moduli Spaces}

\label{section_BFTs_and_CYs}

A remarkable feature of BFTs, which is at the center of their special properties, is that perfect matchings extremely simplify the computation of their moduli space. The moduli spaces are automatically toric and perfect matchings are in one-to-one correspondence with fields in their gauged linear sigma model (GLSM) description. In this section we discuss this calculation, using the example in \fref{toric_Gr24} to illustrate our ideas. Indeed, perfect matchings automatically satisfy F-term equations in the gauge theory. This property was originally identified in the context of dimer models on $T^2$ in \cite{Franco:2005rj}, and a detailed proof of how perfect matchings parametrize the moduli space of the corresponding theories was given in \cite{Franco:2006gc}. We now briefly review the arguments in these papers, which extend without changes to general BFTs.

\bigskip

\subsection{F-Flatness and Perfect Matchings}

\begin{figure}[h]
\begin{center}
\includegraphics[width=8.5cm]{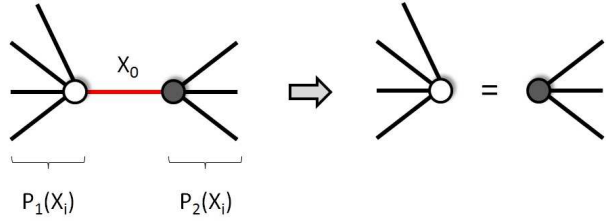}
\caption{Graphic representation of the F-term equations in a BFT.}
\label{F_terms_graphical}
\end{center}
\end{figure}

The map between chiral fields in the BFT and perfect matchings given in \eref{X_pm_map} implies that F-term equations are trivially satisfied, as we now review.  The vanishing of F-terms for fields associated to external legs is not imposed.\footnote{Since these fields appear in a single superpotential term, they would set to zero the product of fields they are coupled to. This special treatment of external legs is motivated by the connection to geometry, which we develop in this section.} For any bifundamental field
$X_0$ associated to an internal edge, we have

\beq
W=X_0 P_1(X_i)- X_0 P_2(X_i) +\ldots,
\eeq
where we have identified the only two terms in the superpotential containing $X_0$. $P_1(X_i)$ and  $P_2(X_i)$ are products of bifundamentals fields. The F-term equation for $X_0$ takes the form

\beq
\partial_{X_0}W =0 \ \ \ \iff \ \ \ P_1(X_i)=P_2(X_i).
\eeq
This equation has a simple graphic representation as shown in \fref{F_terms_graphical}. After removing $X_0$, the product of edges connected to node 1 needs to be equal to the product of edges connected to node 2. Using \eref{X_pm_map}, this becomes

\beq
\prod_{i \in P_1} \prod_\mu p_\mu^{P_{i\mu}}=\prod_{i \in P_2} \prod_\mu p_\mu^{P_{i\mu}}.
\label{F_terms_pms}
\eeq
But this equation is automatically satisfied. Since nodes 1 and 2 are precisely separated by a single edge, every perfect matching that appears on the L.H.S. of \eref{F_terms_pms} also appears on its R.H.S.

\smallskip

\subsection{The Master Space}

\label{section_master_space}

The first step in our discussion of the vacuum structure of BFTs is the concept of {\it master space}, which was introduced for arbitrary $\mathcal{N}=1$ field theories in \cite{Forcella:2008bb}.\footnote{The concept of master space extends to SUSY theories in other dimensions.} The master space is defined as the space of solutions to F-term equations. Since D-terms are not imposed, we can regard the master space as the full moduli space of the gauge theory, including baryonic directions.

Following the discussion in the previous section, the master space of a BFT is naturally parametrized in terms of perfect matchings. For a BFT, the master space is toric, i.e. it can be described by a GLSM. In GLSM language, F-term conditions can be translated to certain $U(1)$ charges of the perfect matchings, which are encoded in a charge matrix $Q_F$ defined as

\beq
Q_F= Ker P .
\eeq
The toric diagram of the master space is given by $Ker \, Q_F$, which is indeed $P$. In other words, the matrix $P$ connecting chiral fields in the quiver to perfect matchings gives the positions of points in the toric diagram of the master space! It is interesting to note that a few months before the general concept of master space was introduced in \cite{Forcella:2008bb}, the same object was constructed in the mathematics literature in \cite{Postnikov_toric}, from a different point of view, for the restricted case of bipartite graphs on a disk.  In that work, the matrix $P$ was referred to as the {\it matching polytope}. Our interpretation of the graph as defining a gauge theory makes the emergence of this geometry absolutely natural and allows its generalization to bipartite graphs on arbitrary Riemann surfaces. 

\begin{figure}[h]
 \centering
 \begin{tabular}[c]{ccc}
 \epsfig{file=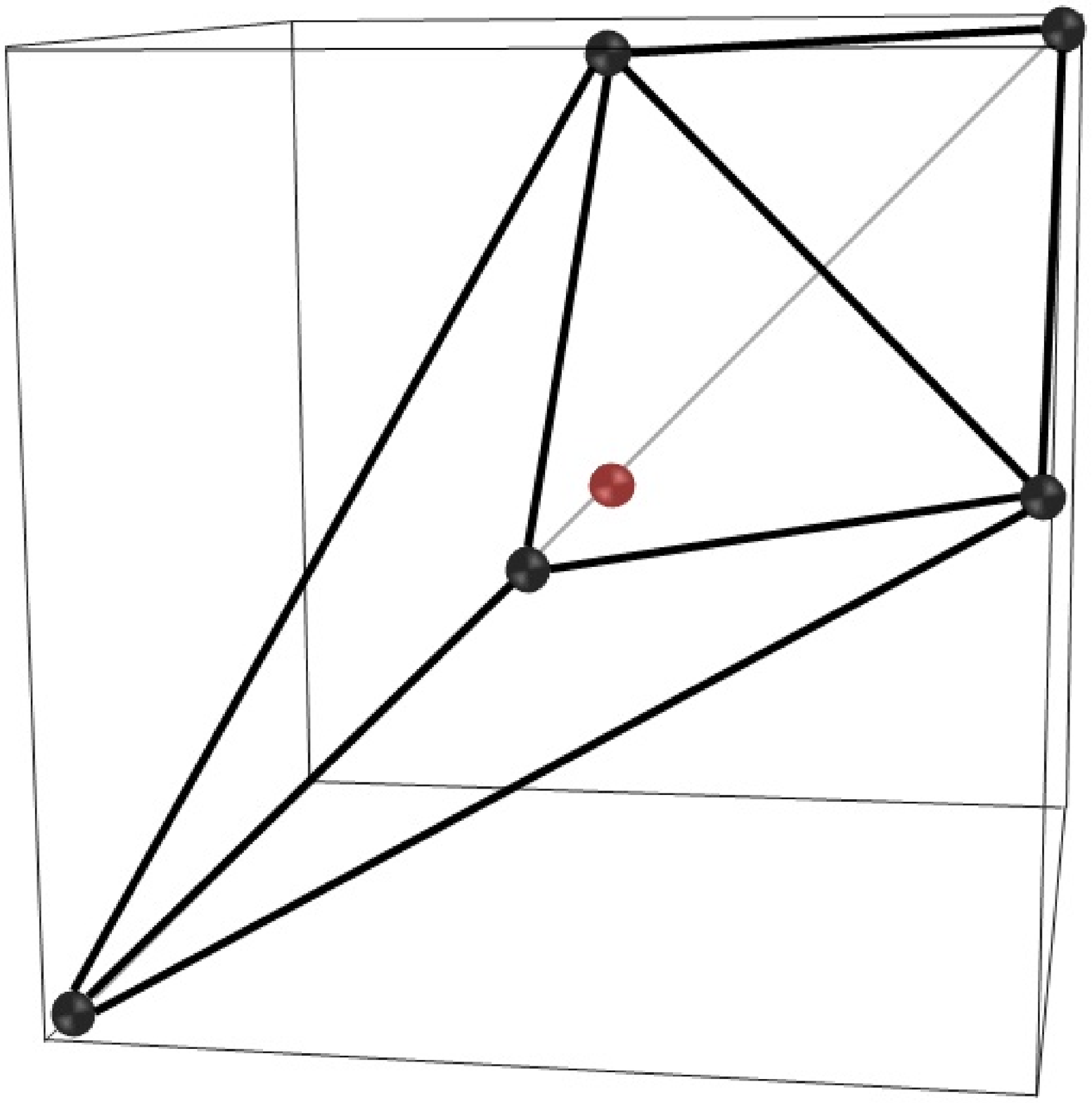,width=0.32\linewidth,clip=} & \ \ \ \ &
\epsfig{file=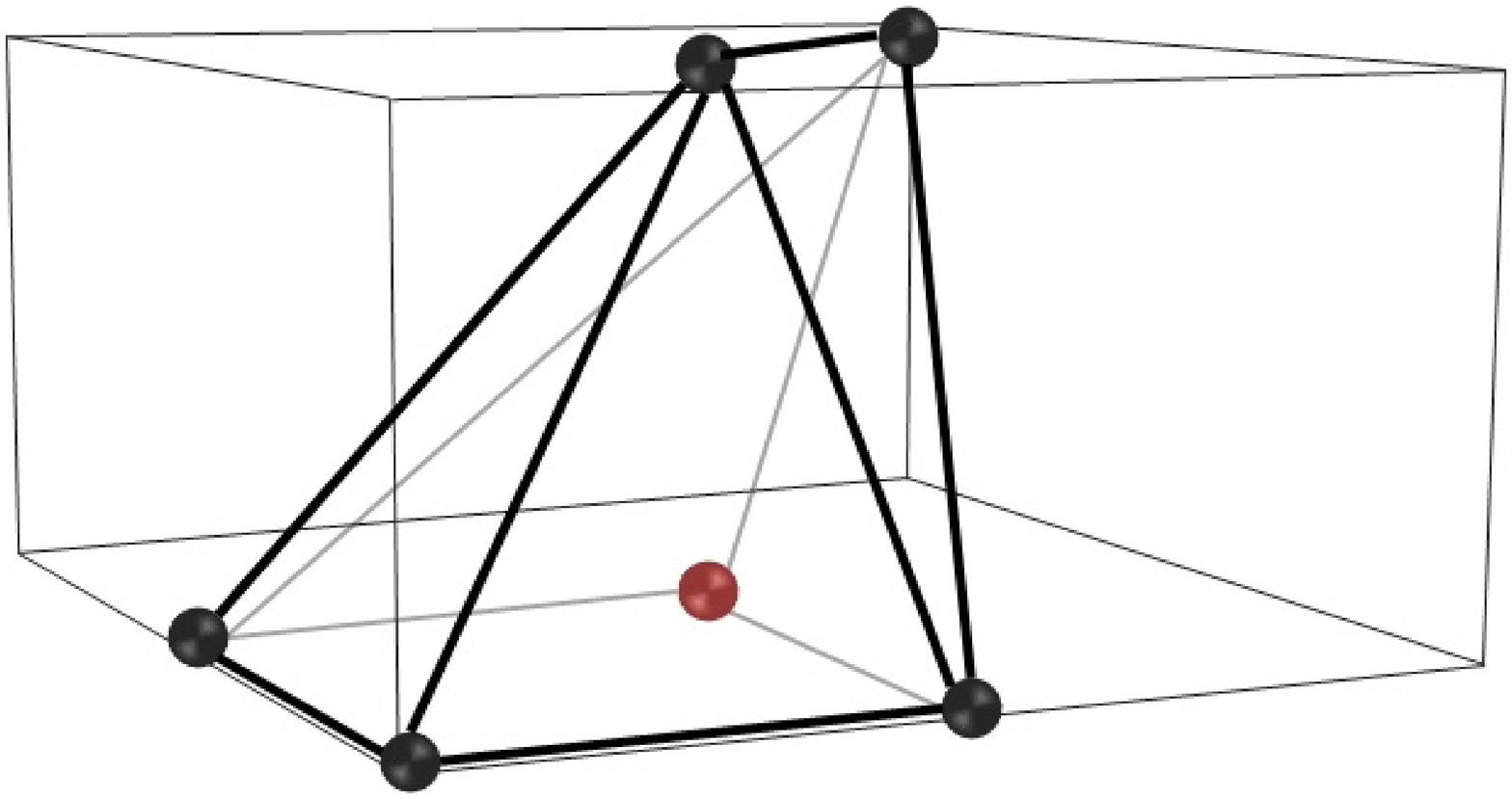,width=0.4\linewidth,clip=} \\ 
\mbox{(a)} & & \mbox{(b)}
 \end{tabular}
\caption{Two projections of the toric diagram for the master space, corresponding to \eref{G_master_Gr24}. Points descending from multiple ones in 5d are shown in red. The projections correspond to keeping the following combinations of rows: a) $(G_{mast,1},G_{mast,2},G_{mast,3})$ and b) $(G_{mast,2},G_{mast,3},G_{mast,4})$.}
\label{toric_master_Gr24} 
\end{figure}

As we have just said, the toric diagram for the master space is given by the $P$ matrix. In order to obtain a better idea of this geometry, it is useful to consider the row-reduced version of $P$ which, for the example at hand, becomes 

\beq
G_{mast}=\left(
\begin{array}{ccccccc}
\ p_1 \ & \ p_2 \ & \ p_3 \ & \ p_4 \ & \ p_5 \ & \ p_6 \ & \ p_7 \ \\ \hline
 1 & 0 & 0 & 0 & 0 & 1 & 0 \\
 0 & 1 & 0 & 0 & 0 & -1 & -1 \\
 0 & 0 & 1 & 0 & 0 & 1 & 1 \\
 0 & 0 & 0 & 1 & 0 & -1 & 0 \\
 0 & 0 & 0 & 0 & 1 & 1 & 1
\end{array}
\right).
\label{G_master_Gr24}
\eeq
We conclude that the master space is a 5-complex dimensional toric geometry with a toric diagram consisting of seven different points. From now on, every time we mention the dimension of a Calabi-Yau manifold, we refer to its complex dimension. Furthermore, the entries in every column of $G_{mast}$ add up to 1, implying the master space is Calabi-Yau. In fact, the Calabi-Yau property will be exhibited by the master spaces of all models considered in this paper. Since the toric diagram lives on a hyperplane at distance 1 from the origin, we can project it down to 4 dimensions by, for example, considering only four of the rows in \eref{G_master_Gr24}. A convenient way of visualizing this 4d toric diagram is by considering different 3d projections, as shown in \fref{toric_master_Gr24}. Different points in the 5d toric diagram might be projected down to the same point in 3d. Such points are indicated in red in \fref{toric_master_Gr24}. 

The interior of the toric diagram of the master space, i.e. of the matching polytope, provides a graphical representation of the corresponding cell in the positive Grassmannian. The BFT interpretation of the lower dimensional sub-cells on its boundary will be discussed in Section \ref{section_boundary_operator}.

Taking the kernel of $P$, or equivalently of $G_{mast}$, we obtain the charge matrix that implements the F-terms

\beq
Q_F=\left(
\begin{array}{ccccccc}
\ p_1 \ & \ p_2 \ & \ p_3 \ & \ p_4 \ & \ p_5 \ & \ p_6 \ & \ p_7 \ \\ \hline
 0 & 1 & -1 & 0 & -1 & 0 & 1 \\
 -1 & 1 & -1 & 1 & -1 & 1 & 0 
\end{array}
\right).
\label{QF_square_4legs}
\eeq

\bigskip
\bigskip

\subsection{The Mesonic Moduli Space}

The {\it mesonic moduli space}, is another natural geometry associated to any BFT. For shortness, we will refer to it simply as the {\it moduli space} from now on. The moduli space of any gauge theory is the vacuum space of solutions of both vanishing F and D-terms. It is thus a projection of the master space onto the subspace of vanishing D-terms. 

There is a D-term contribution for each gauge group in the BFT i.e., by means of the dictionary introduced in Section \ref{section_dictionary}, for every internal face in the bipartite graph. It is convenient to define the charge matrix $\Delta$ of the BFT, as the matrix encoding how every chiral field transforms under the gauge symmetries.\footnote{For the purpose of this paper, it is sufficient to proceed as if every gauge group was $U(1)$.} The matrix $\Delta$ is an $n_{fields} \times n_{gauge} \equiv n_{edges} \times n_{int. \ faces}$ matrix in which rows correspond to chiral fields and columns correspond to gauge groups. For the row associated to $X_{ij}$, the non-zero entries are a $1$ for the $i^{th}$ column and a $-1$ for the $j^{th}$ column. All entries are zero in rows associated to adjoint fields $X_{ii}$.  D-terms can then be encoded in a charge matrix $Q_D$ giving the charge of perfect matchings under the gauge groups. This means that $Q_D$ is defined such that

\beq
P \cdot Q_D^T = \Delta .
\label{Q_D_general}
\eeq 
It is clear that \eref{Q_D_general} does not determine $Q_D$ uniquely. Any solution to this equation is equivalent for the purpose of determining the moduli space.

For our example, there is a single gauge group associated to face 1. Under it, $X_{13}$ and $X_{15}$ have charge $1$,  $X_{21}$ and $X_{41}$ have charge $-1$, and all other fields are neutral. It is straightforward to verify that the following $Q_D$ does the right job

\beq
Q_D=\left(
\begin{array}{ccccccc}
\ p_1 \ & \ p_2 \ & \ p_3 \ & \ p_4 \ & \ p_5 \ & \ p_6 \ & \ p_7 \ \\ \hline
 0 & 0 & -1 & 1 & -1 & 1 & 0
\end{array}
\right).
\label{QD_square_4legs}
\eeq

The next step in the determination of the moduli space is to concatenate $Q_F$ and $Q_D$ into a single charge matrix $Q$
\beq
Q=\left(\begin{array}{c} Q_F \\ Q_D \end{array}\right).
\eeq
The toric diagram of the moduli space is thus encoded in a matrix $G$ such that

\beq
G=Ker \, Q.
\eeq

Let us consider our example. From \eref{QF_square_4legs} and \eref{QD_square_4legs}, we obtain

\beq
G=\left(
\begin{array}{ccccccc}
\ p_1 \ & \ p_2 \ & \ p_3 \ & \ p_4 \ & \ p_5 \ & \ p_6 \ & \ p_7 \ \\ \hline
 -1 & -1 & 0 & 0 & 0 & 0 & 1 \\
 1 & 1 & 1 & 0 & 0 & 1 & 0 \\
 0 & 0 & -1 & 0 & 1 & 0 & 0 \\
 1 & 1 & 1 & 1 & 0 & 0 & 0 
\end{array}
\right).
\label{G_square_4legs}
\eeq

The moduli space is a 4d toric manifold. As for the master space, the entries in every column add up to 1, implying the moduli space is also a Calabi-Yau manifold. This will also be the case for all the examples considered in the paper which, together with our previous observation regarding the master space, leads us to conjecture that

\bigskip
\begin{center}
\begin{tabular}{|c|} \hline 
{\bf Conjecture:}  \\
\ \ \ The master and moduli spaces of every BFT are toric Calabi-Yau manifolds. \ \ \ \\ \hline
\end{tabular}
\end{center}
\bigskip
We expect the existence of a simple proof of this statement based on the combinatorics of $Q_F$ and $Q_D$.
 
We observe an interesting phenomenon: there can be non-trivial multiplicities of perfect matchings associated to the same point in the toric diagram. In particular, we see that the point $(-1,1,0,1)$ corresponds to both $p_1$ and $p_2$. Such multiplicities are generic in BFTs. For example, for the case of D3-branes probing toric CY 3-folds, trying to understand them has been an important factor leading to the correspondence between the associated quivers and dimer models \cite{Hanany:2005ve}. The role of multiplicities in the generalized context of BFTs is certainly an interesting question that deserves further investigation.

\begin{figure}[h]
\begin{center}
\includegraphics[width=7cm]{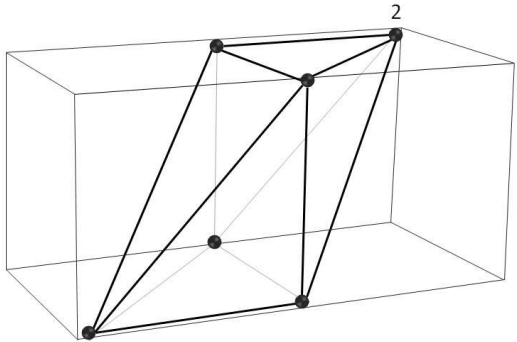}
\caption{Toric diagram for the CY 4-fold that is the moduli space of the 4-leg, 1-loop model. This CY 4-fold is the real cone over $Q^{1,1,1}$. We indicate the non-trivial perfect matching multiplicity of the $(-1,1,0,1)$ point in the toric diagram with a number.}
\label{toric_Gr241}
\end{center}
\end{figure}

Multiplicities can arise for both internal and external points of a toric diagram. We would like to note that the toric diagram in \fref{toric_Gr241} contains corners with multiplicity different from one. In the specific case of gauge theories on D3-branes probing toric CY 3-folds (i.e. BFTs on $T^2$ with no boundaries), this feature has been identified as an indication of an inconsistency. In fact, for this class of theories, this behavior is directly connected to the existence of self-intersecting zig-zag paths \cite{Hanany:2005ss}. This is clearly not the case here, as we can verify by explicit determination of the zig-zag paths. At this time, we are not aware of any pathology signaled by this behavior.\footnote{This diagnostic quite probably does not apply to other classes of theories beyond those on $T^2$. For instance, there are known examples of gauge theories in 2+1 dimensions whose moduli space have toric diagrams with corner multiplicities and that do not have any known problem \cite{Aganagic:2009zk}.} In any case, we are confident that BFTs without this feature exist. Such models can be analyzed with exactly the same methods we have applied here.

Since the toric diagram lives on a hyperplane at distance 1 from the origin, it can be projected down to three dimensions by, for example, considering any three of the rows in \eref{G_square_4legs}. \fref{toric_Gr241} shows the toric diagram for this model. This is indeed a well-known geometry, the real cone over the 7-dimensional Sasaki-Einstein manifold $Q^{1,1,1}$. This CY 4-fold has been extensively investigated in connection to other types of gauge theories associated to M2-branes \cite{Franco:2008um,Franco:2009sp,Aganagic:2009zk,Benini:2009qs}.

The moduli space of a gauge theory is invariant under Seiberg duality. More abstractly, in graph theoretic language, this means that the moduli space of a BFT provides a natural geometry associated to a bipartite graph on a Riemann surface that, by construction, is invariant under square moves. In Section \ref{section_Seiberg_duality}, we will discuss in detail the implications and applications of this fact.

\smallskip

\section{Additional Examples: Increasing Boundaries and Genus}

\label{section_examples}

BFTs associated to graphs without boundaries have been extensively studied in the literature. Tilings of $T^2$ describe the gauge theories on D3-branes over toric CY 3-folds \cite{Franco:2005rj,Franco:2005sm} and tilings on higher genus Riemann surfaces arise when acting on them with the untwisting map as discussed on Section \ref{section_BFTs_everywhere} \cite{Feng:2005gw}. For this reason, we emphasize in this section the novel case of BFTs with boundaries, which are also the ones that are relevant for scattering amplitudes. We start discussing a model on the disk and soon move to theories that have never been studied before: models with multiple boundaries and higher genus. It is natural to expect such configurations to be relevant for leading singularities beyond the planar limit. Further studies of non-planar graphs will appear in \cite{FGS}.

We will put special emphasis in the geometry of the corresponding master and moduli spaces. They can be determined in terms of perfect matchings following the general procedure introduced in Section \ref{section_BFTs_and_CYs}.

\subsection{Another Example on the Disk: The Hexagon-Square Model}

\label{section_hexagon-square}

\begin{figure}[h]
 \centering
 \begin{tabular}[c]{ccc}
 \epsfig{file=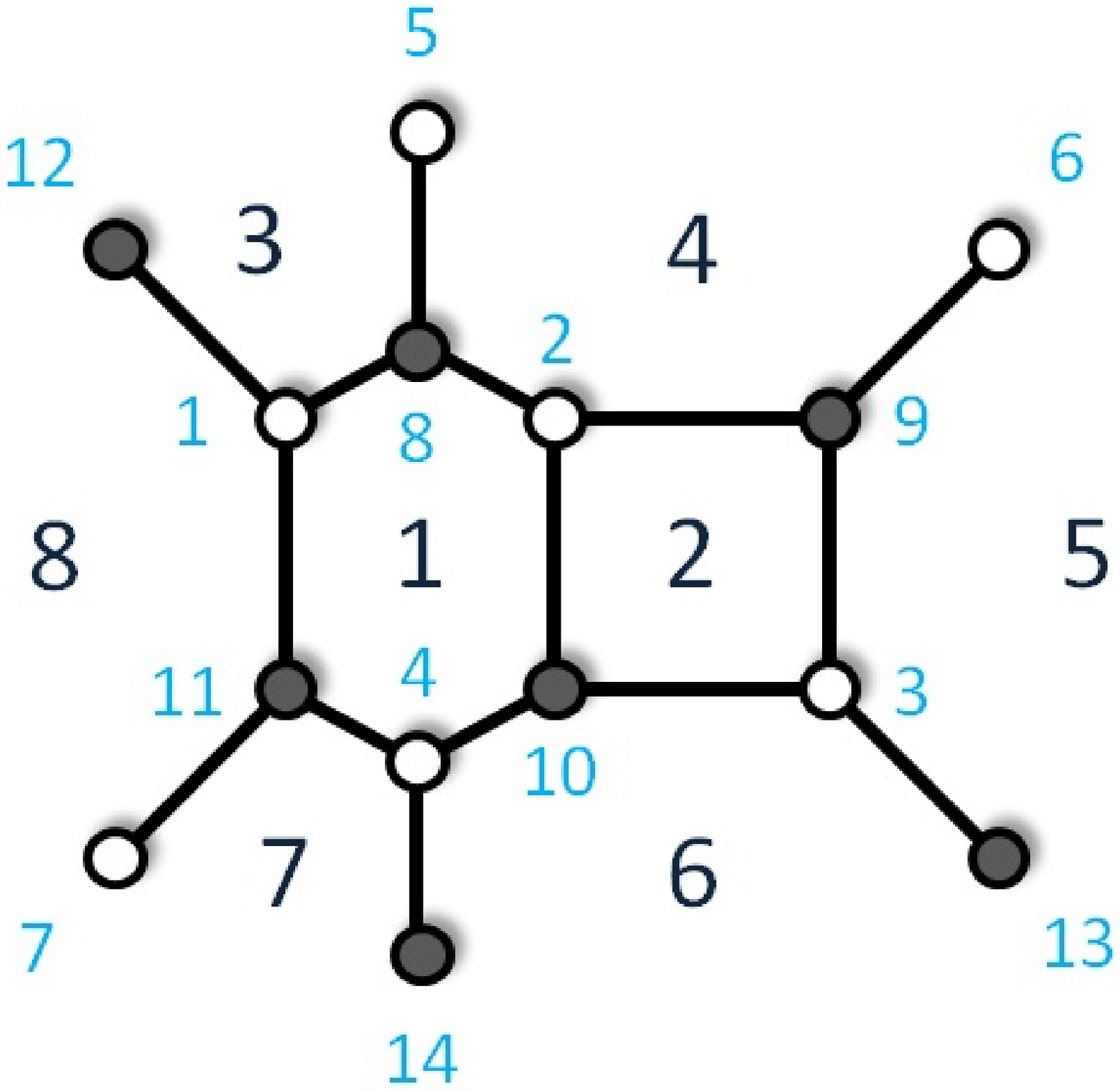,width=0.4\linewidth,clip=} & \ \ \ \ \ &
\epsfig{file=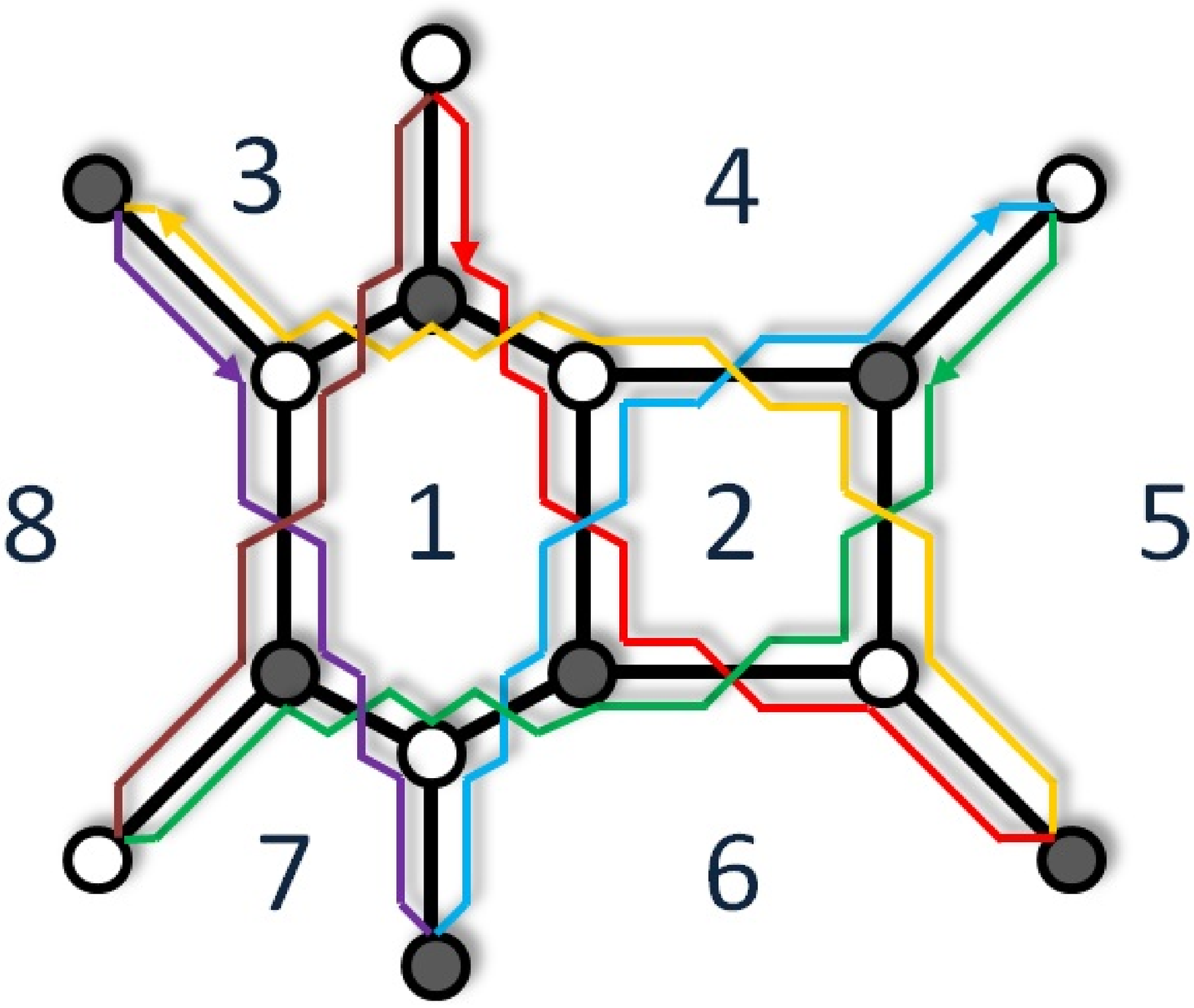,width=0.37\linewidth,clip=} \\ 
\mbox{(a)} & & \mbox{(b)}
 \end{tabular}
\caption{a)	Bipartite graph for the hexagon-square model. It contains two internal and six external faces. b) The six zig-zag paths for this model. We see that it does not have any self-intersecting zig-zag path.}
\label{tiling_hexagon_square} 
\end{figure}

Let us consider the 2-loop graph, shown in \fref{tiling_hexagon_square}, corresponding to the scattering of 3 negative helicity and 3 positive helicity gluons. For pedagogical reasons, we present the full details of the calculation of its master and moduli spaces in Appendix \ref{appendix_4legs_2loops}. The treatment of other examples in the paper will be briefer and will only emphasize the main results. The master Kasteleyn matrix for this model is

{\small
\beq
K_0 = \left(\begin{array}{c|cccc|ccc} 
 & \ \ 8 \ \ & \ \ 9 \ \ & \ \ 10 \ \ & \ \ 11 \ \ & \ \ 12 \ \ & \ \ 13 \ \ & \ \ 14 \ \ \\ \hline
\ 1 \ \ & \ X_{31} \ & 0 & 0 & \ X_{18} \ & \ X_{83} \ & 0 & 0 \\
\ 2 \ \ & \ X_{14} \ & \ X_{42} \ & \ X_{21} \ & 0 & 0 & 0 & 0 \\
\ 3 \ \ & 0 & X_{25} & X_{62} & 0 & 0 & \ X_{56} \ & 0 \\
\ 4 \ \ & 0 & 0 & X_{16} & X_{71} & 0 & 0 & \ X_{67} \ \\ \hline
\ 5 \ \ & X_{43} & 0 & 0 & 0 & 0 & 0 & 0 \\
\ 6 \ \ & 0 & X_{54} & 0 & 0 & 0 & 0 & 0 \\
\ 7 \ \ & 0 & 0 & 0 & X_{87} & 0 & 0 & 0
\end{array}\right).
\label{K0_hexagon_square}
\eeq}

The theory has 25 perfect matchings. The master space is an 8d toric CY. Its toric diagram is given by the matrix

{\scriptsize
\beq
G_{mast}=\left(
\begin{array}{ccccccccccccccccccccccccc}
\ 1\ & \ 0\ & \ 0\ & -1& \ 0\ & -1& \ 0\ & \ 0\ & -1& \ 0\ & \ 0\ & \ 0\ & -1& -1& \ 0\ & \ 0\ & \ 1\ & -1& \ 0\ & \ 0\ & \ 1\ & -2& -1& -1& \ 0 \ \\
0& 1& 0& 1& 0& 0& 0& 0& 0& 0& 0& -1& 1& 0& 0& 0& 0& -1& -1& -1& -1& 1& 1& 1& 1 \\
0& 0& 1& 1& 0& 1& 0& 0& 1& 0& 0& 0& 0& 1& 0& 0& -1& 1& 0& 0& -1& 1& 0& 0& -1 \\
0& 0& 0& 0& 1& 1& 0& 0& 0& 0& 0& 1& 0& 0& -1& 0& -1& 1& 0& 1& 0& 0& -1& 0& -1 \\
0& 0& 0& 0& 0& 0& 1& 0& 0& 0& 0& 0& 0& 0& 1& 0& 1& 0& 1& 0& 1& 0& 1& 0& 1 \\
0& 0& 0& 0& 0& 0& 0& 1& 1& 0& 0& 1& 0& 0& 0& -1& -1& 1& 1& 0& 0& 0& 0& -1& -1 \\
0& 0& 0& 0& 0& 0& 0& 0& 0& 1& 0& 0& 0& 0& 0& 1& 1& 0& 0& 1& 1& 0& 0& 1& 1 \\
0& 0& 0& 0& 0& 0& 0& 0& 0& 0& 1& 0& 1& 1& 1& 1& 1& 0& 0& 0& 0& 1& 1& 1& 1
\end{array}
\right).
\eeq
}

The moduli space is a 6d toric CY. The 25 perfect matchings give rise to 18 different points in its toric diagram, whose positions are captured by the following matrix

\begin{figure}[h]
 \centering
 \begin{tabular}[c]{ccc}
 \epsfig{file=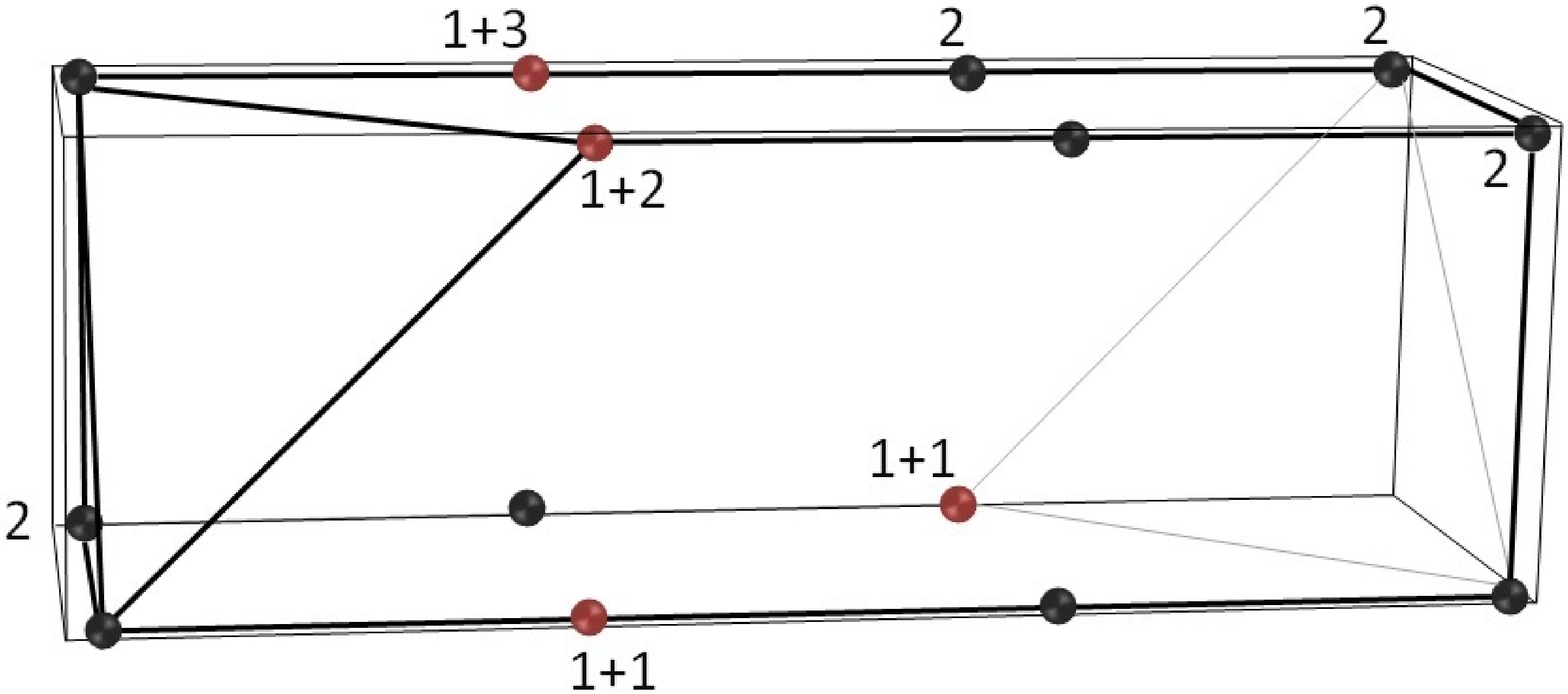,width=0.4\linewidth,clip=} & \ \ \ \ &
\epsfig{file=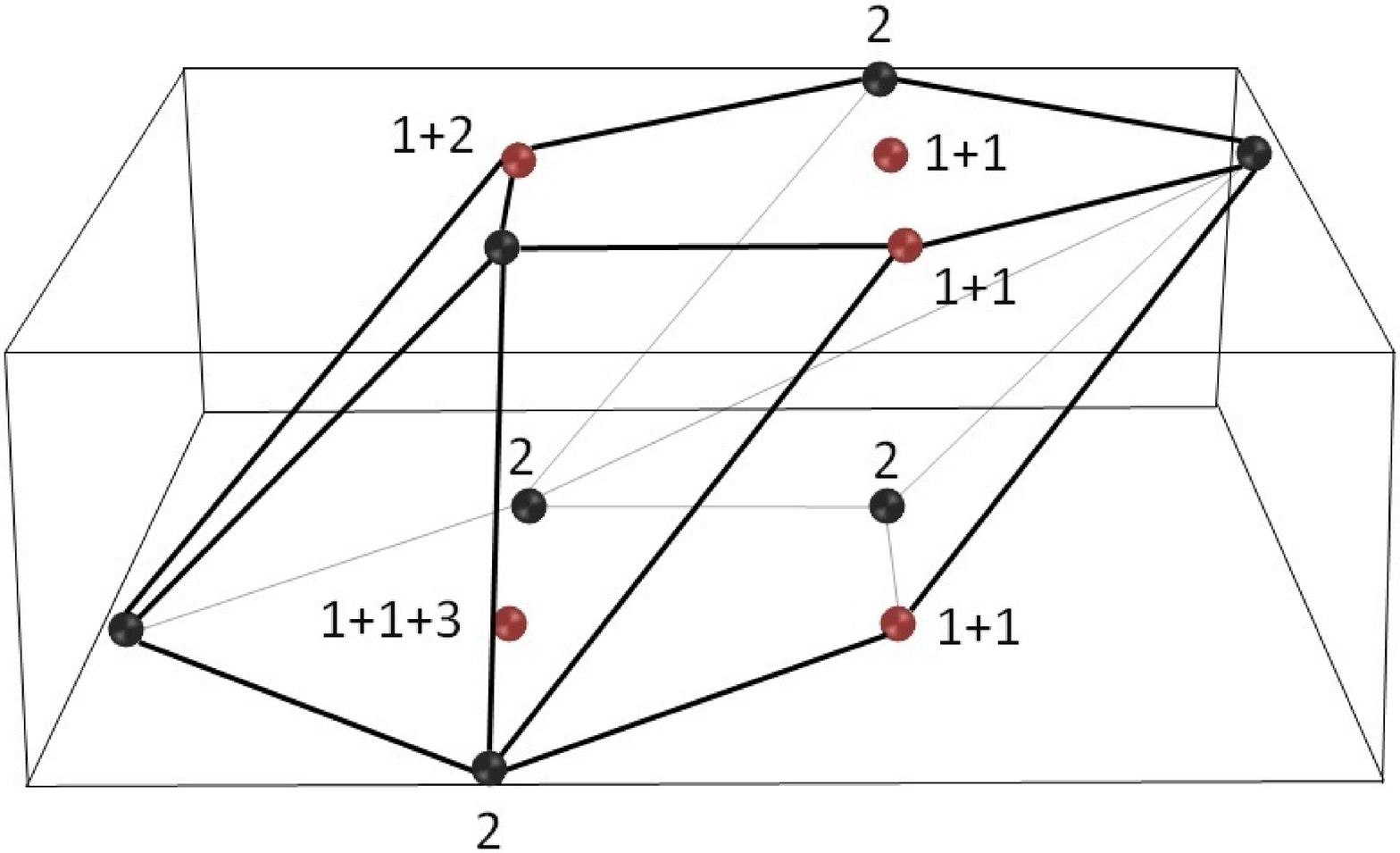,width=0.4\linewidth,clip=} \\ 
\mbox{(a)} & & \mbox{(b)}
 \end{tabular}
\caption{Two projections of the toric diagram corresponding to \eref{G_matrix_multiplicities_hexagon_square}. Points descending from multiple ones in 6d are shown in red. The numbers indicate the non-trivial multiplicity of perfect matchings. The projections correspond to keeping the following combinations of rows: a) $(G_1-G_2+G_3,G_3,G_4+G_6)$ and b) $(G_1,G_2,G_3)$.}
\label{toric_hexagon_square_Model_1} 
\end{figure}

{\footnotesize
\beq
G=\left(\begin{array}{cccccccccccccccccc}
0& 1& -1& -2& -1& -1& -1& 0& 0& -1& 0& 0& -1& 0& 0& -1& 0& 0 \\
0& 0& 0& 1& 0& 1& 1& 1& 1& 2& 1& -1& 0& 0& 0& 1& 0& 1 \\
0& 0& 0& 1& 1& 0& 1& 1& 0& 1& 1& 0& 1& 0& 0& 1& 1& 0 \\
1& 0& 1& 1& 1& 1& 1& 0& 1& 1& 1& 0& 0& 0& 0& 0& 0& 0 \\
0& 0& 1& 0& 0& 0& -1& -1& 0& -1& -1& 1& 0& 0& 1& 0& 0& 0 \\
0& 0& 0& 0& 0& 0& 0& 0& -1& -1& -1& 1& 1& 1& 0& 0& 0& 0 \\ \hline
\ \bf{3} \ & \ \bf{2} \ & \ \bf{2} \ & \ \bf{2} \ & \ \bf{2} \ & \ \bf{2} \ & \ \bf{1} \ & \ \bf{1} \ & \ \bf{1} \ & \ \bf{1} \ & \ \bf{1} \ & \ \bf{1} \ & \ \bf{1} \ & \ \bf{1} \ & \ \bf{1} \ & \ \bf{1} \ & \ \bf{1} \ & \ \bf{1} \ \\ \hline
\end{array}\right),
\label{G_matrix_multiplicities_hexagon_square}
\eeq
}

\noindent where we have introduced a compact notation for the $G$ matrix, in which the last row indicates the multiplicity of perfect matchings for each point in the toric diagram. As already done in Section \ref{section_BFTs_and_CYs}, a useful way of visualizing this 6d toric diagram is by considering different 3d projections, as shown in \fref{toric_hexagon_square_Model_1}.

\subsection{Two Boundaries: the Cylinder}

\begin{figure}[h]
 \centering
 \begin{tabular}[c]{ccc}
 \epsfig{file=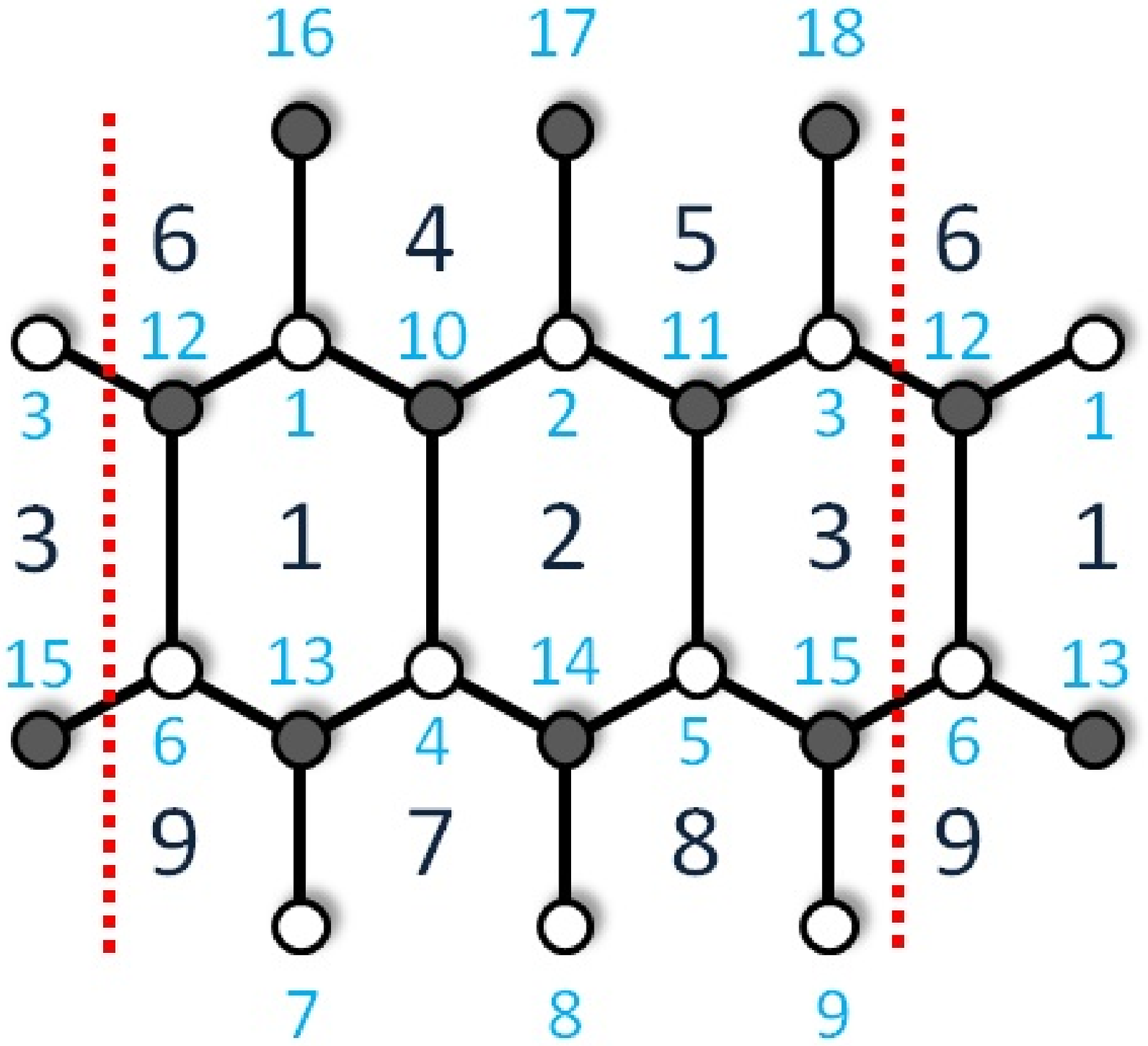,width=0.37\linewidth,clip=} & \ \ \ \ \ &
\epsfig{file=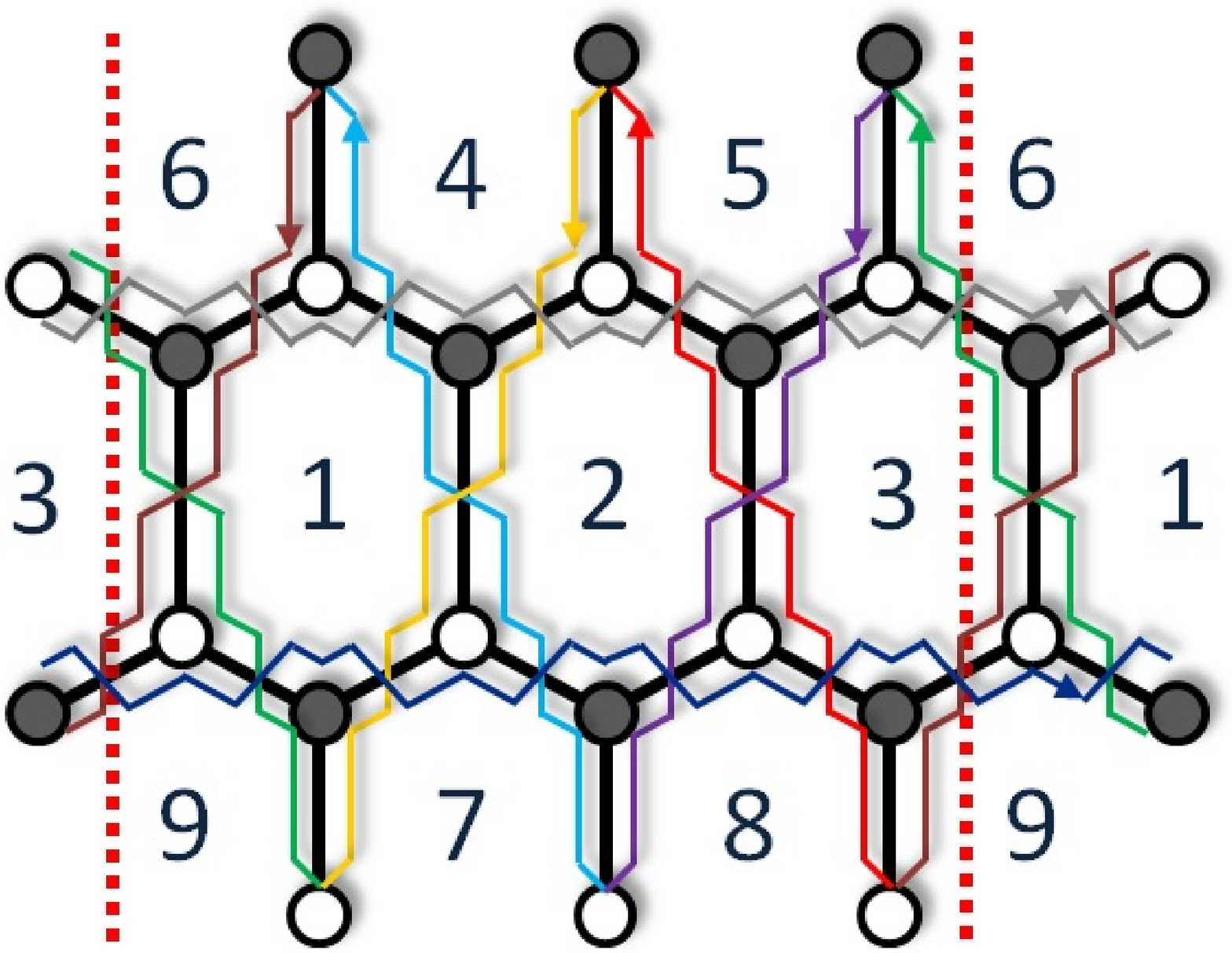,width=0.38\linewidth,clip=} \\ 
\mbox{(a)} & & \mbox{(b)}
 \end{tabular}
\caption{a)	Bipartite graph for the 6-leg, 3-loop model on the cylinder. It contains three internal and six external faces. b) The eight zig-zag paths for this model.}
\label{tiling_6legs_3loops_cylinder} 
\end{figure}

We now study a model with more than one boundary. Let us consider the example in \fref{tiling_6legs_3loops_cylinder}. It has 6 external legs distributed on 2 boundaries, and 3-loops. 

The master Kasteleyn matrix is

{\footnotesize
\beq
K_0 = \left(\begin{array}{c|cccccc|ccc} 
 & \ \ 10 \ \ & \ \ 11 \ \ & \ \ 12 \ \ & \ \ 13 \ \ & \ \ 14 \ \ & \ \ 15 \ \ & \ \ 16 \ \ & \ \ 17 \ \ & \ \ 18 \ \ \\ \hline
\ 1 \ \ & X_{41}& 0& X_{16}& 0& 0& 0& X_{64}& 0& 0 \\
\ 2 \ \ & X_{24}& X_{52}& 0& 0& 0& 0& 0& X_{45}& 0 \\
\ 3 \ \ & 0& X_{35}& X_{63}& 0& 0& 0& 0& 0& X_{56} \\
\ 4 \ \ & X_{12}& 0& 0& X_{71}& X_{27}& 0& 0& 0& 0 \\
\ 5 \ \ & 0& X_{23}& 0& 0& X_{82}& X_{38}& 0& 0& 0 \\
\ 6 \ \ & 0& 0& X_{31}& X_{19}& 0& X_{93}& 0& 0& 0 \\ \hline
\ 7 \ \ & 0& 0& 0& X_{79}& 0& 0& 0& 0& 0 \\
\ 8 \ \ & 0& 0& 0& 0& X_{87}& 0& 0& 0& 0 \\
\ 9 \ \ & 0& 0& 0& 0& 0& X_{98}& 0& 0& 0
\end{array}\right).
\label{K0_3loop_cylinder}
\eeq}
This model has 44 perfect matchings. The master space is a 10d CY. The moduli space is a 7d CY. The 44 perfect matchings organize into 28 distinct points in the toric diagram, which are given by the following matrix

\begin{figure}[h]
 \centering
 \begin{tabular}[c]{ccc}
 \epsfig{file=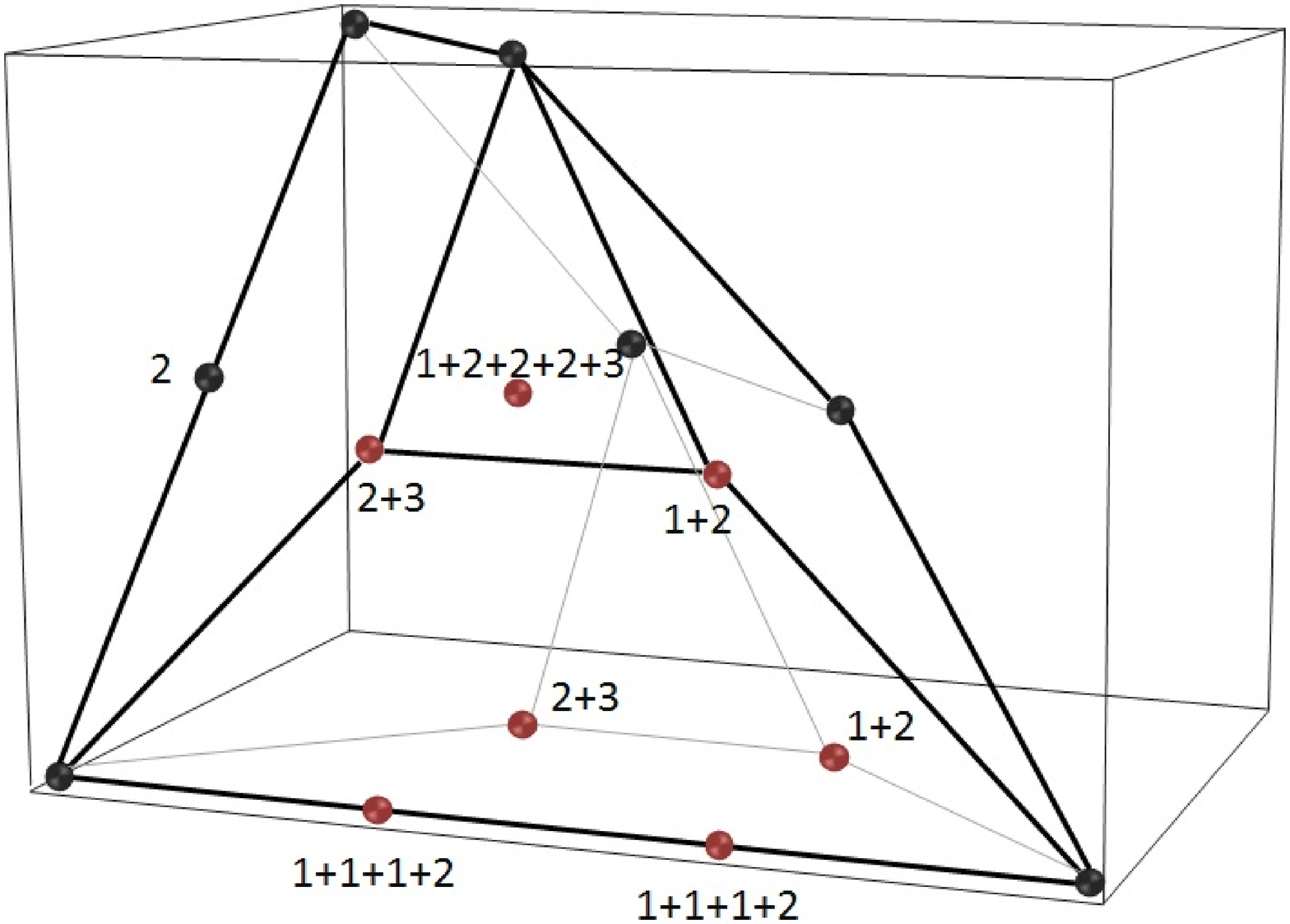,width=0.42\linewidth,clip=} & \ \ \ \ &
\epsfig{file=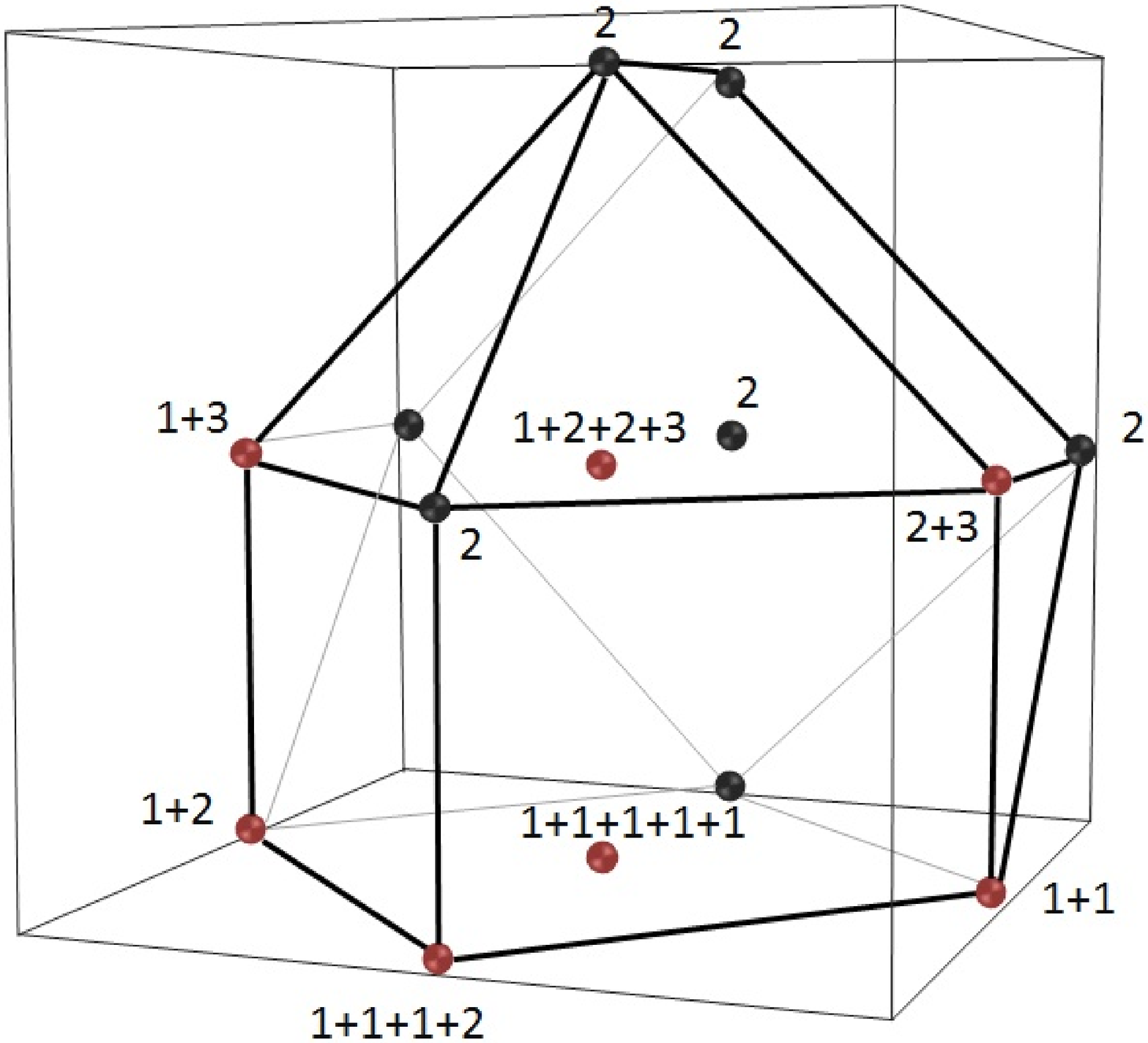,width=0.35\linewidth,clip=} \\ 
\mbox{(a)} & & \mbox{(b)}
 \end{tabular}
\caption{Two projections of the toric diagram corresponding to \eref{G_3loop_cylinder}. Points descending from multiple ones in 7d are shown in red. The numbers indicate the non-trivial multiplicity of perfect matchings. The projections correspond to keeping the following combinations of rows: a) $(G_1,G_2,G_3)$ and b) $(G_2-G_3,G_4,G_5)$.}
\label{toric_3loop_cylinder} 
\end{figure}

{\scriptsize
\beq
G=\left(\begin{array}{cccccccccccccccccccccccccccc}
-1& -1& -1& -2& -1& -1& 0& -1& -1& 0& -1& -1& 0& -2& -1& -1& 0& -2& -1& -1& 0& -1& 0& -1& 0& 0& 0& 1 \\
0& 1& 1& 1& 1& 0& 0& 0& 1& 0& 1& 1& 1& 2& 1& 2& 1& 0& 0& 0& 0& 1& 0& 0& 0& 1& 0& 0 \\
1& 0& 1& 1& 1& 0& 0& 1& 1& 1& 1& 0& 0& 2& 1& 1& 0& 0& 0& 0& 0& 2& 1& 0& 0& 1& 0& 0 \\
1& 1& 0& 1& 1& 1& 1& 0& 1& 0& 1& 0& 0& 2& 2& 1& 1& 0& 0& 0& 0& 1& 1& 0& 0& 0& 0& 0 \\
0& 0& 0& 0& -1& 1& 0& 0& 0& 0& 0& 0& 0& -1& 0& -1& 0& 1& 0& 1& 0& -1& 0& 1& 0& -1& 1& 0 \\
0& 0& 0& 0& 0& 0& 0& 0& 0& 0& -1& 1& 0& -1& -1& 0& 0& 1& 1& 0& 0& -1& -1& 1& 1& 0& 0& 0 \\
0& 0& 0& 0& 0& 0& 0& 1& -1& 0& 0& 0& 0& -1& -1& -1& -1& 1& 1& 1& 1& 0& 0& 0& 0& 0& 0& 0
\\ \hline
\ \bf{3} \ & \ \bf{3} \ & \ \bf{3} \ & \ \bf{2} \ & \ \bf{2} \ & \ \bf{2} \ & \ \bf{2} \ & \ \bf{2} \ & \ \bf{2} \ & \ \bf{2} \ & \ \bf{2} \ & \ \bf{2} \ & \ \bf{2} \ & \ \bf{1} \ & \ \bf{1} \ & \ \bf{1} \ & \ \bf{1} \ & \ \bf{1} \ & \ \bf{1} \ & \ \bf{1} \ & \ \bf{1} \ & \ \bf{1} \ & \ \bf{1} \ & \ \bf{1} \ & \ \bf{1} \ & \ \bf{1} \ & \ \bf{1} \ & \ \bf{1} \ \\ \hline
\end{array}\right) .
\label{G_3loop_cylinder}
\eeq
}
\fref{toric_3loop_cylinder} shows two possible 3d projections of this toric diagram.

\subsection{One Boundary on $T^2$}

\label{section_1_boundary_T2}

\begin{figure}[h]
 \centering
 \begin{tabular}[c]{ccc}
 \epsfig{file=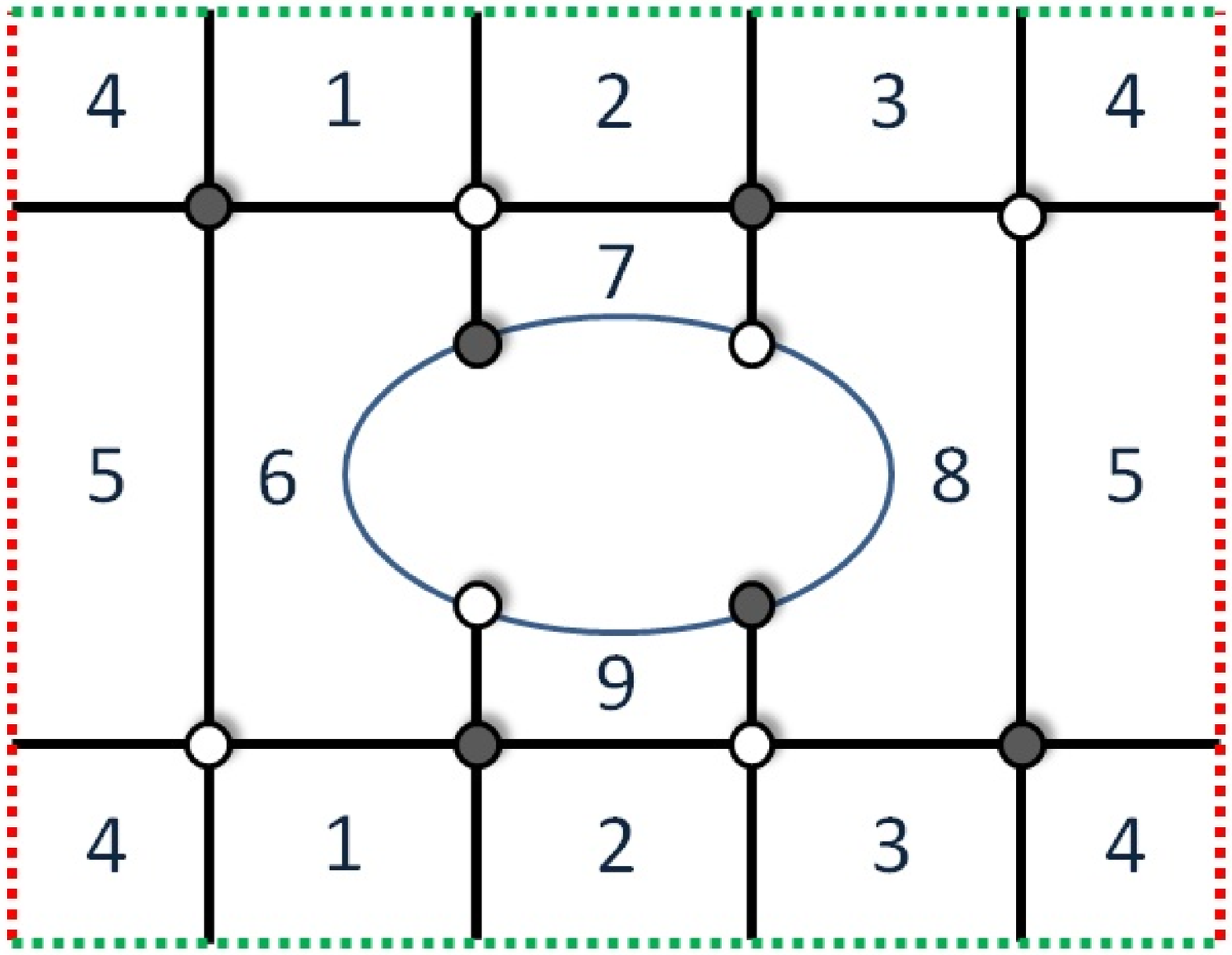,width=0.4\linewidth,clip=} & \ \ \ \ \ &
\epsfig{file=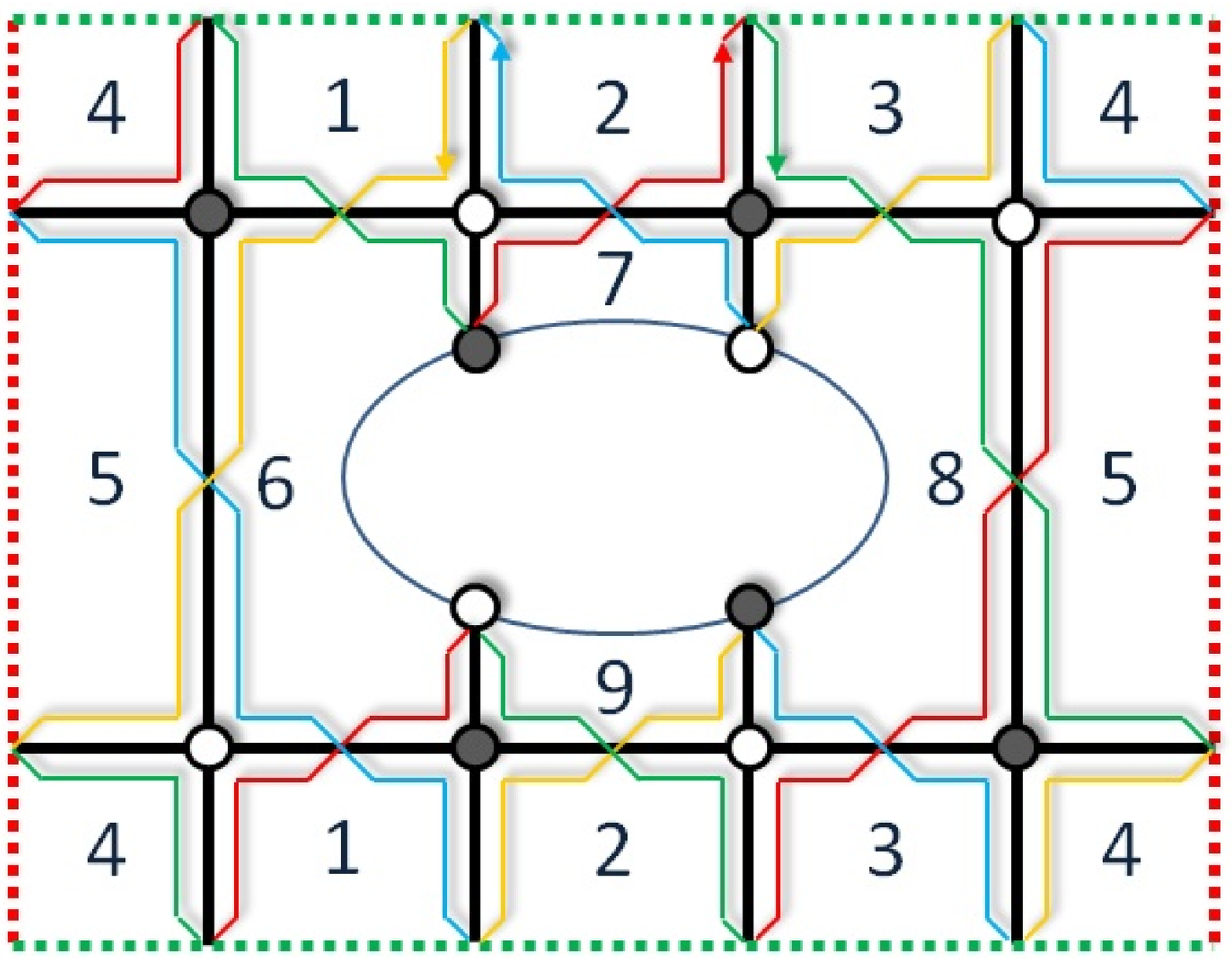,width=0.4\linewidth,clip=} \\ 
\mbox{(a)} & & \mbox{(b)}
 \end{tabular}
\caption{a)	Bipartite graph for model 1. It lives on a 2-torus and has four external nodes on a boundary. It contains five internal and four external faces. b) The four zig-zag paths for this model.}
\label{tiling_1boundary_T2_Model1} 
\end{figure}

Let us move to higher genus and consider a model on a 2-torus, with 4 external legs terminating on a single boundary. From a scattering amplitude perspective, we can regard this diagram as a non-planar, 5-loop contribution to the scattering of 2 negative helicity and 2 positive helicity gluons. We refer to this theory as model 1 and we show the corresponding graph in \fref{tiling_1boundary_T2_Model1}. 

The master Kasteleyn matrix for this model is

{\small
\beq
K_0= \left(\begin{array}{c|cccc|cc} 
& \ \ 7 \ \ & \ \ 8 \ \ & \ \ 9 \ \ & \ \ 10 \ \  & \ \ 11 \ \ & \ \ 12 \ \ \\  \hline
1 & X_{61} & X_{27} & X_{12} & 0 & X_{76} & 0 \\
2 & X_{45} & X_{83} & 0 & \ X_{34} + X_{58} \ & 0 & 0 \\
3 & \ X_{14} + X_{56} \ & 0 & Y_{61} & Y_{45} & 0 & 0 \\ 
4 & 0 & X_{32} & Y_{29} & Y_{83} & 0 & X_{98} \\ \hline
5 & 0 & X_{78} & 0 & 0 & 0 & 0 \\
6 & 0 & 0 & X_{96} & 0 & 0 & 0
\end{array}\right).
\eeq
}

The theory has 48 perfect matchings and the master space is an 11d CY. The perfect matchings give rise to 22 different points in the toric diagram of the moduli space, with positions summarized by the following matrix

{\footnotesize
\beq
\left(\begin{array}{cccccccccccccccccccccc}
1 & 0 & 0 & 1 & 1 & -1 & -1 & 1 & 0 & 0 & 0 & 0 & -1 & -1 & 1 & -1 & -1 & 1 & 0 & 0 & 2 & 0 \\
0 & 0 & 0 & 0 & 0 & 0 & 1 & -1 & 0 & 0 & 0 & 0 & 0 & 1 & -1 & 0 & 1 & -1 & 0 & 0 & -1 & 1 \\
0 & 0 & 1 & 0 & 0 & 1 & 0 & 1 & 0 & 1 & 0 & 1 & 1 & 0 & 1 & 1 & 0 & 1 & 0 & 1 & 0 & 0 \\
0 & 0 & 1 & 0 & 0 & 1 & 1 & 0 & 0 & 1 & 0 & 1 & 1 & 1 & 0 & 1 & 1 & 0 & 1 & 0 & 0 & 0 \\
0 & 0 & 0 & 1 & -1 & 0 & 0 & 0 & 1 & -1 & -1 & 1 & 1 & 1 & 1 & -1 & -1 & -1 & 0 & 0 & 0 & 0 \\
0 & 1 & -1 & -1 & 1 & 0 & 0 & 0 & 0 & 0 & 2 & -2 & -1 & -1 & -1 & 1 & 1 & 1 & 0 & 0 & 0 & 0 \\ \hline
\ \bf{9} \ & \ \bf{4} \ & \ \bf{4} \ & \ \bf{3} \ & \ \bf{3} \ & \ \bf{3} \ & \ \bf{3} \ & \ \bf{3} \ & \ \bf{2} \ & \ \bf{2} \ & \ \bf{1} \ & \ \bf{1} \ & \ \bf{1} \ & \ \bf{1} \ & \ \bf{1} \ & \ \bf{1} \ & \ \bf{1} \ & \ \bf{1} \ & \ \bf{1} \ & \ \bf{1} \ & \ \bf{1} \ & \ \bf{1} \ \\ \hline
\end{array}\right).
\label{G_matrix_multiplicities_Model_1}
\eeq
}

\begin{figure}[h]
 \centering
 \begin{tabular}[c]{ccc}
 \epsfig{file=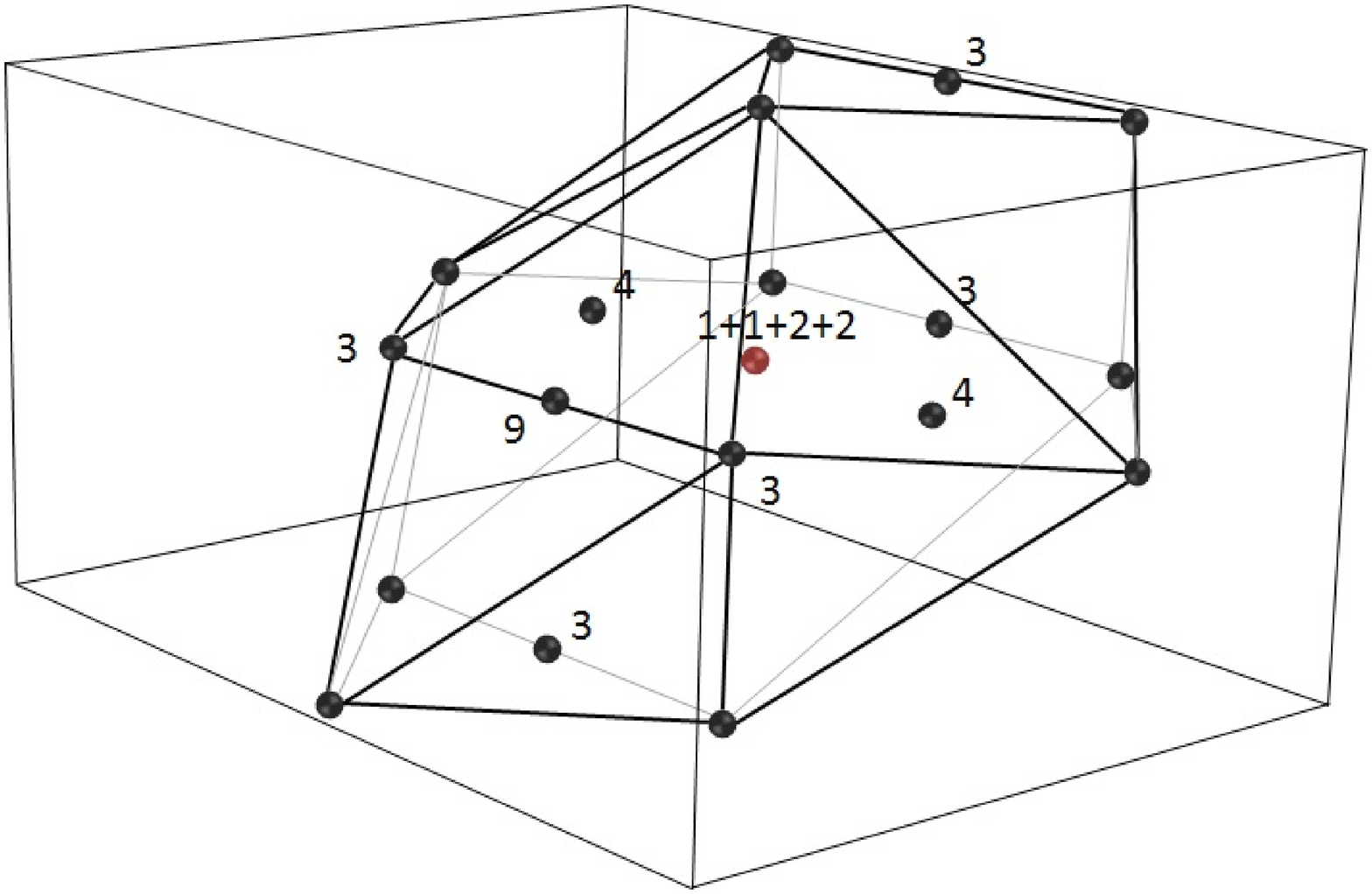,width=0.45\linewidth,clip=} & \ \ \ \ &
\epsfig{file=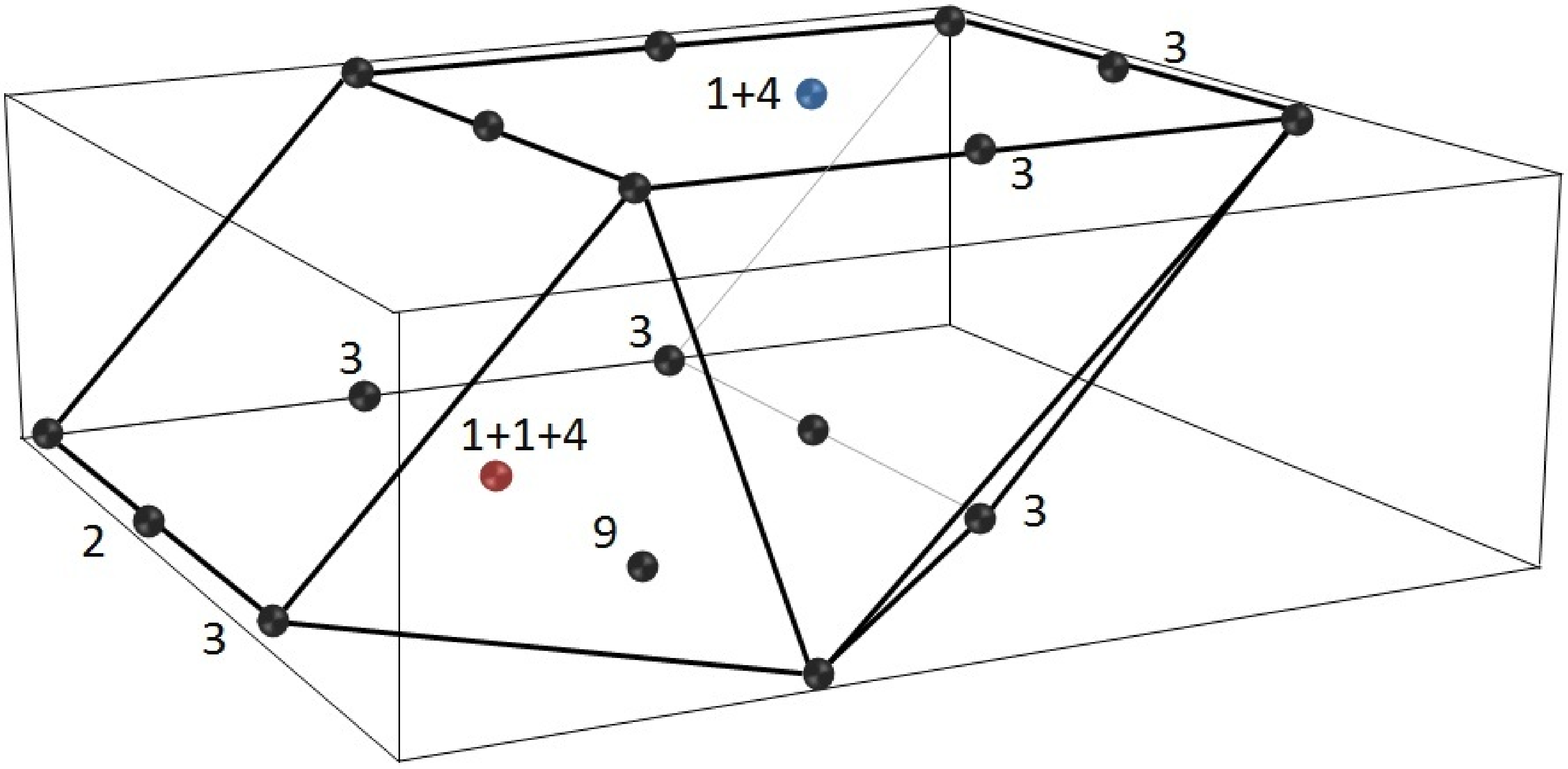,width=0.45\linewidth,clip=} \\ 
\mbox{(a)} & & \mbox{(b)}
 \end{tabular}
\caption{Two projections of the toric diagram corresponding to \eref{G_matrix_multiplicities_Model_1}. We show in red the points descending from multiple ones in 6d. The numbers indicate the non-trivial multiplicity of perfect matchings. The projections correspond to keeping the following combinations of rows: a) $(G_1,G_2,G_6)$ and b) $(G_1,G_2,G_5-G_3)$.}
\label{toric_1boundary_T2_Model1_1} 
\end{figure}

\noindent We see that the moduli space is a 6d toric CY. In \fref{toric_1boundary_T2_Model1_1}, we show two projections of the toric diagram down to three dimensions. They have been chosen in order to minimize the overlap of distinct points after the projections. In both cases, the 22 points of original 6d toric diagram are mapped to 19 points. 

\section{Square Moves and Geometry, or Seiberg Duality and Moduli Spaces}

\label{section_Seiberg_duality}

Let us investigate the effect of square moves on BFT theories. As we have explained in Section \ref{section_Seiberg_duality}, they correspond to Seiberg dualities on certain gauge groups of the BFTs. The moduli space of the theories is, by construction, invariant under Seiberg duality.\footnote{As shown in \cite{Forcella:2008ng}, this is not the case for the master space which, in the case of plabic graphs, is the toric geometry associated to the matching polytope \cite{Postnikov_toric}.} As a result, the moduli space is an ideally suited object for identifying theories connected by square moves. This problem becomes rather non-trivial for large graphs, multiple square moves, multiple boundaries and/or higher genus Riemann surfaces. 

\subsection{The Dual of the Hexagon-Square Model}

\begin{figure}[h]
 \centering
 \begin{tabular}[c]{ccc}
 \epsfig{file=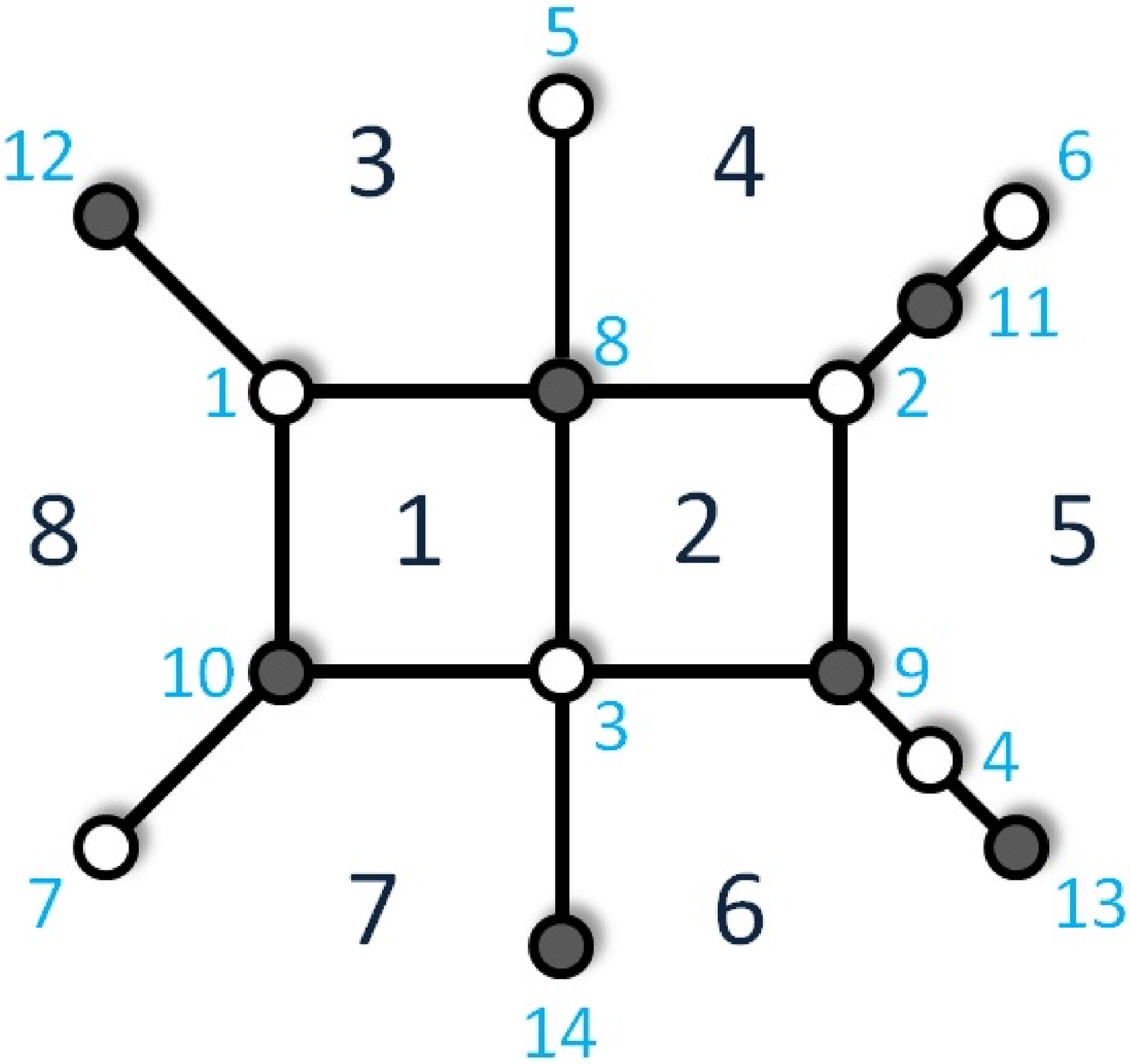,width=0.38\linewidth,clip=} & \ \ \ \ \ &
\epsfig{file=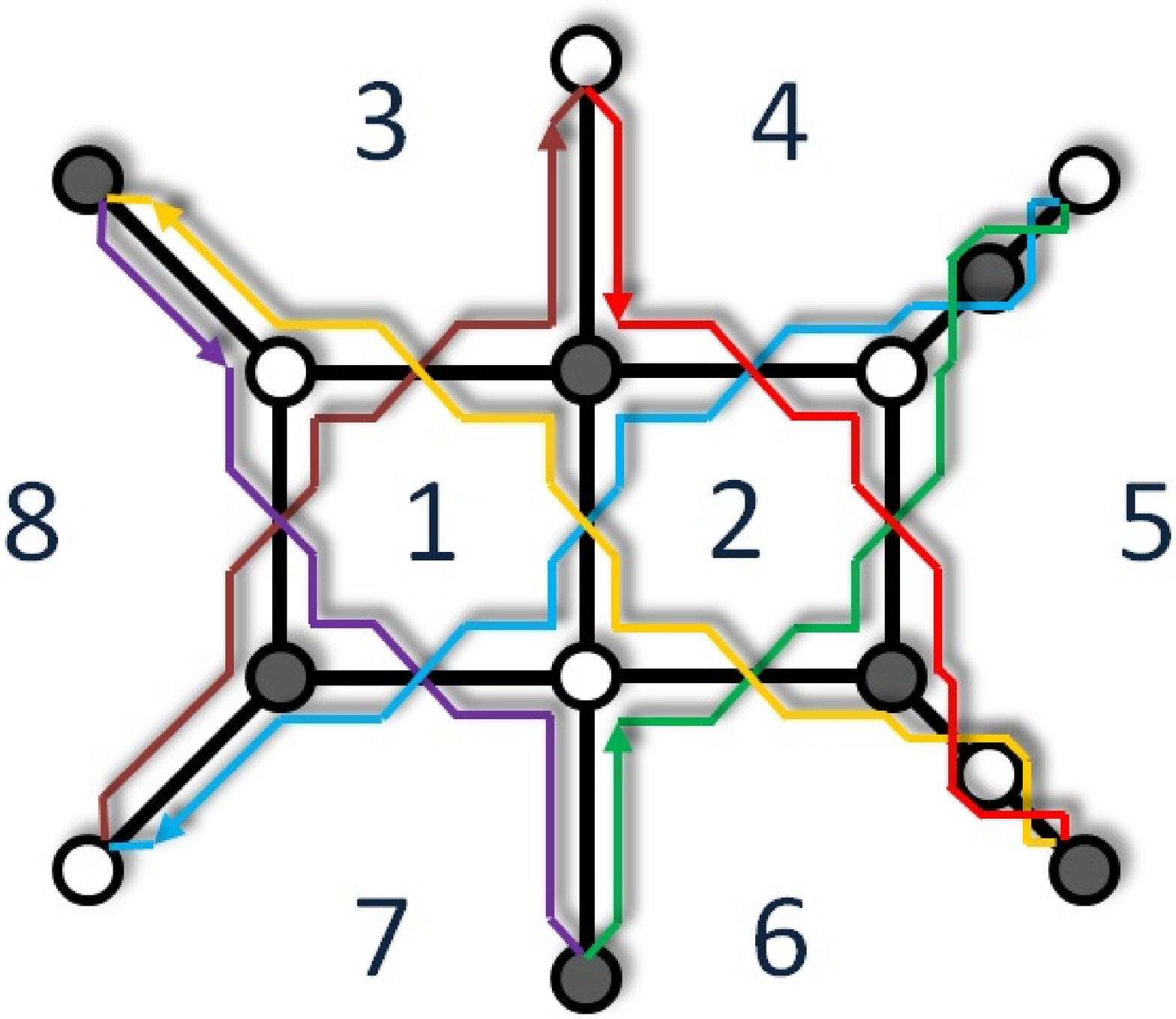,width=0.38\linewidth,clip=} \\ 
\mbox{(a)} & & \mbox{(b)}
 \end{tabular}
\caption{a)	Bipartite graph obtained by Seiberg dualizing the gauge group associated to face 2 of the hexagon-square model. It contains two internal and six external faces. b) The six zig-zag paths for this model.}
\label{hexagon-square_dual_tiling} 
\end{figure}

Let us consider the model shown in \fref{hexagon-square_dual_tiling}, which is obtained from the hexagon-square model discussed in Section \ref{section_hexagon-square} by Seiberg dualizing the gauge group associated to face 2. Four 2-valent nodes, i.e. mass terms in the BFT, are generated by the duality. We have only integrated out the massive fields associated to two of them, in order to preserve the external legs connected to nodes 6 and 13. 

The master Kasteleyn matrix is

{\small
\beq
K_0= \left(\begin{array}{c|cccc|ccc} 
& \ \ 8 \ \ &\ \ 9 \ \ & \ \ 10 \ \ & \ \ 11 \ \ & \ \ 12 \ \  & \ \ 13 \ \ & \ \ 14 \ \ \\ \hline
1 & \ X_{31} \ & 0 & \ X_{18} \ & 0 & \ X_{83} \ & 0 & 0 \\
2 & X_{24} & \ X_{52} \ & 0 & \ X_{45} \ & 0 & 0 & 0 \\
3 & X_{12} & X_{26} & X_{71} & 0 & 0 & 0 & \ X_{67} \ \\
4 & 0 & X_{65} & 0 & 0 & 0 & \ X_{56} \ & 0 \\ \hline
5 & X_{43} & 0 & 0 & 0 & 0 & 0 & 0 \\
6 & 0 & 0 & 0 & X_{54} & 0 & 0 & 0 \\
7 & 0 & 0 & X_{87} & 0 & 0 & 0 & 0
\end{array}\right).
\eeq
}
The theory has 22 perfect matchings and the master space is an 8d toric CY. The moduli space is a 6d CY, with toric diagram given by

\begin{figure}[h]
 \centering
 \begin{tabular}[c]{ccc}
 \epsfig{file=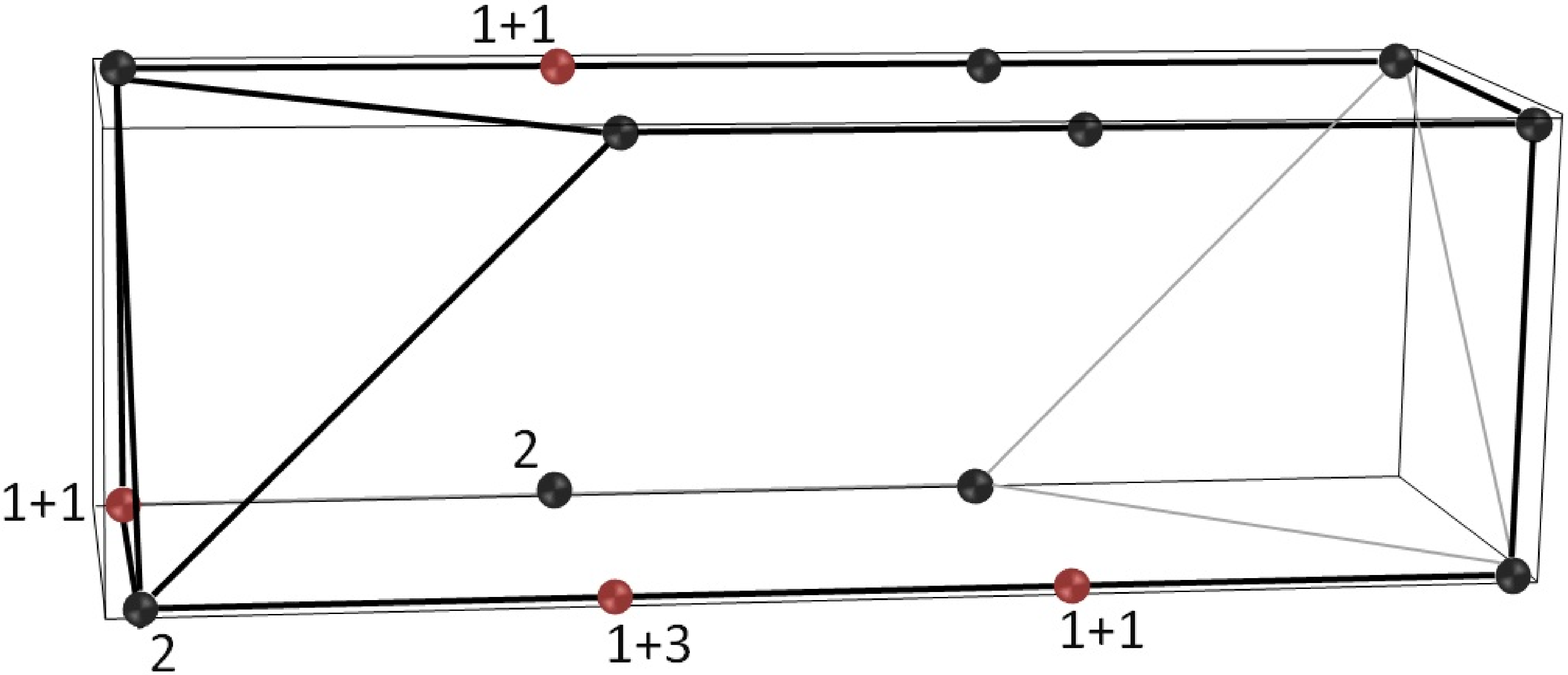,width=0.4\linewidth,clip=} & \ \ \ \ &
\epsfig{file=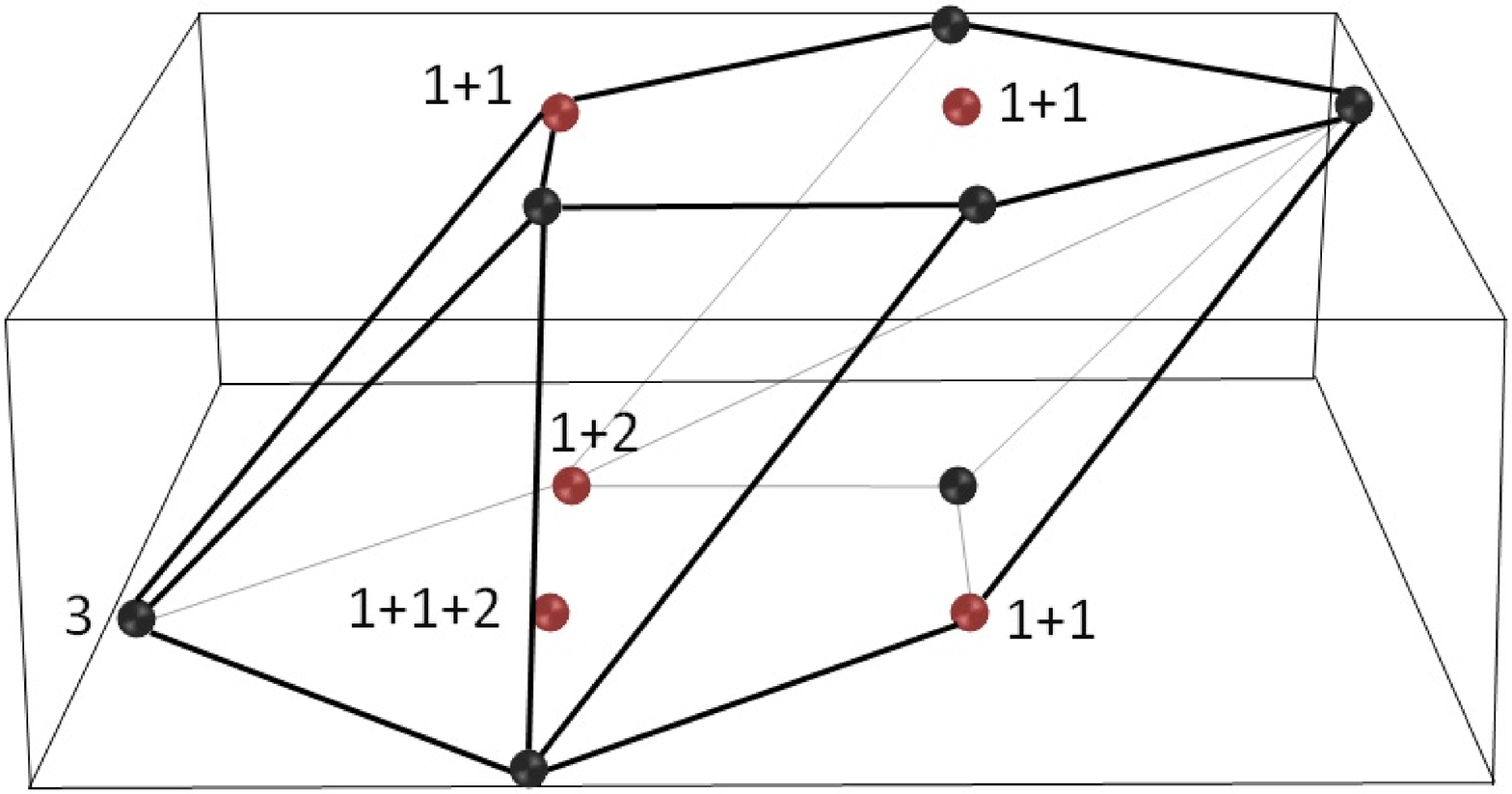,width=0.4\linewidth,clip=} \\ 
\mbox{(a)} & & \mbox{(b)}
 \end{tabular}
\caption{Two projections of the toric diagram corresponding to \eref{G_matrix_multiplicities_hexagon_square_dual}. Points descending from multiple ones in 6d are shown in red. The numbers indicate the non-trivial multiplicity of perfect matchings. The projections correspond to keeping the following combinations of rows: a) $(G_2-G_4,G_5,G_6)$ and b) $(G_2-G_3,G_4,G_6)$.}
\label{toric_hexagon_square_Model_2} 
\end{figure}

{\footnotesize
\beq
G=\left(\begin{array}{cccccccccccccccccc}
 0 & 0 & 0 & -1 & -1 & -1 & 0 & -1 & -1 & -1 & 0 & -1 & 0 & 0 & -1 & -1 & 0 & 1 \\
 0 & 0 & 0 & 0 & 1 & 1 & 1 & 0 & 0 & 1 & 0 & 1 & 0 & 0 & 1 & 1 & 1 & 0 \\
 1 & 0 & 0 & 0 & 0 & 1 & 1 & -1 & 0 & -1 & -1 & 0 & 0 & -1 & 0 & 1 & 0 & 0 \\
 0 & 0 & 1 & 0 & -1 & -1 & -1 & 1 & 1 & 0 & 0 & 0 & 0 & 1 & 0 & 0 & 0 & 0 \\
 0 & 1 & 0 & 1 & 1 & 0 & 0 & 1 & 1 & 1 & 1 & 0 & 0 & 1 & 1 & 0 & 0 & 0 \\
 0 & 0 & 0 & 1 & 1 & 1 & 0 & 1 & 0 & 1 & 1 & 1 & 1 & 0 & 0 & 0 & 0 & 0 \\ \hline
\ \bf{3} \ & \ \bf{2} \ & \ \bf{2} \ & \ \bf{1} \ & \ \bf{1} \ & \ \bf{1} \ & \ \bf{1} \ & \ \bf{1} \ & \ \bf{1} \ & \ \bf{1} \ & \ \bf{1} \ & \ \bf{1} \ & \ \bf{1} \ & \ \bf{1} \ & \ \bf{1} \ & \ \bf{1} \ & \ \bf{1} \ & \ \bf{1} \ \\ \hline
\end{array}\right)
\label{G_matrix_multiplicities_hexagon_square_dual}
\eeq
}
This moduli space is the same as the one for the original hexagon-square model. It is indeed possible to find an $SL(6,\mathbb{Z})$ transformation that takes \eref{G_matrix_multiplicities_hexagon_square_dual} into \eref{G_matrix_multiplicities_hexagon_square}. Instead of giving the explicit transformation, we show two projections of the toric diagram in \fref{toric_hexagon_square_Model_2}, which are identical to those shown in \fref{toric_hexagon_square_Model_1} for the Seiberg dual theory. 

Comparing \eref{G_matrix_multiplicities_hexagon_square} and \eref{G_matrix_multiplicities_hexagon_square_dual}, we see that the original hexagon-square model and its Seiberg dual differ in the multiplicity of perfect matchings associated to each point in the toric diagram of the moduli space. This is a generic feature of dual theories that will be also encountered in the examples that follow. Different multiplicities are a manifestation, in the context of toric geometry, of the action of {\it cluster transformations}. They relate the partition functions for perfect matching of dual models and, in particular, produce the perfect matching multiplicities for any of the two theories in terms of those for the other one. The central role of cluster transformations, which leave the boundary measurement invariant \cite{Postnikov_plabic}, in the study of leading singularities has been investigated in \cite{Nima}. A more intuitive understanding of the role of cluster transformations in the BFT will be presented in future work \cite{cluster_future}. 

\subsection{Non-Planar Duals}

\begin{figure}[h]
 \centering
 \begin{tabular}[c]{ccc}
 \epsfig{file=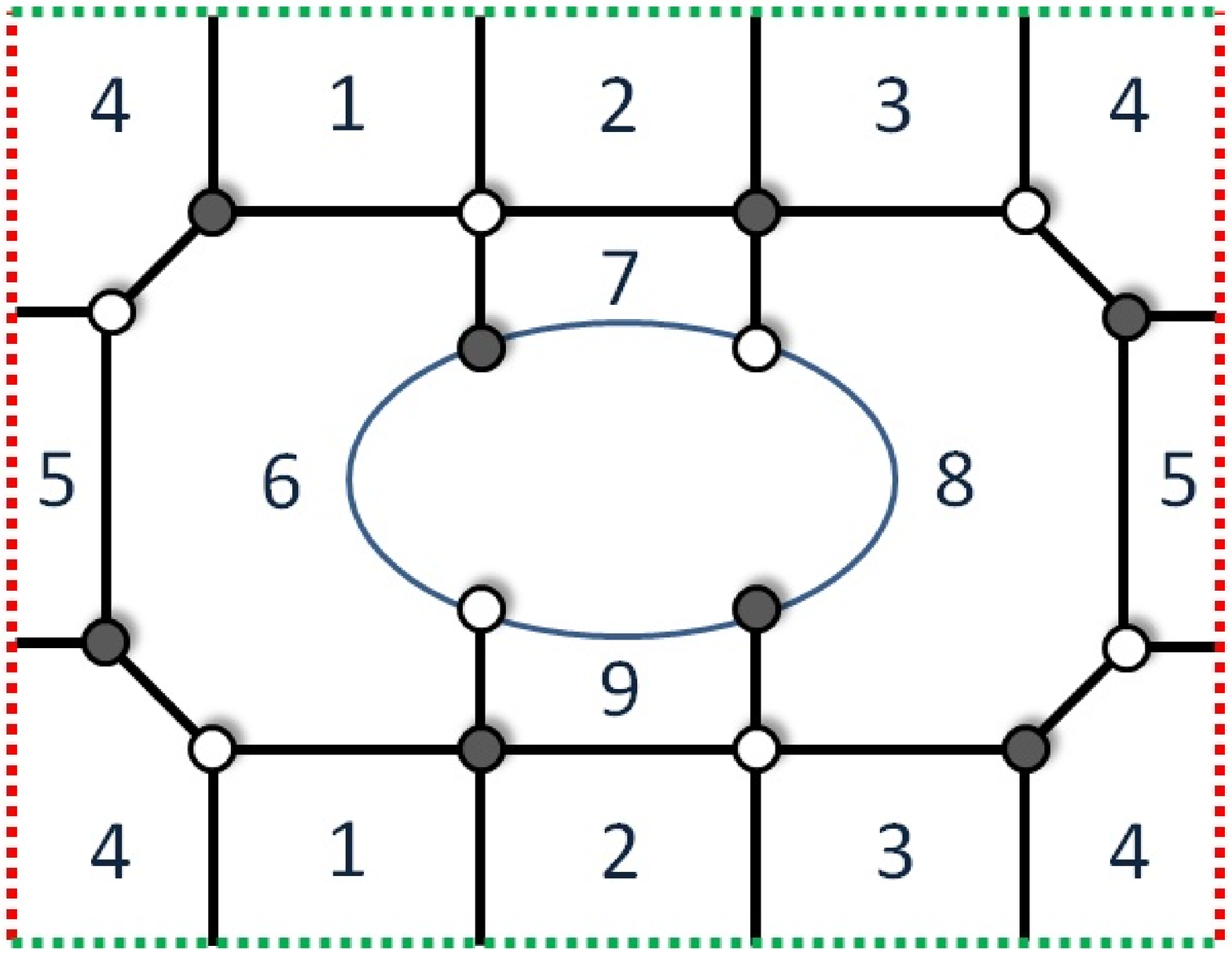,width=0.4\linewidth,clip=} & \ \ \ \ \ &
\epsfig{file=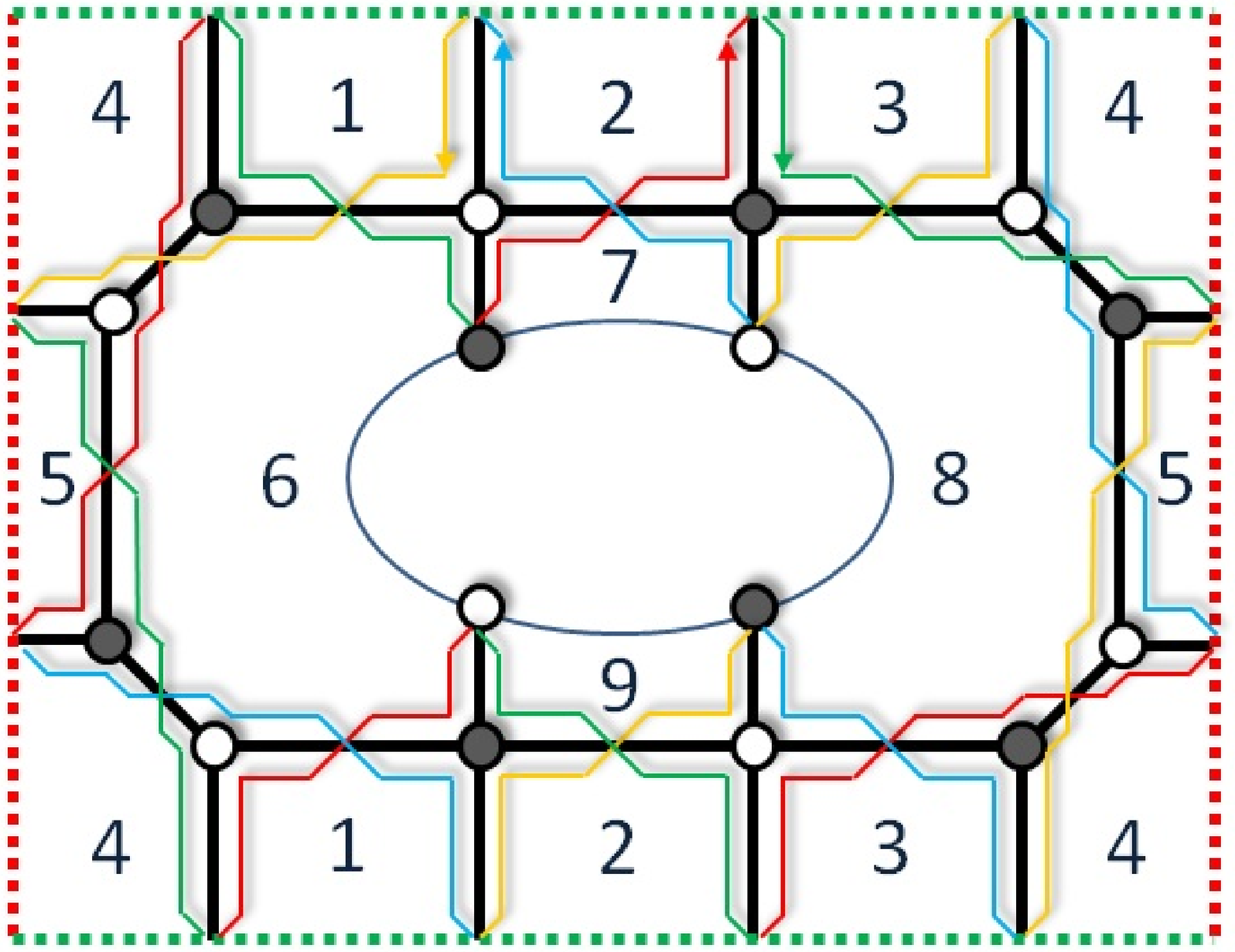,width=0.4\linewidth,clip=} \\ 
\mbox{(a)} & & \mbox{(b)}
 \end{tabular}
\caption{a)	Bipartite graph for model 2, obtained from model 1 by Seiberg dualizing face 4. It lives on a 2-torus and has four external nodes on a boundary. It contains five internal and four external faces. b) The four zig-zag paths for this model.}
\label{tiling_1boundary_T2_Model2} 
\end{figure}

Let us now consider a theory, that we call model 2, obtained from model 1 in Section \ref{section_1_boundary_T2} by Seiberg dualizing face 4. The resulting graph is shown in \fref{tiling_1boundary_T2_Model2}.

The master Kasteleyn matrix is

{\small
\beq
K_0= \left(\begin{array}{c|cccccc|cc} 
& \ \ 9 \ \ & \ \ 10 \ \ & \ \ 11 \ \ & \ \ 12 \ \  & \ \ 13 \ \ & \ \ 14 \ \ & \ \ 15 \ \ & \ \ 16 \ \ \\ \hline
1 & X_{61} & X_{27} & 0 & 0 & X_{12} & 0 & X_{76} & 0 \\
2 & 0 & X_{83} & X_{48} & 0 & 0 & X_{34} & 0 & 0 \\
3 & X_{46} & 0 & X_{54} & X_{65} & 0 & 0 & 0 & 0 \\ 
4 & 0 & 0 & X_{85} & Y_{54} & 0 & Y_{48} & 0 & 0 \\
5 & X_{14} & 0 & 0 & Y_{46} & Y_{61} & 0 & 0 & 0 \\
6 & 0 & X_{32} & 0 & 0 & Y_{29} & Y_{83} & 0 & X_{98} \\ \hline
7 & 0 & X_{78} & 0 & 0 & 0 & 0 & 0 & 0 \\
8 & 0 & 0 & 0 & 0 & X_{96} & 0 & 0 & 0
\end{array}\right).
\eeq
}

The theory has 53 perfect matchings, which turn into 22 distinct points in the toric diagram of the moduli space. We already see that this number agrees with its Seiberg dual. The moduli space is a 6d toric CY whose toric diagram is given by

{\footnotesize
\beq
\left(\begin{array}{cccccccccccccccccccccc}
1 & 0 & 0 & 1 & 1 & -1 & -1 & 1 & 0 & 0 & 0 & 0 & -1 & -1 & 1 & -1 & -1 & 1 & 0 & 0 & 2 & 0 \\
0 & 0 & 0 & 0 & 0 & 0 & 1 & -1 & 0 & 0 & 0 & 0 & 0 & 1 & -1 & 0 & 1 & -1 & 0 & 0 & -1 & 1 \\
0 & 0 & 1 & 1 & -1 & 1 & 0 & 1 & 1 & 0 & -1 & 2 & 2 & 1 & 2 & 0 & -1 & 0 & 0 & 1 & 0 & 0 \\
0 & 0 & 1 & 1 & -1 & 1 & 1 & 0 & 1 & 0 & -1 & 2 & 2 & 2 & 1 & 0 & 0 & -1 & 1 & 0 & 0 & 0 \\
0 & 0 & 0 & -1 & 1 & 0 & 0 & 0 & -1 & 1 & 1 & -1 & -1 & -1 & -1 & 1 & 1 & 1 & 0 & 0 & 0 & 0 \\
0 & 1 & -1 & -1 & 1 & 0 & 0 & 0 & 0 & 0 & 2 & -2 & -1 & -1 & -1 & 1 & 1 & 1 & 0 & 0 & 0 & 0 \\ \hline
\ \bf{12} \ & \ \bf{5} \ & \ \bf{5} \ & \ \bf{3} \ & \ \bf{3} \ & \ \bf{3} \ & \ \bf{3} \ & \ \bf{3} \ & \ \bf{2} \ & \ \bf{2} \ & \ \bf{1} \ & \ \bf{1} \ & \ \bf{1} \ & \ \bf{1} \ & \ \bf{1} \ & \ \bf{1} \ & \ \bf{1} \ & \ \bf{1} \ & \ \bf{1} \ & \ \bf{1} \ & \ \bf{1} \ & \ \bf{1} \ \\ \hline
\end{array}\right).
\label{G_matrix_multiplicities_Model_2}
\eeq
}

\begin{figure}[h]
 \centering
 \begin{tabular}[c]{ccc}
 \epsfig{file=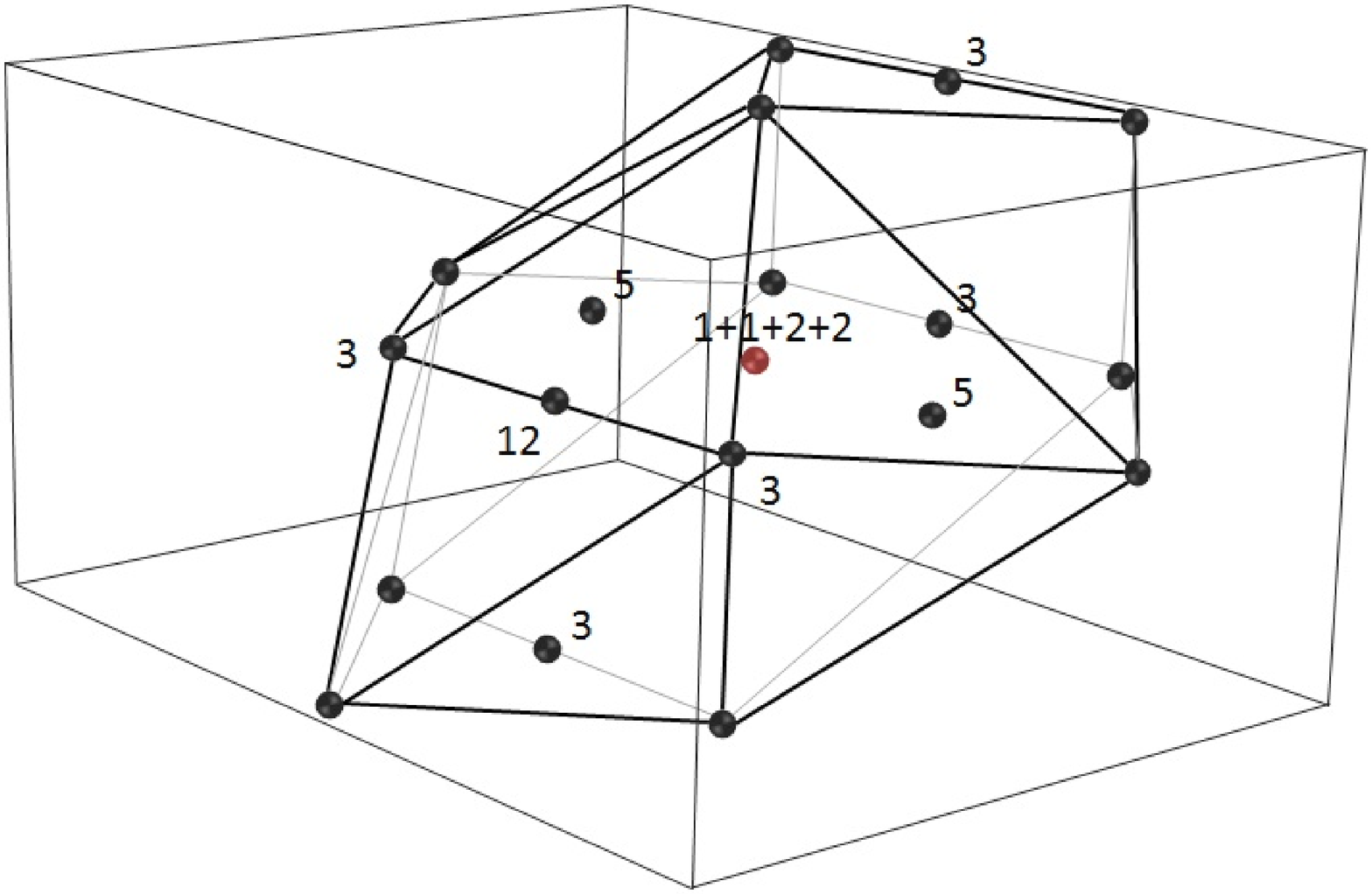,width=0.45\linewidth,clip=} & \ \ \ \ &
\epsfig{file=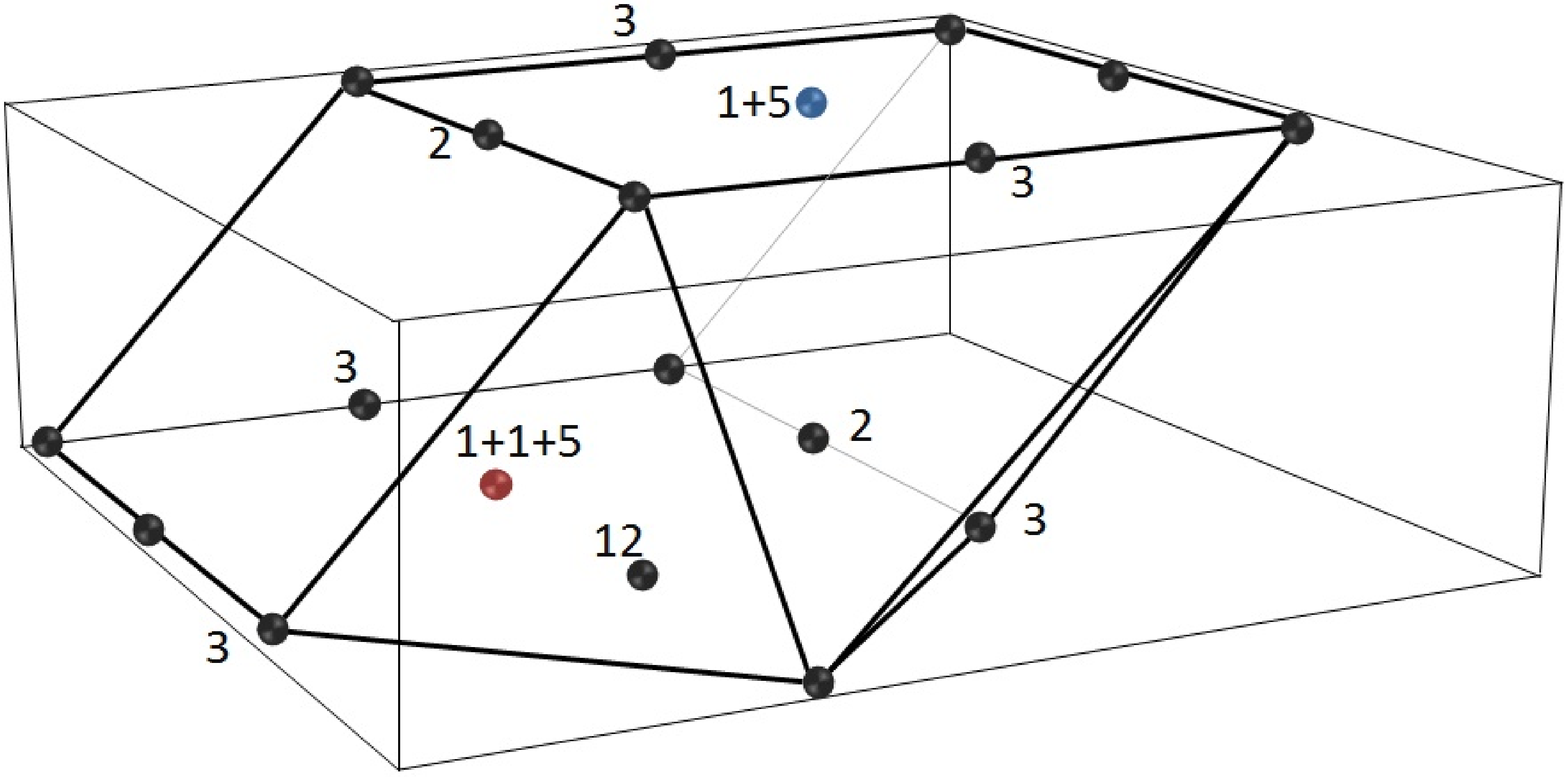,width=0.45\linewidth,clip=}  \\ 
\mbox{(a)} & & \mbox{(b)}
 \end{tabular}
\caption{Two projections of the toric diagram of model 2, given by \eref{G_matrix_multiplicities_Model_2}. Points descending from multiple ones in 6d are shown in red. The numbers indicate the non-trivial multiplicity of perfect matchings. The projections correspond to keeping the following combinations of rows: a) $(G_1,G_2,G_6)$ and b) $(G_1,G_2,G_5+G_3)$. Models 1 and 2 have the same moduli space.}
\label{toric_1boundary_T2_Model2_1} 
\end{figure}

This moduli space is identical to the one for model 1. As in the previous example, instead of providing the explicit $SL(6,\mathbb{Z})$ transformation connecting the two toric diagrams, we present some 3d projections in \fref{toric_1boundary_T2_Model2_1}, which match those in \fref{toric_1boundary_T2_Model1_1}. 

Despite their simplicity, the examples considered in this section show how powerful the concept of moduli space is for identifying models connected by Seiberg duality, i.e. configurations related by square moves in the graph. The moduli space serves as a practical and sharp diagnostic even for large graphs, complicated topologies and/or theories related by a chain of multiple Seiberg dualities.

\bigskip



\section{Loop Reduction and Calabi-Yau Geometry}

\label{section_loop_reduction}

\begin{figure}[h]
\begin{center}
\includegraphics[width=8.5cm]{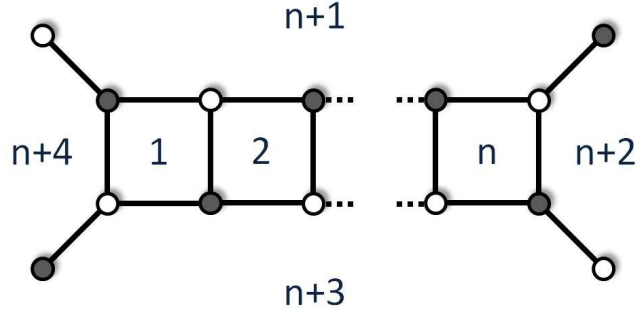}
\caption{The $n$-loop ladder diagram for the scattering of 2 negative and 2 positive helicity particles generalizing the 1-loop model discussed in Section \ref{section_BFTs_and_CYs}.}
\label{tiling_Gr24_nloops}
\end{center}
\end{figure}

In this section we study some nice behavior exhibited by the moduli spaces associated to multi-loop diagrams. For concreteness, we consider the scattering of 2 negative and 2 positive helicity gluons and focus on the multi-loop ladder diagrams given in \fref{tiling_Gr24_nloops}, which generalize the model studied in Section \ref{section_BFTs_and_CYs}.\footnote{This example was also independently considered by the authors of \cite{Nima}.}

For BFTs on a disk, the dimension of the master space is equal to the total number of faces of the graph \cite{Postnikov_toric}. To determine the moduli space, we further impose the D-term equations associated to internal faces, so the dimension of the moduli space is equal to the number of external faces. Applying this general discussion to the class of models given by \fref{tiling_Gr24_nloops}, we conclude that while the master space of the $n$-loop theory is $(n+4)$-dimensional, the moduli space is a CY 4-fold for every $n$. We will soon see that the agreement between moduli spaces goes beyond just the number of dimensions.

\medskip

\subsection*{Two Loops}

\begin{figure}[h]
 \centering
 \begin{tabular}[c]{ccc}
 \epsfig{file=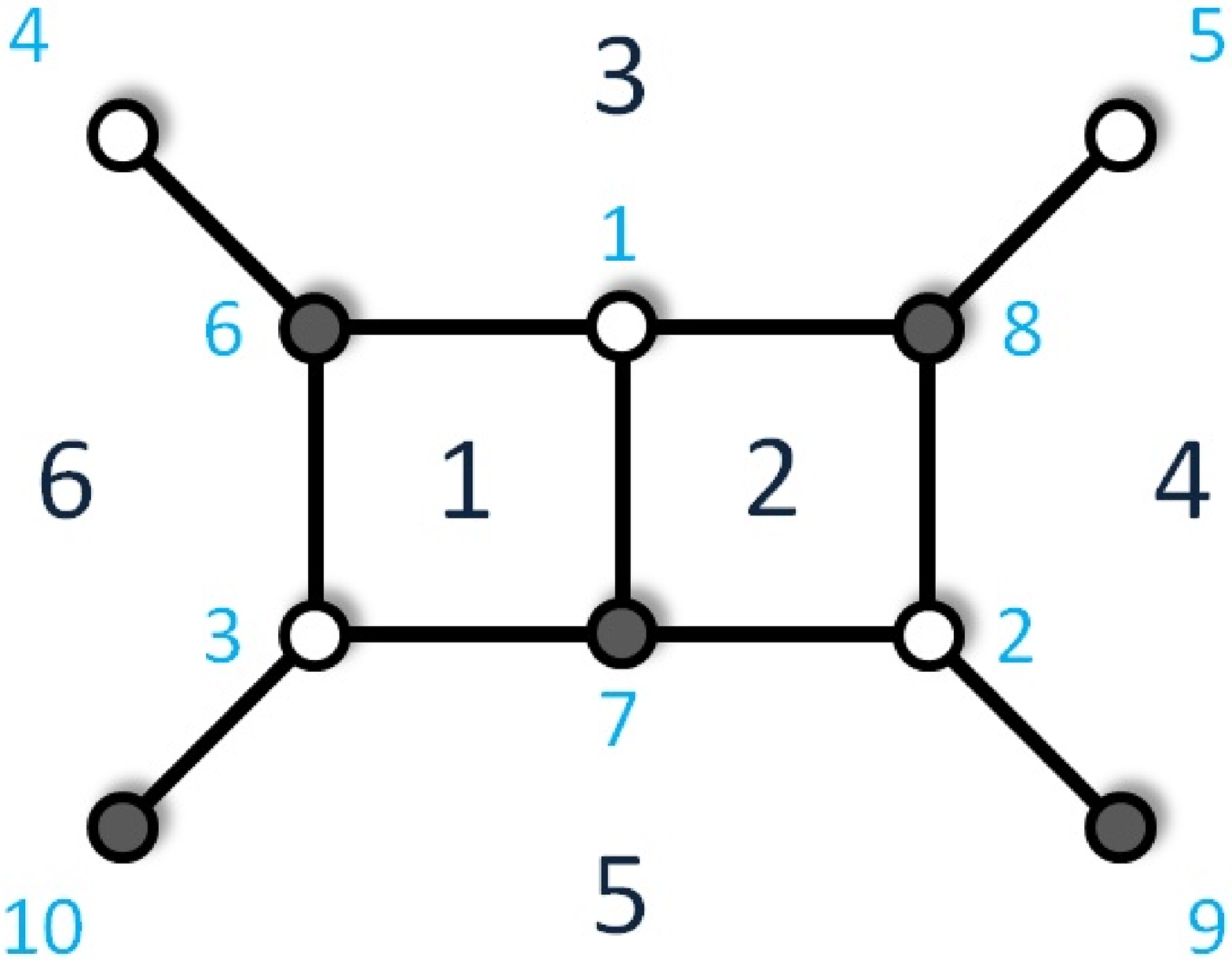,width=0.37\linewidth,clip=} & \ \ \ \ \ &
\epsfig{file=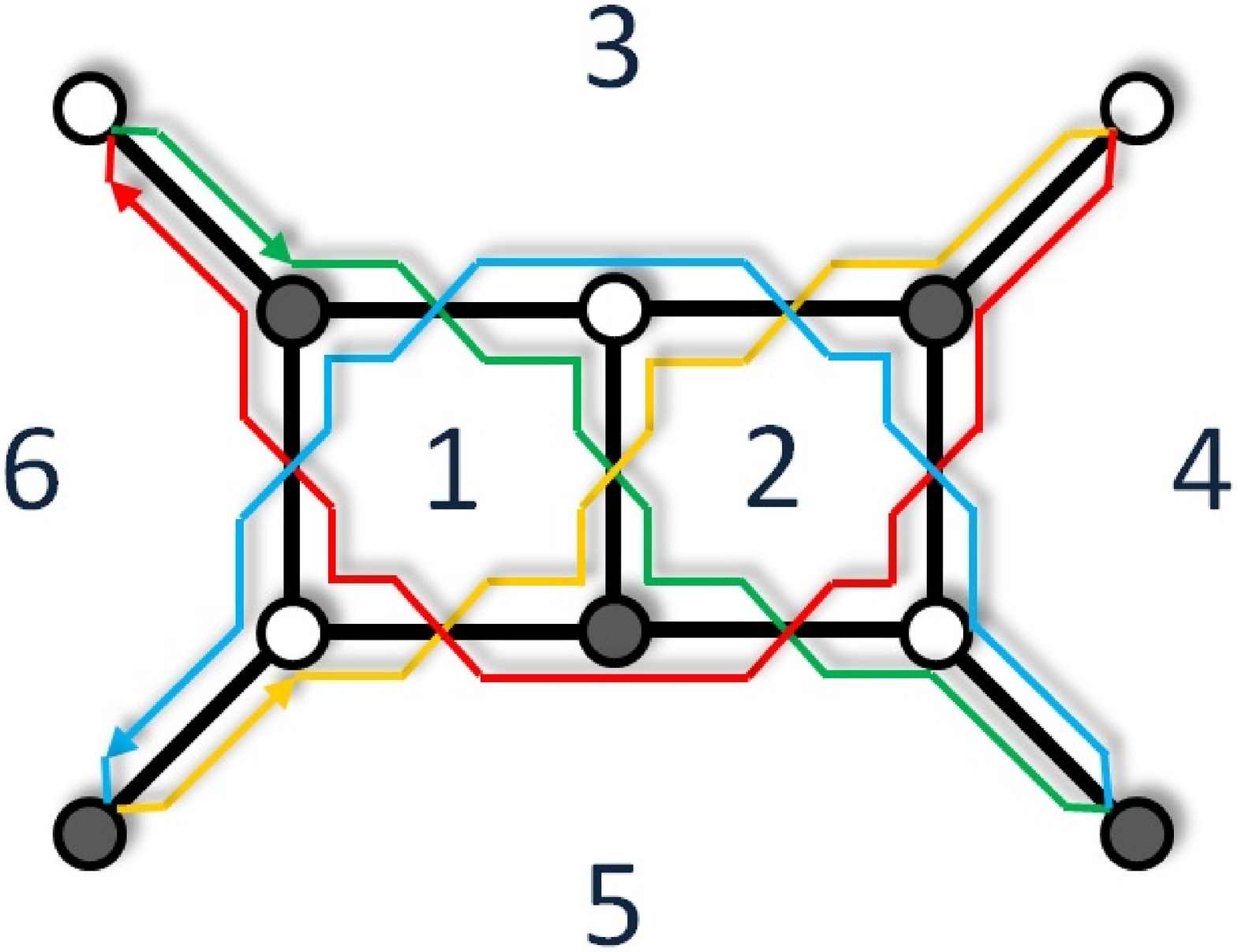,width=0.36\linewidth,clip=} \\ 
\mbox{(a)} & & \mbox{(b)}
 \end{tabular}
\caption{a)	Bipartite graph for the 4-leg, 2-loop model. It contains two internal and four external faces. b) The four zig-zag paths for this model.}
\label{toric_Gr24_2loops} 
\end{figure}

Let us begin with the 2-loop diagram shown in \fref{toric_Gr24_2loops}. The master Kasteleyn matrix for this model is
{\small
\beq
K_0 = \left(\begin{array}{c|ccc|cc} 
 & \ \ 6 \ \ & \ \ 7 \ \ & \ \ 8 \ \ & \ \ 9 \ \  & \ \ 10 \ \ \\ \hline
\ 1 \ & \ X_{13} \ & \ X_{21} \ & \ X_{32} \ &  0 & 0 \\
\ 2 \ & 0 & X_{52} & X_{24} & \ X_{45} \ & 0 \\
\ 3 \ & X_{61} & X_{15} & 0 & 0 &  \ X_{56} \ \\ \hline
\ 4 \ & X_{36} & 0 & 0 & 0 & 0 \\ 
\ 5 \ & 0 & 0 & X_{43} & 0 & 0
\end{array}\right).
\eeq}
The model has 10 perfect matchings and the master space is, as anticipated, a toric 6d CY. This theory has two gauge groups, associated to faces 1 and 2. Performing the further quotient by these symmetries we obtain the moduli space, which is a toric 4d CY. Its toric diagram is given by the matrix

{\small
\beq
G=\left(\begin{array}{cccccc}
0& 1& -1& 0& 0& 0 \\
0& 0& 1& 1& 0& 1 \\
1& 0& 1& 1& 0& 0 \\
0& 0& 0& -1& 1& 0 \\ \hline
\ \bf{3} \ & \ \bf{2} \ & \ \bf{2} \ & \ \bf{1} \ & \ \bf{1} \ & \ \bf{1} \ \\ \hline
\end{array}\right).
\label{G_matrix_multiplicities_Gr24_2loops}
\eeq
}

\begin{figure}[h]
\begin{center}
\includegraphics[width=7cm]{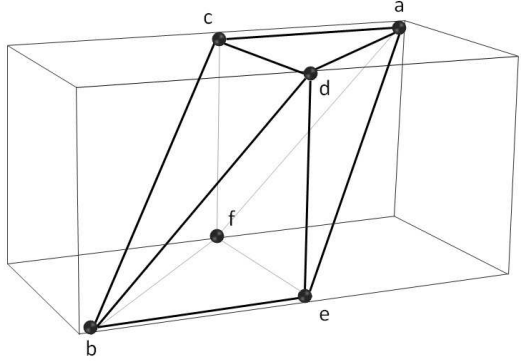}
\caption{General toric diagram for the moduli space of the 4-leg, 2, 3 and 4-loop models. The moduli space is in the three cases the same as the one for the 1-loop model, i.e. the real cone over $Q^{1,1,1}$. Letters indicate the non-trivial perfect matching multiplicities of points in the toric diagram, which depend on the number of loops.}
\label{toric_Gr24_nloops}
\end{center}
\end{figure}

The associated toric diagram is presented in \fref{toric_Gr24_nloops}. Interestingly, the moduli space is exactly the same as for the 1-loop model considered in Section \ref{section_BFTs_and_CYs}, although with different perfect matching multiplicities $(a,b,c,d,e,f)=(2,2,1,1,1,3)$. 

\subsection*{Three Loops}

\begin{figure}[h]
 \centering
 \begin{tabular}[c]{ccc}
 \epsfig{file=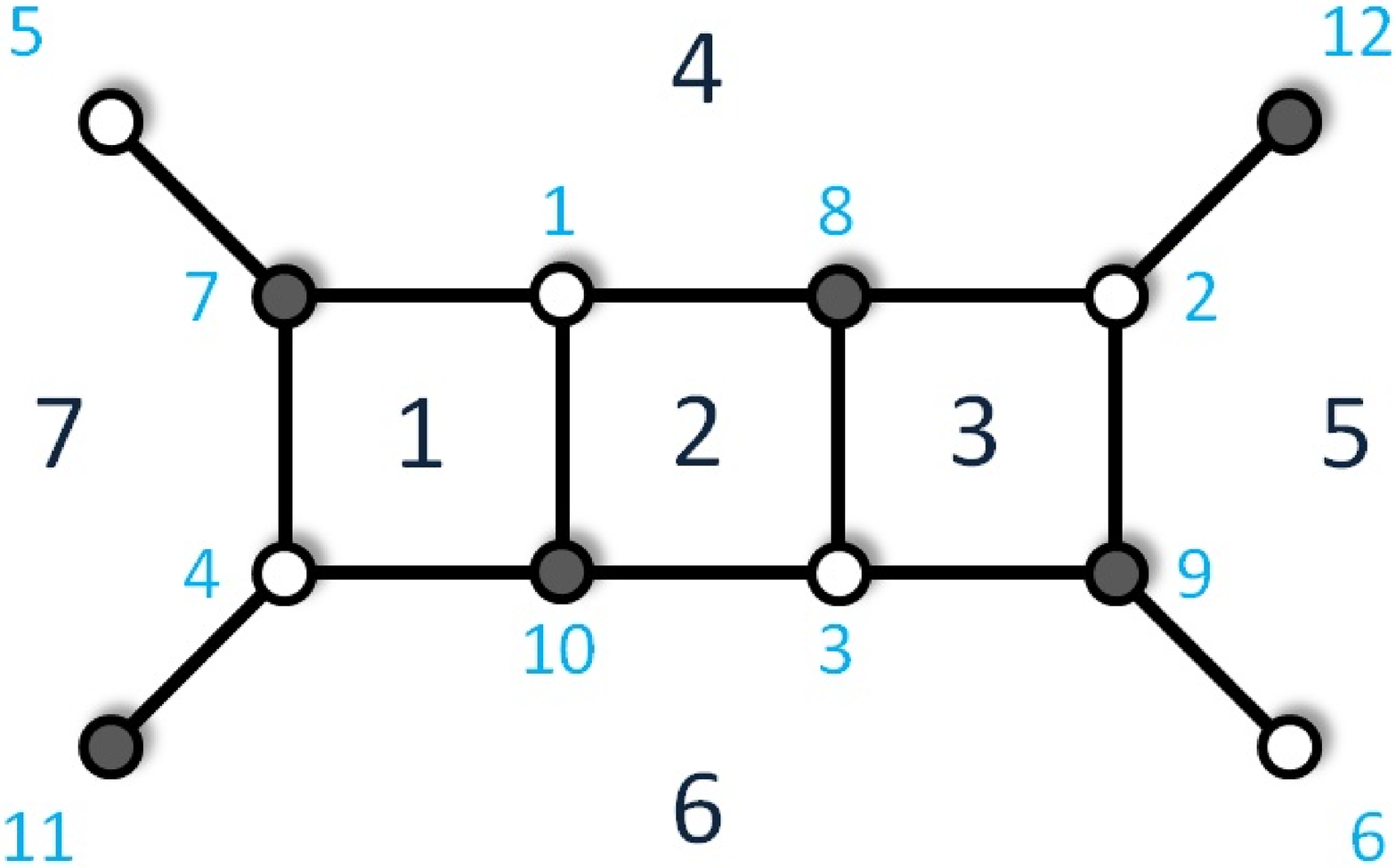,width=0.37\linewidth,clip=} & \ \ \ \ \ &
\epsfig{file=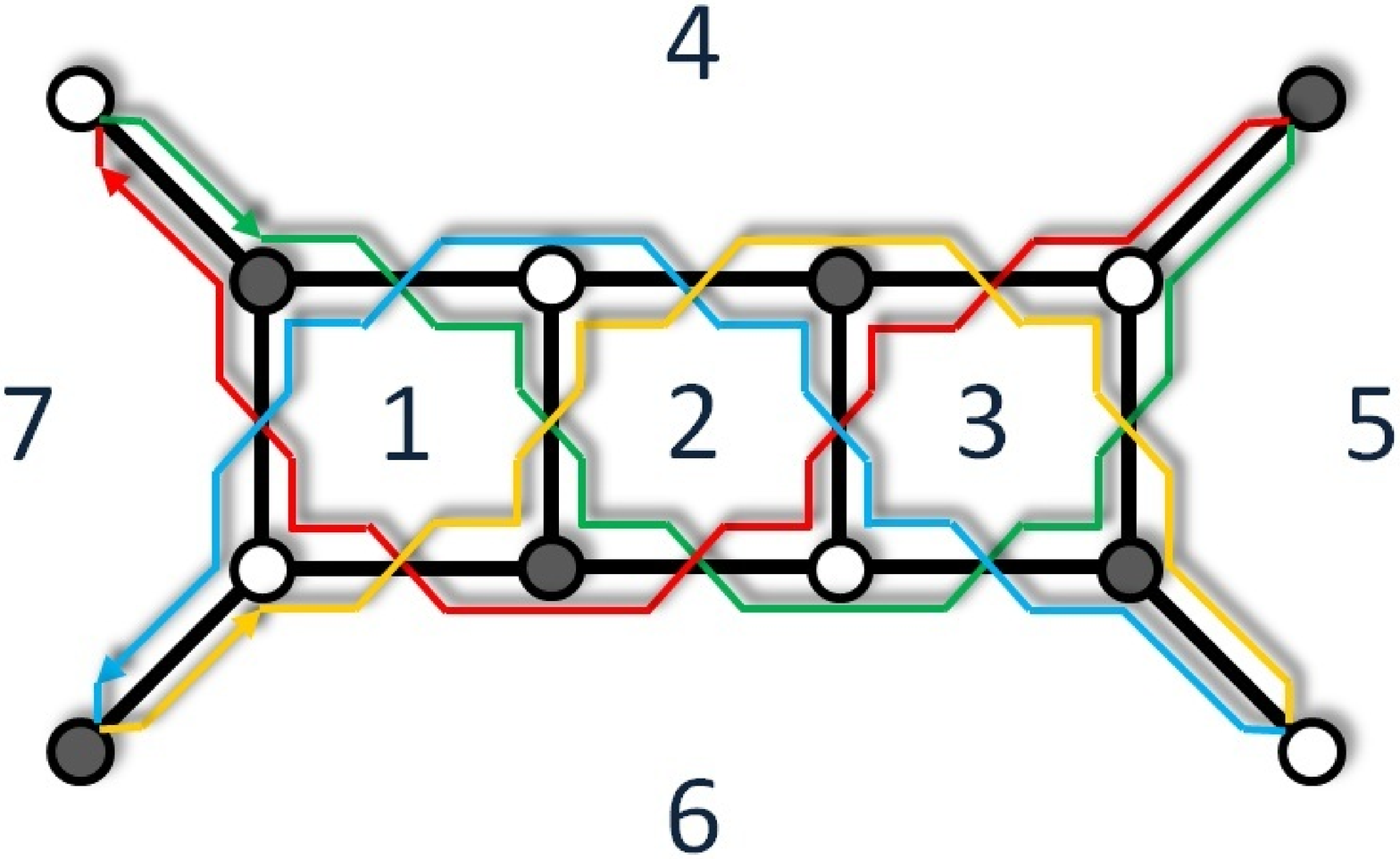,width=0.35\linewidth,clip=} \\ 
\mbox{(a)} & & \mbox{(b)}
 \end{tabular}
\caption{a)	Bipartite graph for the 4-leg, 3-loop model. It contains three internal and four external faces. b) The four zig-zag paths for this model.}
\label{tiling_Gr24_3loops} 
\end{figure}

Let us now quickly analyze the 3-loop model, given by the graph in \fref{tiling_Gr24_3loops}. The master Kasteleyn matrix is
{\small
\beq
K_0 = \left(\begin{array}{c|cccc|cc} 
 & \ \ 7 \ \ & \ \ 8 \ \ & \ \ 9 \ \  & \ \ 10 \ \ & \ \ 11 \ \  & \ \ 12 \ \ \\ \hline
\ 1 \ & \ X_{14} \ & \ X_{42} \ & 0 & \ X_{21} \ & 0 & 0 \\
\ 2 \ & 0 & X_{34} & \ X_{53} \ & 0 & 0 & \ X_{45} \ \\
\ 3 \ & 0 & X_{23} & X_{36} & X_{62} & 0 & 0 \\
\ 4 \ & X_{71} & 0 & 0 & X_{16} & \ X_{67} \ & 0 \\ \hline
\ 5 \ & X_{47} & 0 & 0 & 0 & 0 & 0 \\
\ 6 \ & 0 & 0 & X_{65} & 0 & 0 & 0
\end{array}\right).
\eeq}

The model has 15 perfect matchings and a master space that is a CY 7-fold. After imposing the D-terms associated to the three gauge groups, we see that the moduli space is a CY 4-fold with toric diagram given by
{\small
\beq
G=\left(\begin{array}{cccccc}
1& 0& 0& -1& 0& 0 \\
0& 0& 1& 1& 1& 0 \\
0& 1& 0& 1& 1& 0 \\
0& 0& 0& 0& -1& 1 \\ \hline
\ \bf{5} \ & \ \bf{3} \ & \ \bf{3} \ & \ \bf{2} \ & \ \bf{1} \ & \ \bf{1} \ \\ \hline
\end{array}\right).
\label{G_matrix_multiplicities_Gr24_3loops}
\eeq
}
We conclude that the moduli space of the 3-loop graph is identical to the one for 1 and 2-loops, shown in \fref{toric_Gr24_nloops}, with multiplicities $(a,b,c,d,e,f)=(5,2,1,3,1,3)$.

\subsection*{Four Loops}

\begin{figure}[h]
 \centering
 \begin{tabular}[c]{ccc}
 \epsfig{file=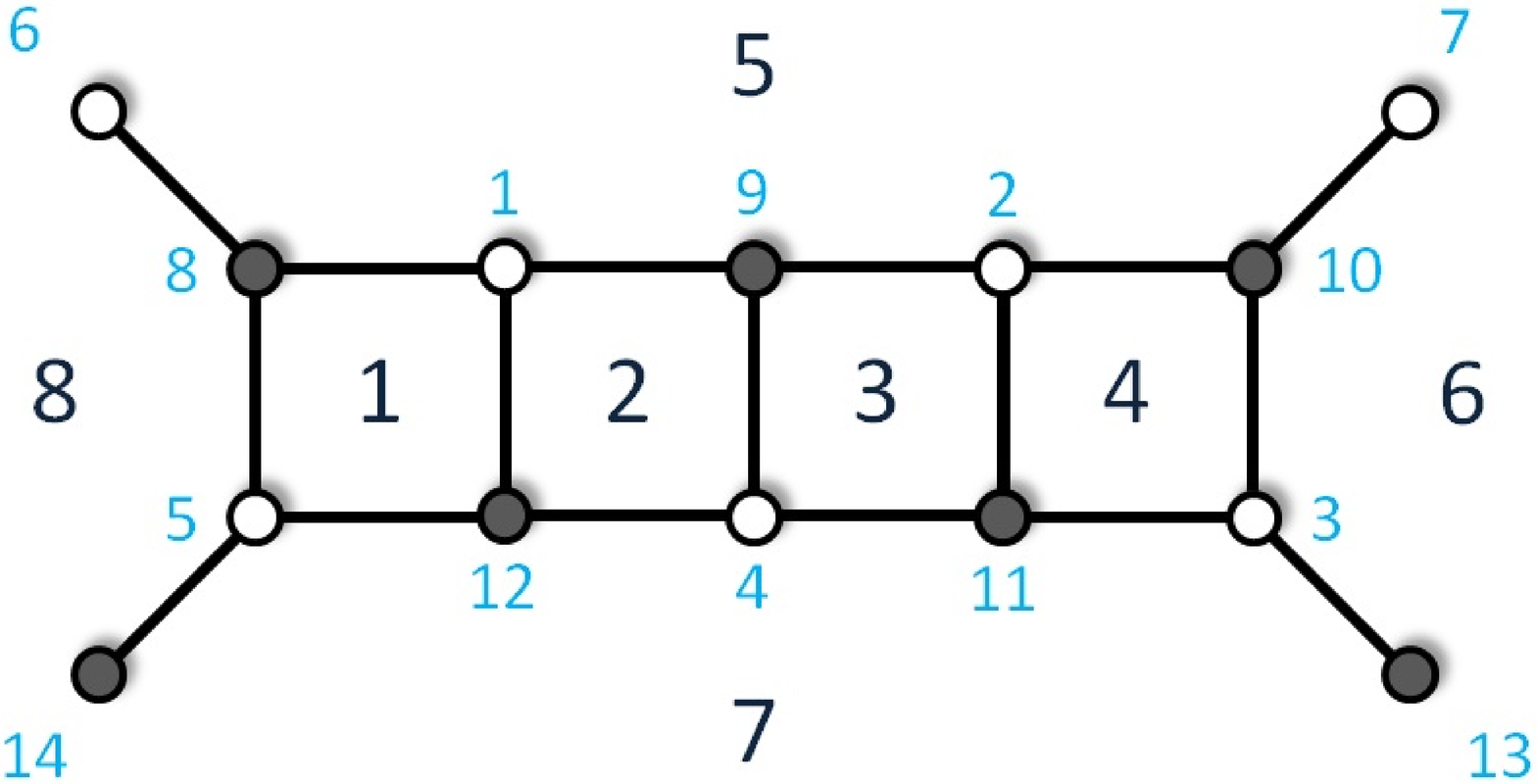,width=0.4\linewidth,clip=} & \ \ \ \ \ &
\epsfig{file=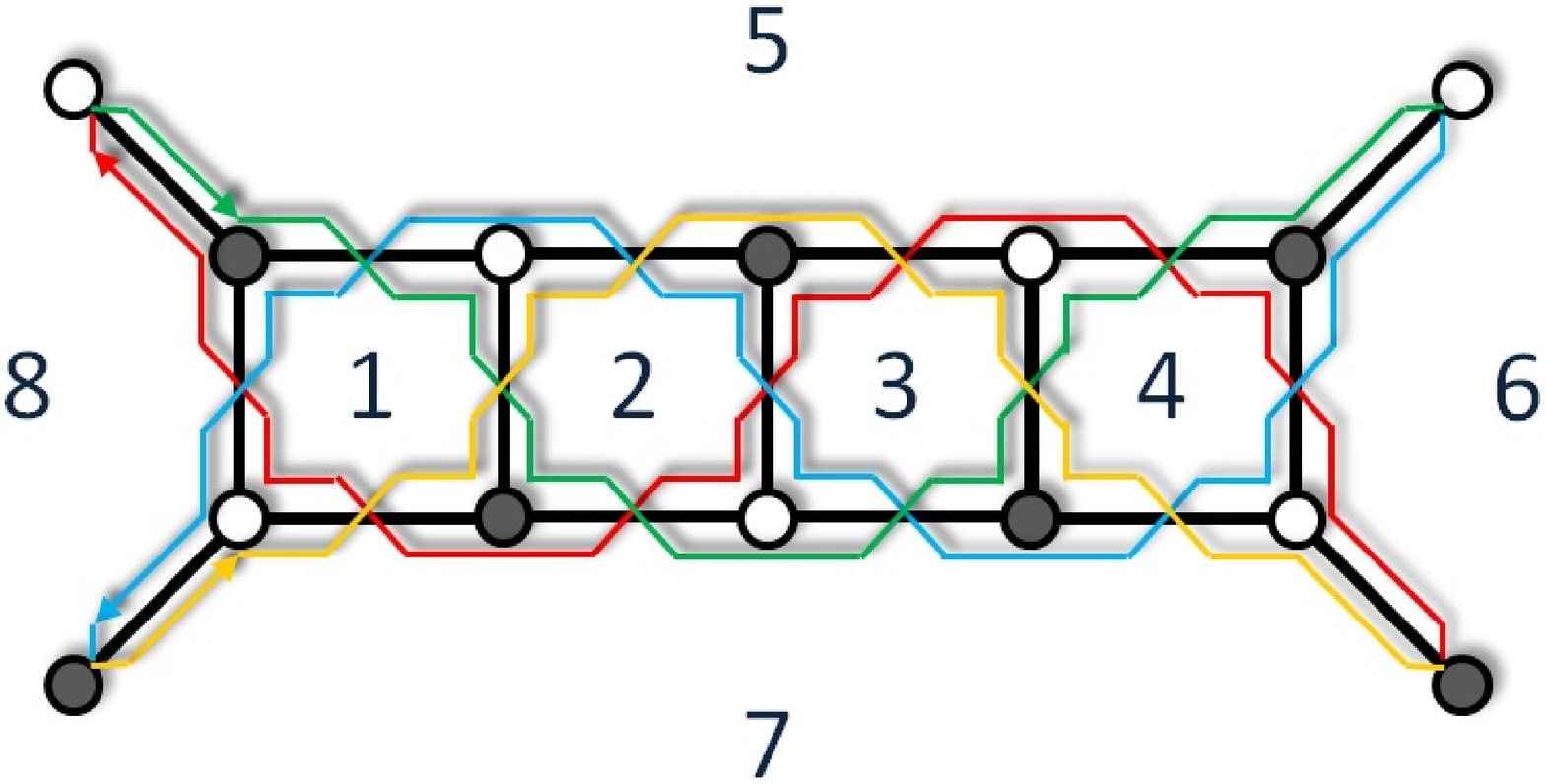,width=0.37\linewidth,clip=} \\ 
\mbox{(a)} & & \mbox{(b)}
 \end{tabular}
\caption{a)	Bipartite graph for the 4-leg, 4-loop model. It contains four internal and four external faces. b) The four zig-zag paths for this model.}
\label{tiling_Gr24_4loops} 
\end{figure}

The 4-loop model corresponds to the graph in \fref{tiling_Gr24_4loops}. The master Kasteleyn matrix is
{\small
\beq
K_0 = \left(\begin{array}{c|ccccc|cc} 
 & \ \ 8 \ \ & \ \ 9 \ \  & \ \ 10 \ \ & \ \ 11 \ \  & \ \ 12 \ \  & \ \ 13 \ \  & \ \ 14 \ \ \\ \hline
\ 1 \ & \ X_{15} \ & \ X_{52} \ & 0& 0& X_{21}& 0& 0 \\
\ 2 \ & 0& X_{35}& \ X_{54} \ & \ X_{43} \ & 0& 0& 0 \\
\ 3 \ & 0& 0& X_{46}& X_{74}& 0& \ X_{67} \ & 0 \\
\ 4 \ & 0& X_{23}& 0& X_{37}& \ X_{72} \ & 0& 0 \\ 
\ 5 \ & X_{81}& 0& 0& 0& X_{17}& 0& \ X_{78} \ \\ \hline
\ 6 \ & X_{58}& 0& 0& 0& 0& 0& 0 \\
\ 7 \ & 0& 0& X_{65}& 0& 0& 0& 0
\end{array}\right).
\eeq}

The model has 23 perfect matchings and a master space that is a CY 8-fold. The moduli space is a CY 4-fold with toric diagram given by

{\small
\beq
G=\left(\begin{array}{cccccc}
1& 0& 0& -1& 0& 0 \\
0& 0& 1& 1& 1& 0 \\
0& 1& 0& 1& 1& 0 \\
0& 0& 0& 0& -1& 1 \\ \hline
\ \bf{8} \ & \ \bf{5} \ & \ \bf{5} \ & \ \bf{3} \ & \ \bf{1} \ & \ \bf{1} \ \\ \hline
\end{array}\right),
\label{G_matrix_multiplicities_Gr24_4loops}
\eeq
\noindent which, once again, precisely agrees with the moduli spaces for the lower loop models, but with multiplicties $(a,b,c,d,e,f)=(8,3,1,5,1,5)$. 

\bigskip

It is natural to expect that the moduli space remains the same, up to perfect matching multiplicities, for arbitrary number of loops. Below we show that this is indeed the case by proving that different loops are connected by Seiberg duality. We conjecture that this behavior is a geometric manifestation of the fact that for a given set of scattered particles the number of leading singularities is finite and determines the scattering amplitude to an arbitrary number of loops. It would be interesting to investigate this phenomenon for other multi-loop diagrams.

\subsection{Loop Reduction and Seiberg Duality}

The fact that BFTs associated to the ladder diagrams with an arbitrary number of loops share the same moduli space suggests that they are connected by Seiberg duality. Since a different number of loops maps to a different number of gauge groups in the BFT, Seiberg duality clearly needs to be supplemented with some additional dynamics, as we now explain.

\begin{figure}[h]
\begin{center}
\includegraphics[width=10.5cm]{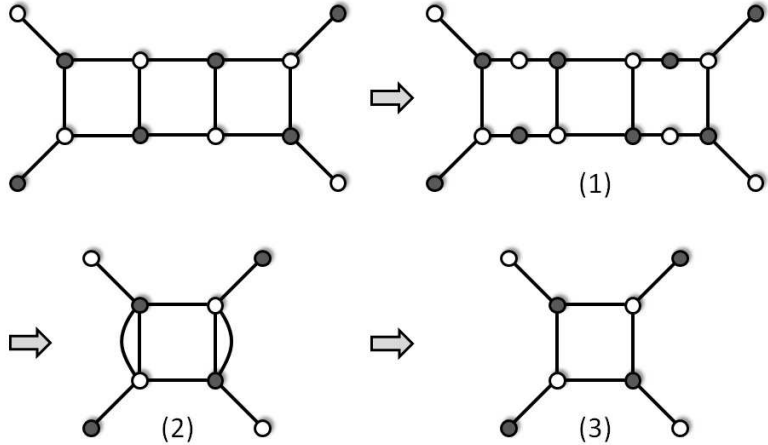}
\caption{An $n$-loop diagram can be turned into an $(n-2)$-loop one by a sequence of steps that in the BFT correspond to: 1) Seiberg duality, 2) Integrating out massive fields, and 3) Confinement of $N_f=N_c$ gauge groups and formation of mesons.}
\label{loop_reduction_Seiberg_duality}
\end{center}
\end{figure}

The $n$-loop diagram is connected to the $(n-2)$-loop one by the sequence of steps shown in \fref{loop_reduction_Seiberg_duality}. In terms of gauge theory dynamics, these steps have the following meaning:

\begin{itemize}
\item[1)] We perform a Seiberg duality on some of the internal faces that are not at the endpoints of the ladder. This transformation is implemented by a square move in the graph and generates four 2-valent nodes.
\item[2)] The 2-valent nodes generated in the previous step correspond to mass terms in the superpotential. We integrate out the massive fields, which maps to condensation of nodes in the graph. When doing so, the number of sides of each of the two faces adjacent to the dualized one is reduced to two.
\item[3)] Internal faces with two sides correspond to $SU(N_c)$ gauge groups with $N_f=N_c$. At low energies, such gauge groups confine and their dynamics is expressed in terms of gauge invariant (under the confined gauge group) mesons and baryons. The graphic implementation of the formation of mesons corresponds to combining the two edges on the boundary of these faces into a single one. This process makes the faces disappear, in agreement with confinement.
\end{itemize}
Iterating this process we can show that all diagrams with an even number of loops give rise to dual gauge theories. Similarly, theories with an odd number of loops are also dual.

\bigskip

\section{The Boundary Operator as Higgsing}

\label{section_boundary_operator}

In Section \ref{section_leading_singularities}, we reviewed the correspondence between bipartite graphs and cells in the positive Grassmannian. In Section \ref{section_master_space} we explained, in terms of the master space, that a cell takes the form of a convex polytope and its boundary is a collection of lower dimensional cells. Being at a boundary of a cell corresponds to setting some of the entries of the corresponding matrix $C$ to zero. The larger the number of vanishing entries is, the lower dimensional the corresponding boundary cell is.

Setting an entry in $C$ to zero corresponds to eliminating the connectivity between the associated external nodes. This is achieved by removing an internal edge in the graph, disrupting oriented paths between the nodes. \fref{Higgsing_perfect_orientation} shows an example of this process. Removing the edge shown in red results in setting $c_{42}$, $c_{62}$, $c_{72}$ and $c_{73}$ to zero in \eref{C_matrix_example}.
The discussion in Section \ref{section_Higgsing} implies that the boundary operator maps to higgsing in the BFT.

\begin{figure}[h]
\begin{center}
\includegraphics[width=5.4cm]{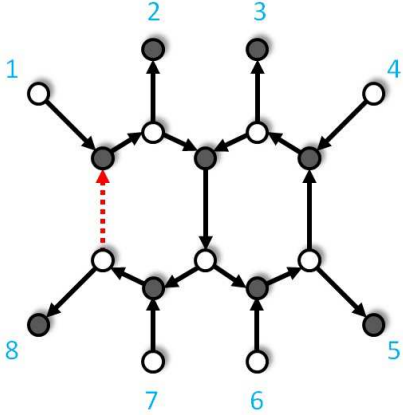}
\caption{An example of higgsing. An edge is removed from the graph, resulting in the disappearance of some oriented paths, associated to a perfect orientation, connecting external nodes.}
\label{Higgsing_perfect_orientation}
\end{center}
\end{figure}

Turning on a non-zero vev for a bifundamentals field $X_0$ determines an energy scale $\langle X_0 \rangle=\Lambda$. Removing the edge associated to $X_0$ from the graph corresponds to considering energies much smaller than $\Lambda$. The surviving graph accurately captures the low energy physics, such as the moduli space of the theory, provided the vevs involved are much smaller than $\Lambda$.

The BFT not only captures the combinatorics of the boundary but also describes the continuous approach to boundary facets.  In fact, identifying expectation values of bifundamental fields with the inverse of edge weights entering \eref{boundary_measure}, we obtain perfect agreement between the higgsing and Grassmannian pictures.\footnote{The inversion of edge weights in this correspondence is not surprising. The structure of F-terms in BFTs is such that there is a trivial $x_e \leftrightarrow x_e^{-1}$ symmetry.} As some expectation value is increased certain entries in $C$ get suppressed, eventually vanishing once the vev is sent to infinity. 

We have discussed how the boundary operator is linked to a simple local operation on the graph: edge removal. Given the map between zig-zag paths and Deligne permutations explained in Section \ref{section_zig_zags}, it is straightforward to see that the boundary operator acts by flipping Deligne permutations \cite{Nima}, according to  
\beq
\begin{array}{ccc}
f_d(a)  & = & b \\ f_d(c)  & = & d
\end{array} \ \ \ 
\underrightarrow{\ \ \partial \ \ } \ \ \ 
\begin{array}{ccc} 
f_d(a)  & = & d \\ f_d(c)  & = & b
\end{array}
\eeq
where $a$, $b$, $c$ and $d$ are the endpoints of two intersecting zig-zag paths. \fref{zig_zag_Higgsing} shows how the permutation flip results from edge removal. Notice that while permutation flip is a global operation in the graph, it is equivalent to removing edges, which is local.

\begin{figure}[h]
\begin{center}
\includegraphics[width=11cm]{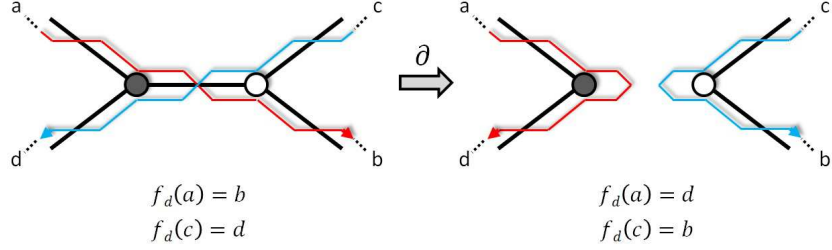}
\caption{Removing an internal edge gives rise to a recombination of zig-zag paths that results in a flip of Deligne permutations.}
\label{zig_zag_Higgsing}
\end{center}
\end{figure}

\subsection{Consistent Higgsing and Untwisting}

We should only consider higgsings that produce consistent graphs, i.e. graphs without self-intersecting zig-zag paths. When an edge is removed, two zig-zags are split at some intermediate points and then recombined as in \fref{zig_zag_Higgsing}. This implies that removing an edge generates self-intersections only when the two zig-zags involved originally have multiple intersections. This situation is sketched in \fref{Higgsing_multiple_intersections}.

\begin{figure}[h]
\begin{center}
\includegraphics[width=11.5cm]{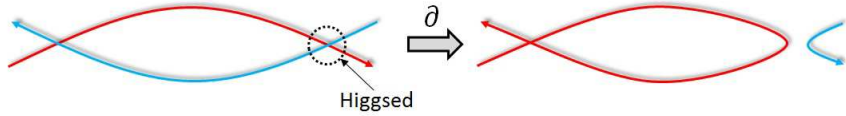}
\caption{Removing an edge between two zig-zag paths with multiple crossings results in a zig-zag with multiple intersections.}
\label{Higgsing_multiple_intersections}
\end{center}
\end{figure}

Zig-zags are not manifest in the graph $G$, and keeping track of them or recomputing them after each higgsing in order to check consistency is rather tedious. This is particularly hard when multiple non-zero vevs are involved. It is then useful to consider the untwisted graph $\tilde{G}$. Zig-zags of $G$ become boundaries of faces (both internal and external) in $\tilde{G}$ and explicit in the graph even after higgsing.\footnote{It is important to emphasize that we do not need to require $\tilde{G}$ to be consistent. In fact, it has self-intersecting zig-zags if the original theory has adjoint fields.} Recombination of zig-zags maps to the recombination of faces and we can efficiently follow them through the process of removing edges.

According to our previous discussion, in order to preserve consistency, edges between zig-zags that intersect more than once cannot be removed. It is then straightforward to identify inconsistent higgsings using $\tilde{G}$: we simply cannot delete edges sitting between faces with multiple intersections. 

Let us illustrate these concepts with an explicit example. Consider \fref{Higgsing_example}, which corresponds to a leading singularity in the scattering of 4 negative and 4 positive helicity gluons at 4-loops. We have labeled edges to facilitate their identification after untwisting.

\begin{figure}[h]
 \centering
 \begin{tabular}[c]{ccc}
 \epsfig{file=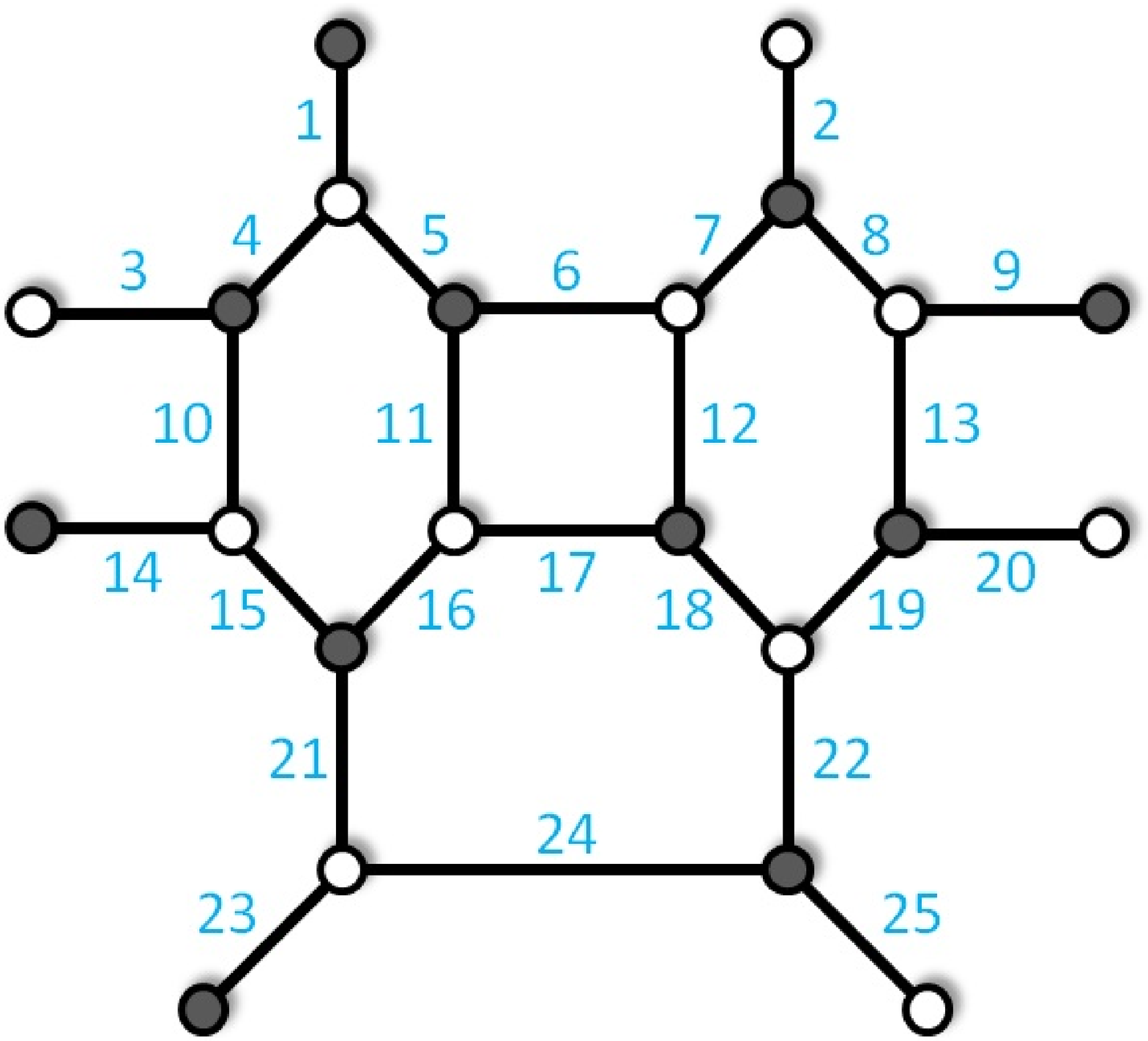,width=0.36\linewidth,clip=} & \ \ \ \ \ &
\epsfig{file=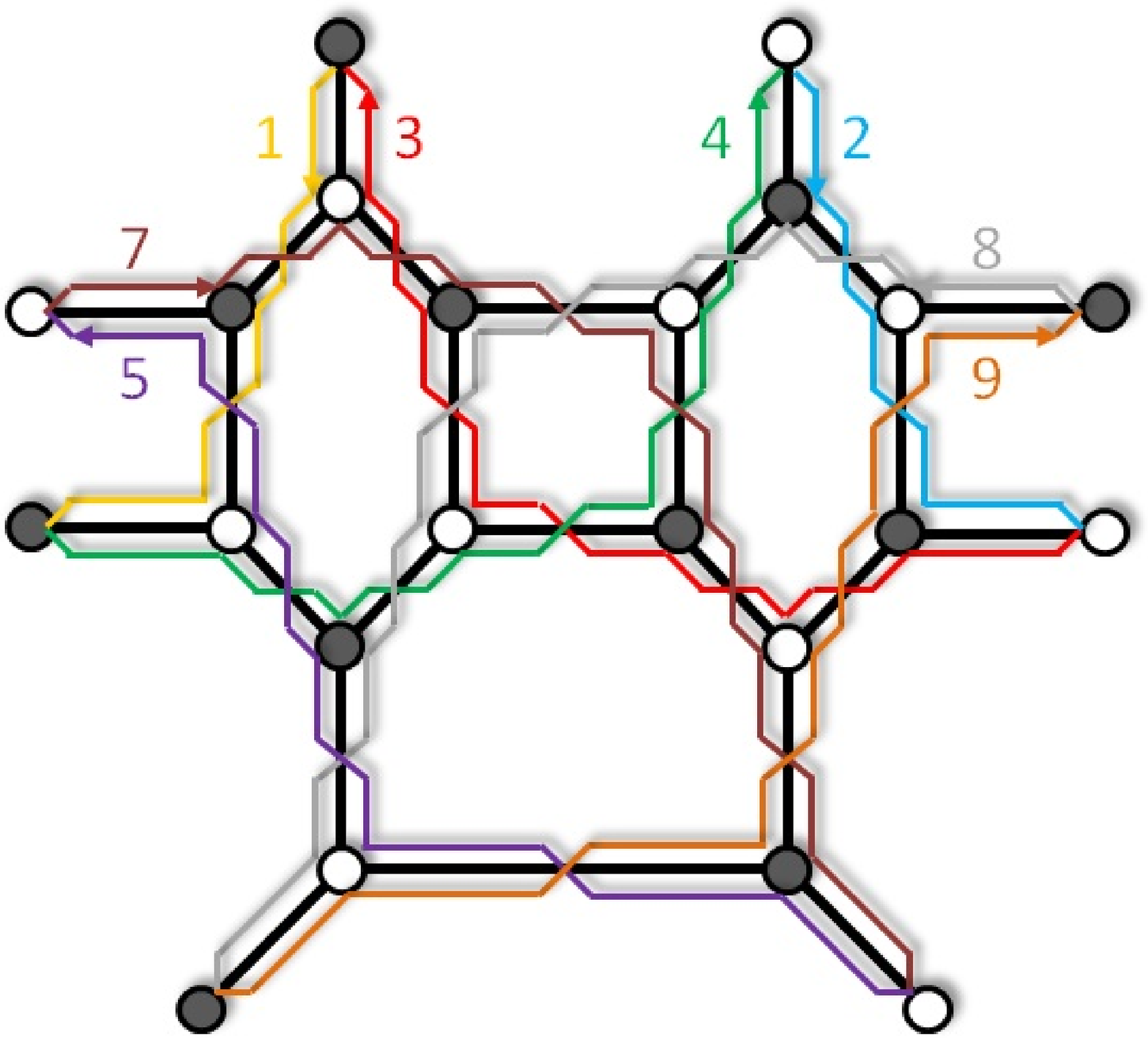,width=0.36\linewidth,clip=} \\ 
\mbox{(a)} & & \mbox{(b)}
 \end{tabular}
\caption{a) A 4-loop diagram associated to the scattering of 4 negative and 4 positive helicity gluons. We have labeled edges in blue. b) The eight zig-zag paths of this model.}
\label{Higgsing_example} 
\end{figure} 

There are eight zig-zag paths: two of length 4, two of length 6, two of length 7 and two of length 8. They are given by the following collections of edges:

\beq
\begin{array}{ccclccccl}
\mbox{{\bf \underline{Length 4}}} & \ & {\bf 1:} & (1,4,10,14) & \ \ \ \ \ \ & \mbox{{\bf \underline{Length 7}}} & \ & {\bf 3:} & (1,5,11,17,18,19,20) \\
& \ & {\bf 2:} & (2,8,13,20) & \ \ \ \ \ \ & & \ & {\bf 4:} & (2,7,12,17,16,15,14) \\ \\
\mbox{{\bf \underline{Length 6}}} & \ & {\bf 5:} & (3,10,15,21,24,25) & \ \ \ \ \ \ & \mbox{{\bf \underline{Length 8}}} & \ & {\bf 7:} & (3,4,5,6,12,18,22,25) \\
& \ & {\bf 6:} & (9,13,19,22,24,23) & \ \ \ \ \ \ & & \ & {\bf 8:} & (9,8,7,6,11,16,21,23) 
\end{array}
\eeq

The untwisted graph $\tilde{G}$ is shown in \fref{Higgsing_example_untwisting}.\footnote{We are extremely grateful to Rak-Kyeong Seong for sharing his expertise in untwisting complicated graphs and for verifying this example.} The eight zig-zag paths of the original graph turn into eight external faces via untwisting. $\tilde{G}$ has three boundaries and lives on a 2-torus. 

\begin{figure}[h]
\begin{center}
\includegraphics[width=9.5cm]{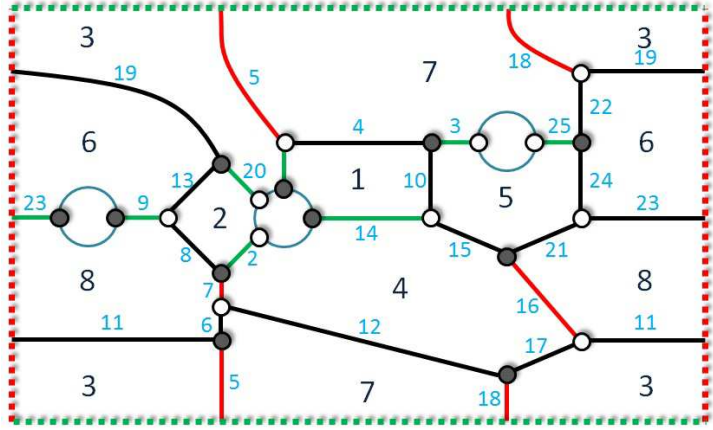}
\caption{The graph $\tilde{G}$ obtained by untwisting the zig-zag paths in \fref{Higgsing_example}. Face labels are those of the corresponding zig-zag paths in the original graph. $\tilde{G}$ contains eight external faces, has three boundaries (represented as blue circles) and lives on a 2-torus.}
\label{Higgsing_example_untwisting}
\end{center}
\end{figure}

As explained above, $\tilde{G}$ can be used to identify, by simple inspection, all edges that cannot be removed without spoiling consistency. They are indicated in color in \fref{Higgsing_example_untwisting}. In green, we show external legs, which correspond to the scattered particles and hence are preserved.  In red, we show the edges that cannot be removed because they sit between faces with more than one intersection. In particular we have edges 5 and 18 between faces 3 and 7, and edges 7 and 16 between faces 4 and 8. While this approach might seem too elaborate for dealing with the removal of single edges, it becomes particularly useful for systematically determining consistent removals of multiple edges.

\subsection{Higgsing and Geometry}

\label{section_Higgsing_and_geometry}

Let us discuss the effect of higgsing on the geometry of the master and moduli spaces. The correspondence between perfect matchings $p_\mu$, which should be regarded as GLSM fields, and a chiral field $X_i$ in the quiver is given by \eref{X_pm_map}, which we reproduce here for convenience
\beq
X_i = \prod_{\mu=1}^c p_\mu^{P_{i\mu}}.
\eeq
This means that, in order for $X_i$ to acquire a non-zero vev, all perfect matchings with $P_{i\mu}=1$, i.e. all the ones that contain the corresponding edge in the graph, need to get a non-zero vev and are eliminated. This removal of perfect matchings results in the disappearance of some points in the toric diagrams of the master and moduli spaces, and in a reduction in the multiplicity of others.

While higgsing removes points from the toric diagrams of the master and moduli spaces, their qualitative behavior is different. Higgsing decreases the number of internal faces in $G$ by one, which implies that the dimensionality of the master space is reduced by the same amount. The toric diagram of the master space of the higgsed theory is, as expected, a lower dimensional sub-cell on the boundary of the original one. On the other hand, the number of D-term equations is also reduced by one when higgsings, which implies that the dimension of the moduli space remains constant.

\section{Conclusions and Outlook}

\label{section_conclusions}

In this paper, we have introduced and started the study of Bipartite Field Theories, a general class of 4d, $\mathcal{N}=1$ quiver gauge theories defined by a partite graph on a Riemann surface. We explained the map between the field theory dynamics and graph modifications, the emergence of toric CY manifolds as the master and moduli spaces of the gauge theory, the connection between Seiberg duality and square moves and interpretation of the boundary operator on cells in the positive Grassmannian as higgsing. While our discussion has been completely general, most of our explicit examples have boundaries, i.e. they were relevant for scattering amplitudes.

We developed specific tools, in the form of generalized Kasteleyn matrix techniques, for the study of general BFTs. We extensively demonstrated in examples how these methods allow the explicit determination of master and moduli spaces, even for complicated graphs. Finding the corresponding CY manifolds is reduced to computing determinants of rather simple matrices.

We have studied models that go beyond the planar limit, by increasing the number of boundaries and the genus of the underlying Riemann surface. From a BFT viewpoint, they are not more involved than planar graphs. As discussed in Section \ref{section_Seiberg_duality}, it is in the context of non-planar graphs that the moduli space of the BFT more strikingly shows its power for identifying theories related by square moves, since they cannot be understood in terms of permutations as in the planar case.

Our work suggests various interesting directions for future investigation, some of which are summarized below:

\begin{itemize}
\item One of the main lessons of our work is the existence of a deep connection between Calabi-Yau manifolds, which appear in the form of master and moduli spaces of BFTs, and leading singularities. As explained in Sections \ref{section_master_space} and \ref{section_boundary_operator}, the master space is the natural object for describing cells in the Grassmannian and their boundaries. But, to us, the moduli space appears to be a more fundamental object, since it is invariant under all equivalence transformations of the graph. It is natural to expect these geometries to play central role in the study of leading singularities, yet to be unveiled in full generality. In the restricted case of graphs on a disk, the toric diagram of the moduli space exhibits striking similarities with the {\it matroid polytope} of \cite{Postnikov_toric}. The connection between these two objects will be explained in \cite{FGS}. 

\item For graphs without boundaries on $T^2$, there is an intimate connection between zig-zag paths and the geometry of the moduli space of the associated BFT, which is a CY 3-fold. Zig-zags are in one-to-one correspondence with $(p,q)$-legs normal to the faces of the toric diagram. It would be interesting to understand whether, and if so how, zig-zags on bipartite graphs on generic Riemann surfaces with boundaries are related to the moduli spaces of the corresponding BFTs. We expect that understanding this connection, in conjunction with the previous point, will lead to a map between invariants, global properties of zig-zags and CY manifolds.

\item A string theory embedding is known for certain classes of BFTs. This is the case for those associated to graphs without boundaries on $T^2$, which arise on the worldvolume of D3-branes probing toric CY 3-folds. In these cases, the graph can be interpreted as a physical web of NS5-branes from which D5-branes are suspended, connected by two T-dualities to the original configuration of D3-branes. In addition, the graph on a (typically) higher genus Riemann surface that is obtained from it by the untwisting map describes a configuration of D6-branes on the mirror manifold. It would be extremely interesting to investigate whether additional sub-classes of BFTs admit a string theory realization.

\item In Section \ref{section_Seiberg_duality}, we discussed how different BFTs can give rise to the same CY manifolds as their moduli spaces. We can regard such equivalence as a generalized version of Toric Duality \cite{Feng:2000mi,Beasley:2001zp,Feng:2001bn,Feng:2002zw}. In Section \ref{section_loop_reduction}, we presented explicit examples showing that even graphs with a different number of loops can lead to the same CY. There is a priori no obvious reason preventing graphs with different numbers of boundaries and/or genus from producing the same CY.\footnote{In fact we know that a similar behavior is possible for the master space: specular duals, whose underlying graphs are related by untwisting zig-zag paths, share the same CY as the master space \cite{Hanany:2012vc}.} It would be interesting to search for explicit examples realizing this behavior and, if they indeed exist, to understand what its physical interpretation is.

\item The Grassmannian $G(k,n)$ also arises as the moduli space of $k$ $U(n)$ vortices. Given the connection between cells in the Grassmannian and bipartite graphs and the fact that any such graph can be reduced to one only involving 2 and 3-valent nodes, it would be interesting to explore whether some natural decomposition of the moduli space, in which $U(2)$ and $U(3)$ vortices are basic building blocks, exist.

\end{itemize}

We foresee BFTs will provide useful insights and guidance and will fuel future developments in the study of systems associated to bipartite graphs, in particular in the area of leading singularities in scattering amplitudes.

\bigskip

\section*{Acknowledgments}

We would like to thank J. Bourjaily, J. Kaplan and L. Williams for useful correspondence and explanations, and D. Galloni and R.-K. Seong for interesting discussions and collaboration on a related project. We are particularly thankful to N. Arkani-Hamed for very useful correspondence. Finally, we thank D. Xie and M. Yamazaki for informing us about their upcoming work \cite{Xie:2012mr}, and sharing a draft of it prior to publication. This work was supported by the US DOE under contract number DE-AC02-76SF00515 and by the U.K. Science and Technology Facilities Council (STFC).

\bigskip


\appendix

\section{Master and Moduli Spaces of the Hexagon-Square Model}

\label{appendix_4legs_2loops}

The bipartite graph for this model is shown in \fref{tiling_hexagon_square}. The master Kasteleyn matrix for this model is
{\small
\beq
K_0 = \left(\begin{array}{c|cccc|ccc} 
 & \ \ 8 \ \ & \ \ 9 \ \ & \ \ 10 \ \ & \ \ 11 \ \  & \ \ 12 \ \ & \ \ 13 \ \ & \ \ 14 \ \ \\ \hline
\ 1 \ & \ X_{31} \ & 0& 0& \ X_{18} \ & \ X_{83} \ & 0& 0 \\
\ 2 \ & X_{14}& \ X_{42} \ & \ X_{21} \ & 0& 0& 0& 0 \\
\ 3 \ & 0& X_{25}& X_{62}& 0& 0& \ X_{56}\ & 0 \\
\ 4 \ & 0& 0& X_{16}& X_{71}& 0& 0& \ X_{67} \ \\ \hline
\ 5 \ & X_{43}& 0& 0& 0& 0& 0& 0 \\
\ 6 \ & 0& X_{54}& 0& 0& 0& 0& 0 \\
\ 7 \ & 0& 0& 0& X_{87}& 0& 0& 0
\end{array}\right),
\eeq}
from which we determine
{\small
\begin{eqnarray}
\mathcal{P} & = & -X_{14} X_{16} X_{18} X_{25} - X_{16} X_{18} X_{42} X_{43} X_{56} + X_{14} X_{16} X_{18} X_{54} X_{56} - 
 X_{18} X_{21} X_{25} X_{43} X_{67} \nonumber \\ & & - X_{18} X_{21} X_{43} X_{54} X_{56} X_{67} + X_{18} X_{42} X_{43} X_{62} X_{67} - X_{14} X_{18} X_{54} X_{62} X_{67} - X_{21} X_{25} X_{31} X_{71} \nonumber \\ & & + X_{21} X_{31} X_{54} X_{56} X_{71} + 
 X_{31} X_{42} X_{62} X_{71} + X_{21} X_{25} X_{43} X_{71} X_{83} - X_{21} X_{43} X_{54} X_{56} X_{71} X_{83} \nonumber \\ & & - 
 X_{42} X_{43} X_{62} X_{71} X_{83} + X_{14} X_{54} X_{62} X_{71} X_{83} + X_{16} X_{31} X_{42} X_{56} X_{87} + 
 X_{21} X_{25} X_{31} X_{67} X_{87} \nonumber \\ & & - X_{21} X_{31} X_{54} X_{56} X_{67} X_{87} - X_{31} X_{42} X_{62} X_{67} X_{87} +
  X_{14} X_{16} X_{25} X_{83} X_{87} - X_{16} X_{42} X_{43} X_{56} X_{83} X_{87} \nonumber \\ & & + X_{14} X_{16} X_{54} X_{56} X_{83} X_{87} - X_{21} X_{25} X_{43} X_{67} X_{83} X_{87} + X_{21} X_{43} X_{54} X_{56} X_{67} X_{83} X_{87} \nonumber \\ & & + X_{42} X_{43} X_{62} X_{67} X_{83} X_{87} - 
 X_{14} X_{54} X_{62} X_{67} X_{83} X_{87}. 
\label{P_polynomial_hexagon_square}
\end{eqnarray}
}
I.e. the model has 25 perfect matchings. The $P$ matrix becomes:

{\scriptsize
\beq
P=\left(
\begin{array}{c|ccccccccccccccccccccccccc}
& \ p_1 \ & \ p_2 \ & \ p_3 \ & \ p_4 \ & \ p_5 \ & \ p_6 \ & \ p_7 \ & \ p_8 \ & \ p_9 \ & p_{10} & p_{11} & p_{12} & p_{13} & p_{14} & p_{15} & p_{16} & p_{17} & p_{18} & p_{19} & p_{20} & p_{21} & p_{22} & p_{23} & p_{24} & p_{25} \\ \hline
\ X_{14} \ & 1& 1& 1& 1& 1& 1& 1& 0& 0& 0& 0& 0& 0& 0& 0& 0& 0& 0& 0& 0& 0& 0& 0& 0& 0 \\
X_{16} & 1& 1& 1& 1& 0& 0& 0& 1& 1& 1& 0& 0& 0& 0& 0& 0& 0& 0& 0& 0& 0& 0& 0& 0& 0 \\
X_{18} & 1& 1& 0& 0& 1& 0& 0& 1& 0& 0& 1& 1& 1& 0& 0& 0& 0& 0& 0& 0& 0& 0& 0& 0& 0 \\
X_{25} & 1& 0& 1& 0& 0& 0& 0& 0& 0& 0& 1& 0& 0& 1& 1& 1& 1& 0& 0& 0& 0& 0& 0& 0& 0 \\
X_{42} & 0& 0& 0& 0& 0& 0& 0& 1& 1& 1& 0& 1& 0& 0& 0& 0& 0& 1& 1& 1& 1& 0& 0& 0& 0 \\
X_{43} & 0& 0& 0& 0& 0& 0& 0& 1& 1& 0& 1& 1& 1& 1& 1& 0& 0& 1& 1& 0& 0& 1& 1& 0& 0 \\
X_{56} & 0& 1& 0& 1& 0& 0& 0& 1& 1& 1& 0& 0& 1& 0& 0& 0& 0& 0& 0& 0& 0& 1& 1& 1& 1 \\
X_{54} & 0& 1& 0& 1& 1& 1& 1& 0& 0& 0& 0& 0& 1& 0& 0& 0& 0& 0& 0& 0& 0& 1& 1& 1& 1 \\
X_{21} & 0& 0& 0& 0& 0& 0& 0& 0& 0& 0& 1& 0& 1& 1& 1& 1& 1& 0& 0& 0& 0& 1& 1& 1& 1 \\
X_{67} & 0& 0& 0& 0& 1& 1& 0& 0& 0& 0& 1& 1& 1& 1& 0& 1& 0& 1& 0& 1& 0& 1& 0& 1& 0 \\ 
X_{62} & 0& 0& 0& 0& 1& 1& 1& 0& 0& 0& 0& 1& 0& 0& 0& 0& 0& 1& 1& 1& 1& 0& 0& 0& 0 \\
X_{31} & 0& 0& 0& 0& 0& 0& 0& 0& 0& 1& 0& 0& 0& 0& 0& 1& 1& 0& 0& 1& 1& 0& 0& 1& 1 \\
X_{71} & 0& 0& 0& 0& 0& 0& 1& 0& 0& 0& 0& 0& 0& 0& 1& 0& 1& 0& 1& 0& 1& 0& 1& 0& 1 \\
X_{83} & 0& 0& 1& 1& 0& 1& 1& 0& 1& 0& 0& 0& 0& 1& 1& 0& 0& 1& 1& 0& 0& 1& 1& 0& 0 \\
X_{87} & 0& 0& 1& 1& 0& 1& 0& 0& 1& 1& 0& 0& 0& 1& 0& 1& 0& 1& 0& 1& 0& 1& 0& 1& 0
\end{array}
\right).
\eeq}

\noindent We can get a better idea of the toric diagram of the master space by considering the row-reduced version of this matrix

{\scriptsize
\beq
G_{mast}=\left(
\begin{array}{ccccccccccccccccccccccccc}
\ p_1 \ & \ p_2 \ & \ p_3 \ & \ p_4 \ & \ p_5 \ & \ p_6 \ & \ p_7 \ & \ p_8 \ & \ p_9 \ & p_{10} & p_{11} & p_{12} & p_{13} & p_{14} & p_{15} & p_{16} & p_{17} & p_{18} & p_{19} & p_{20} & p_{21} & p_{22} & p_{23} & p_{24} & p_{25} \\ \hline
1& 0& 0& -1& 0& -1& 0& 0& -1& 0& 0& 0& -1& -1& 0& 0& 1& -1& 0& 0& 1& -2& -1& -1& 0 \\
0& 1& 0& 1& 0& 0& 0& 0& 0& 0& 0& -1& 1& 0& 0& 0& 0& -1& -1& -1& -1& 1& 1& 1& 1 \\
0& 0& 1& 1& 0& 1& 0& 0& 1& 0& 0& 0& 0& 1& 0& 0& -1& 1& 0& 0& -1& 1& 0& 0& -1 \\
0& 0& 0& 0& 1& 1& 0& 0& 0& 0& 0& 1& 0& 0& -1& 0& -1& 1& 0& 1& 0& 0& -1& 0& -1 \\
0& 0& 0& 0& 0& 0& 1& 0& 0& 0& 0& 0& 0& 0& 1& 0& 1& 0& 1& 0& 1& 0& 1& 0& 1 \\
0& 0& 0& 0& 0& 0& 0& 1& 1& 0& 0& 1& 0& 0& 0& -1& -1& 1& 1& 0& 0& 0& 0& -1& -1 \\
0& 0& 0& 0& 0& 0& 0& 0& 0& 1& 0& 0& 0& 0& 0& 1& 1& 0& 0& 1& 1& 0& 0& 1& 1 \\
0& 0& 0& 0& 0& 0& 0& 0& 0& 0& 1& 0& 1& 1& 1& 1& 1& 0& 0& 0& 0& 1& 1& 1& 1
\end{array}
\right).
\eeq
}

\noindent The master space is hence an 8d toric CY. The charge matrix encoding F-terms is $Q_F = Ker \, P$, which becomes

{\scriptsize
\beq
Q_F=\left(
\begin{array}{ccccccccccccccccccccccccc}
\ p_1 \ & \ p_2 \ & \ p_3 \ & \ p_4 \ & \ p_5 \ & \ p_6 \ & \ p_7 \ & \ p_8 \ & \ p_9 \ & p_{10} & p_{11} & p_{12} & p_{13} & p_{14} & p_{15} & p_{16} & p_{17} & p_{18} & p_{19} & p_{20} & p_{21} & p_{22} & p_{23} & p_{24} & p_{25} \\ \hline
0 & -1 & 1 & 0 & 1 & 0 & -1 & 1 & 0 & -1 & -1 & 0 & 0 & 0 & 0 & 0 & 0 & 0 & 0 & 0 & 0 & 0 & 0 & 0 & 1 \\
1 & -1 & 0 & 0 & 0 & 0 & 0 & 1 & 0 & -1 & -1 & 0 & 0 & 0 & 0 & 0 & 0 & 0 & 0 & 0 & 0 & 0 & 0 & 1 & 0 \\
1 & -1 & 0 & 0 & 1 & 0 & -1 & 0 & 0 & 0 & -1 & 0 & 0 & 0 & 0 & 0 & 0 & 0 & 0 & 0 & 0 & 0 & 1 & 0 & 0 \\
2 & -1 & -1 & 0 & 0 & 0 & 0 & 0 & 0 & 0 & -1 & 0 & 0 & 0 & 0 & 0 & 0 & 0 & 0 & 0 & 0 & 1 & 0 & 0 & 0 \\
-1 & 1 & 1 & 0 & 0 & 0 & -1 & 0 & 0 & -1 & 0 & 0 & 0 & 0 & 0 & 0 & 0 & 0 & 0 & 0 & 1 & 0 & 0 & 0 & 0 \\
0 & 1 & 0 & 0 & -1 & 0 & 0 & 0 & 0 & -1 & 0 & 0 & 0 & 0 & 0 & 0 & 0 & 0 & 0 & 1 & 0 & 0 & 0 & 0 & 0 \\
0 & 1 & 0 & 0 & 0 & 0 & -1 & -1 & 0 & 0 & 0 & 0 & 0 & 0 & 0 & 0 & 0 & 0 & 1 & 0 & 0 & 0 & 0 & 0 & 0 \\
1 & 1 & -1 & 0 & -1 & 0 & 0 & -1 & 0 & 0 & 0 & 0 & 0 & 0 & 0 & 0 & 0 & 1 & 0 & 0 & 0 & 0 & 0 & 0 & 0 \\
-1 & 0 & 1 & 0 & 1 & 0 & -1 & 1 & 0 & -1 & -1 & 0 & 0 & 0 & 0 & 0 & 1 & 0 & 0 & 0 & 0 & 0 & 0 & 0 & 0 \\
0 & 0 & 0 & 0 & 0 & 0 & 0 & 1 & 0 & -1 & -1 & 0 & 0 & 0 & 0 & 1 & 0 & 0 & 0 & 0 & 0 & 0 & 0 & 0 & 0 \\
0 & 0 & 0 & 0 & 1 & 0 & -1 & 0 & 0 & 0 & -1 & 0 & 0 & 0 & 1 & 0 & 0 & 0 & 0 & 0 & 0 & 0 & 0 & 0 & 0 \\
1 & 0 & -1 & 0 & 0 & 0 & 0 & 0 & 0 & 0 & -1 & 0 & 0 & 1 & 0 & 0 & 0 & 0 & 0 & 0 & 0 & 0 & 0 & 0 & 0 \\
1 & -1 & 0 & 0 & 0 & 0 & 0 & 0 & 0 & 0 & -1 & 0 & 1 & 0 & 0 & 0 & 0 & 0 & 0 & 0 & 0 & 0 & 0 & 0 & 0 \\
0 & 1 & 0 & 0 & -1 & 0 & 0 & -1 & 0 & 0 & 0 & 1 & 0 & 0 & 0 & 0 & 0 & 0 & 0 & 0 & 0 & 0 & 0 & 0 & 0 \\
1 & 0 & -1 & 0 & 0 & 0 & 0 & -1 & 1 & 0 & 0 & 0 & 0 & 0 & 0 & 0 & 0 & 0 & 0 & 0 & 0 & 0 & 0 & 0 & 0 \\
1 & 0 & -1 & 0 & -1 & 1 & 0 & 0 & 0 & 0 & 0 & 0 & 0 & 0 & 0 & 0 & 0 & 0 & 0 & 0 & 0 & 0 & 0 & 0 & 0 \\
1 & -1 & -1 & 1 & 0 & 0 & 0 & 0 & 0 & 0 & 0 & 0 & 0 & 0 & 0 & 0 & 0 & 0 & 0 & 0 & 0 & 0 & 0 & 0 & 0 
\end{array}
\right).
\eeq
}

This theory has two gauge groups. The matrix associated to D-terms can be chosen to be:

{\scriptsize
\beq
Q_D=\left(
\begin{array}{ccccccccccccccccccccccccc}
\ p_1 \ & \ p_2 \ & \ p_3 \ & \ p_4 \ & \ p_5 \ & \ p_6 \ & \ p_7 \ & \ p_8 \ & \ p_9 \ & p_{10} & p_{11} & p_{12} & p_{13} & p_{14} & p_{15} & p_{16} & p_{17} & p_{18} & p_{19} & p_{20} & p_{21} & p_{22} & p_{23} & p_{24} & p_{25} \\ \hline
0 & 0 & 0 & 0 & 0 & 0 & 1 & 0 & 0 & 1 & 0 & 0 & 1 & 0 & 0 & 0 & 0 & 0 & 0 & 0 & -1 & 0 & -1 & -1 & 0 \\
 0 & 0 & 0 & 0 & 0 & 0 & 0 & 0 & 0 & 0 & 0 & 0 & 0 & 0 & 0 & 0 & 1 & 0 & 0 & 0 & -1 & 0 & 0 & 0 & 0 
\end{array}
\right).
\eeq
}

Combining $Q_F$ and $Q_D$ into the total charge matrix $Q$, we obtain the matrix $G=Ker \, Q$ defining the toric diagram of the moduli space

{\scriptsize
\beq
G=\left(
\begin{array}{ccccccccccccccccccccccccc}
\ p_1 \ & \ p_2 \ & \ p_3 \ & \ p_4 \ & \ p_5 \ & \ p_6 \ & \ p_7 \ & \ p_8 \ & \ p_9 \ & p_{10} & p_{11} & p_{12} & p_{13} & p_{14} & p_{15} & p_{16} & p_{17} & p_{18} & p_{19} & p_{20} & p_{21} & p_{22} & p_{23} & p_{24} & p_{25} \\ \hline
0 & 1 & -1 & 0 & 0 & -1 & 0 & 0 & -1 & 0 & -1 & -1 & 0 & -2 & -1 & -1 & 0 & -2 & -1 & -1 & 0 & -1 & 0 & 0 & 1 \\
0 & 0 & 1 & 1 & 1 & 2 & 1 & -1 & 0 & 0 & 0 & 0 & 0 & 1 & 0 & 1 & 0 & 1 & 0 & 1 & 0 & 1 & 0 & 1 & 0 \\
0 & 0 & 1 & 1 & 0 & 1 & 1 & 0 & 1 & 0 & 0 & 0 & 0 & 1 & 1 & 0 & 0 & 1 & 1 & 0 & 0 & 1 & 1 & 0 & 0 \\
1 & 0 & 1 & 0 & 1 & 1 & 1 & 0 & 0 & 0 & 1 & 1 & 0 & 1 & 1 & 1 & 1 & 1 & 1 & 1 & 1 & 0 & 0 & 0 & 0 \\
0 & 0 & -1 & -1 & 0 & -1 & -1 & 1 & 0 & 0 & 1 & 1 & 1 & 0 & 0 & 0 & 0 & 0 & 0 & 0 & 0 & 0 & 0 & 0 & 0 \\
0 & 0 & 0 & 0 & -1 & -1 & -1 & 1 & 1 & 1 & 0 & 0 & 0 & 0 & 0 & 0 & 0 & 0 & 0 & 0 & 0 & 0 & 0 & 0 & 0 
\end{array}
\right).
\eeq
}
In \eref{G_matrix_multiplicities_hexagon_square}, we presented a reduced version of this matrix in which we only show the different column vectors and indicate their multiplicity.


\end{document}